\documentclass[nofootinbib,preprintnumbers,superscriptaddress,twocolumn,aps,prc,10pt]{revtex4-2}
\usepackage{hyperref}
\usepackage{amsmath}
\usepackage{amssymb}
\usepackage[T1]{fontenc}
\usepackage{graphicx}
\usepackage{mathrsfs}
\usepackage{bm}
\usepackage{color}
\usepackage{dcolumn}
\usepackage{ulem}
\usepackage{xcolor}

\def\ve#1{{\bm{#1}}}
\def\nuc#1#2#3{{}^{#2}_{#3}\mathrm{#1}}
\def\urm#1{\scriptstyle{\text{\textrm{\textmd{\textup{#1}}}}}}
\def\uurm#1{\scriptscriptstyle{\text{\textrm{\textmd{\textup{#1}}}}}}
\def\avr#1{\left\langle{#1}\right\rangle}
\def\Nabla{\bm{\nabla}}
\def\ca#1{{\mathcal{#1}}}

\begin{document}
\preprint{RIKEN-iTHEMS-Report-25}
\title{Charge symmetry breaking effects of $\omega$-$\rho^0$ mixing in relativistic mean-field model}
\author{Yusuke Tanimura}
\affiliation{
  Department of Physics and Origin of Matter and Evolution of Galaxies Institute,
  Soongsil University, Seoul 06978, Korea}
\author{Tomoya Naito}
\affiliation{
  RIKEN Center for Interdisciplinary Theoretical and Mathematical Sciences (iTHEMS),
  Wako 351-0198, Japan}
\affiliation{
  Department of Physics, Graduate School of Science, The University of Tokyo,
  Tokyo 113-0033, Japan}
\author{Hiroyuki Sagawa}
\affiliation{
  Center for Mathematics and Physics, University of Aizu,
  Aizu-Wakamatsu 965-8560, Japan}
\affiliation{
  RIKEN Nishina Center, Wako 351-0198, Japan}
\author{Myung-Ki Cheoun}
\affiliation{
  Department of Physics and Origin of Matter and Evolution of Galaxies Institute,
  Soongsil University, Seoul 06978, Korea}
\date{\today}
\begin{abstract}
  We present a relativistic mean-field model that incorporates charge symmetry breaking (CSB) 
  of nuclear force via $ \omega $-$ \rho^0 $ meson mixing, along with corrections 
  to the electromagnetic interaction including the nucleon form factors, 
  first-order vacuum polarization, and Coulomb exchange and pairing terms. 
  The model parameters are refitted using the mass differences of $ T = 1/2 $ mirror nuclei 
  and ground-state properties of magic nuclei, yielding DD-ME-CSB parameter set. 
  The DD-ME-CSB parameter set reproduces the mass 
  differences of mirror nuclei reasonably well up to $ T = 2 $, demonstrating the importance of $ \omega $-$ \rho^0 $ mixing. 
  A connection of the present model to a Skyrme-type CSB interaction is also established 
  through a gradient expansion of the energy density functional.
\end{abstract}
\maketitle
\section{Introduction}
\par
The isospin symmetry is one of the important properties in nuclear physics.
The isospin symmetry of nucleons almost holds and thus, protons and neutrons share common properties.
That of nuclear interaction also almost holds and thus, the proton-proton, neutron-neutron, and the $ T = 1 $ channel of the proton-neutron nuclear interactions are almost identical.
However, the isospin symmetry of nuclear interaction is slightly broken
because of the mass difference of the up- and down-quarks and electromagnetic interaction.
The isospin symmetry breaking in nuclear structure has been paid attention for decades
in many aspects,
such as mass and radii differences between a pair of mirror nuclei~\cite{
  Okamoto1964Phys.Lett.11_150,
  Nolen1969Annu.Rev.Nucl.Sci.19_471,
  Hatsuda1990Phys.Rev.C42_2212,
  Hatsuda1991Phys.Rev.Lett.66_2851,
  Auerbach1992Phys.Lett.B282_263,
  Shahnas1994Phys.Rev.C50_2346,
  Saito1994Phys.Lett.B335_17,
  Brown00,
  Baczyk2018Phys.Lett.B778_178,
  Dong2018Phys.Rev.C97_021301,
  Baczyk2019J.Phys.G46_03LT01,
  Sagawa2022Phys.Lett.B829_137072,  
  Naito2022Phys.Rev.C105_L021304,
  Naito2022Phys.Rev.C106_L061306,
  Naito2023Phys.Rev.C107_064302,
  SNRH24,
  Naito:2025qub},
isobaric analog energy~\cite{
  Suzuki1993Phys.Rev.C47_R1360,
  Roca-Maza2018Phys.Rev.Lett.120_202501},
and the superallowed beta decay~\cite{
  Sagawa1996Phys.Rev.C53_2163,
  Liang2009Phys.Rev.C79_064316,
  Satula2011Phys.Rev.Lett.106_132502,
  Satula2012Phys.Rev.C86_054316,
  Rafalski2012Phys.Scr.T150_014032,
  Kaneko2017Phys.Lett.B773_521,
  Hardy2020Phys.Rev.C102_045501,
  Xayavong2022Phys.Rev.C105_044308}.
Such a study is also important for the unitarity test of the Cabbibo-Kobayashi-Maskawa matrix~\cite{
  Hardy2013Ann.Phys.525_443,
  Porter2016Prog.Part.Nucl.Phys.91_101,
  pdg24}.
\par
The isospin symmetry breaking of the nuclear interaction can be classified into two classes:
the charge independence breaking (CIB) and charge symmetry breaking (CSB) interactions,
which are often referred to as the class II and III interactions, respectively~\cite{
  Henley1969IsospininNuclearPhysics_15,
  Miller2006Annu.Rev.Nucl.Part.Sci.56_253}.
The CIB interaction corresponds to the difference between the like-particle interaction and the different-particle one,
which mainly originates from the mass difference of pions~\cite{
  Coon1982Phys.Rev.C26_2402}.
The CSB interaction corresponds to the difference between the proton-proton interaction and the neutron-neutron one,
which mainly originates from the $ \omega $-$ \rho^0 $ and $ \pi $-$ \eta $ mixings in the meson exchange picture~\cite{
  Coon1977Nucl.Phys.A287_381,
  Coon1982Phys.Rev.C26_562,
  Coon1987Phys.Rev.C36_2189}
and
the mass difference of nucleons in medium in the picture of the quantum chromodynamics sum rule~\cite{
  Hatsuda1990Phys.Rev.C42_2212,
  Hatsuda1991Phys.Rev.Lett.66_2851,
  SNRH24}.
Recently, the isospin symmetry breaking interaction has also been studied in the chiral effective field theory~\cite{
  Kolck1995Few-BodySyst.Suppl.9_444,
  Kolck1996Phys.Lett.B371_169,
  Epelbaum2005Phys.Rev.C72_044001,
  Meissner2008Eur.Phys.J.A36_37}.
\par
The density functional theory (DFT) is a powerful tool to calculate nuclear properties among the nuclear chart~\cite{
  Hohenberg1964Phys.Rev.136_B864,
  Kohn1965Phys.Rev.140_A1133,
  Vautherin1972Phys.Rev.C5_626,
  Kohn1999Rev.Mod.Phys.71_1253,
  Bender2003Rev.Mod.Phys.75_121,
  Colo2020Adv.Phys.X5_1740061}.
The starting point of the DFT calculation is an energy density functional (EDF).
In most calculation, an EDF has been obtained with the momentum expansion method without a microscopic background and
an \textit{ab initio} determination of an EDF is an important topic in nuclear physics community~\cite{
  Naito2019J.Phys.B52_245003,
  Accorto2020Phys.Rev.C101_024315,
  Marino2021Phys.Rev.C104_024315}.
So far,
the Skyrme Hartree-Fock calculation~\cite{
  Skyrme1958Nucl.Phys.9_615,
  Vautherin1972Phys.Rev.C5_626}
has been used in 
most study of isospin symmetry breaking in nuclear DFT,
where the EDF form used is obtained phenemenologically or based on QCD sum rule.
In the covariant density functional theory (CDFT), which is another branch of the nuclear DFT,
one starts from the Lagrangian of the meson exchange picture~\cite{
  Meng2006Prog.Part.Nucl.Phys.57_470,
  Niksic2011Prog.Part.Nucl.Phys.66_519}.
Since the origin of the isospin asymmetric nuclear interaction is well understood in the meson exchange picture, 
the study of the isospin symmetry breaking using the CDFT should be a promissing approach.
\par
In this paper, we consider the $ \omega $-$ \rho^0 $ mixing process in the relativistic Hartree-Bogoliubov (RHB) model,
which is a branch of the CDFT.
Although the Fock term is neglected and thus the pion cannot be considered explicitly,
the numerical cost of the RHB calculation is low compared to the relativistic Hartree-Fock-Bogoliubov calculation.
The CSB interaction usually contributes to nuclear properties stronger than the CIB one~\cite{
  Naito2023Phys.Rev.C107_064302},
while it has been known that the bare CIB interaction is approximately one order of the magnitude stronger than the CSB one in the bare interaction~\cite{AV18,Miller2006Annu.Rev.Nucl.Part.Sci.56_253}.
In addition, since the $ \omega $-$ \rho^0 $ mixing is the dominant origin of the CSB interaction in the meson exchange picture,
the RHB calculation with the $ \omega $-$ \rho^0 $ mixing is enough as the first step.
We start from the DD-ME2 Lagrangian~\cite{
  ddme2},
which is one of the successful relativistic mean-field Lagrangians with the density-dependent meson-exchange model,
and consider the $ \omega $-$ \rho^0 $ mixing explicitly.
Since the size of the $ \omega $-$ \rho^0 $ mixing is rather arbitrary~\cite{
  Coon1977Nucl.Phys.A287_381,
  Coon1987Phys.Rev.C36_2189,
  PhysRevC.47.R2462},
we fix the size of the mixing using the mass difference of the mirror nuclei.
Using the optimized parameter sets for selected observables,
we perform the systematic calculations of mass differences of mirror nuclei and discuss also the properties of the equation of state of nuclear matter.
\par
The paper is organized as follows. 
In Sec.~\ref{sec:model}, we introduce the relativistic mean-field model with 
$ \omega $-$ \rho^0 $ mixing. 
In Sec.~\ref{sec:results}, we discuss the results obtained by the model for 
properties of nuclear matter and finite nuclei. We also present the connection 
between the present model and the Skyrme-type CSB model. 
The summary is given in Sec.~\ref{sec:summary}. 
\section{Model setup}
\label{sec:model}
\subsection{Relativistic Hartree-Bogoliubov model}
\par
In this study, we employ the RHB model.
For the particle-hole ($ ph $) channel, we adopt a density-dependent meson-exchange interaction of the DD-ME2 type~\cite{ddme2} which is optimized to incorporate $ \omega $-$ \rho^0 $ mixing.
In the particle-particle ($ pp $) channel, the Gogny D1S interaction~\cite{BGG91,YGB19} is used, as in the original DD-ME2 model.
Further details of the RHB model and the DD-ME2 parameter set can be found in 
Refs.~\cite{ddme2,ddme1,dirhb,LVPR98,SeRi02,KuRi91}.
\par
In addition to the DD-ME2 model, we incorporate the following effects: 
$ \omega $-$ \rho^0 $ mixing in the vector-meson mass term, nucleon electromagnetic form factors, 
and vacuum polarization (VP) effects due to an electron loop in the photon propagator.
\subsubsection{Model Lagrangian}
\par
The Lagrangian density for the $ ph $ channel interaction is expressed in terms of the nucleon field $ \psi $,
the meson fields $ \sigma $, $ \omega^{\mu} $, and $ \vec{\rho}^{\mu} $, and the electromagnetic field $ A^{\mu} $ as
\begin{align}
  \ca{L}
  & =
    \overline{\psi}
    \left( i \gamma^{\mu} \partial_{\mu} - m_0 - \tau_3 m_1 \right)
    \psi
    \notag \\
  & \quad
    +
    \frac{1}{2}
    \left( \partial_{\mu} \sigma \right)
    \left( \partial^{\mu} \sigma \right)
    -
    \frac{1}{2}
    m_{\sigma}^2
    \sigma^2
    \notag \\
  & \quad 
    -
    \frac{1}{4}
    \Omega^{\mu \nu} \Omega_{\mu \nu}
    -
    \frac{1}{4}
    \vec{R}^{\mu \nu} \cdot \vec{R}_{\mu \nu}
    +
    \frac{1}{2}
    \begin{pmatrix}
      \omega & \vec{\rho}^{\dagger} 
    \end{pmatrix}^\mu
               M_{\urm{v}}^2
               \begin{pmatrix}
                 \omega \\
                 \vec{\rho}
               \end{pmatrix}_{\mu}
  \notag \\
  & \quad 
    -
    \frac{1}{4}
    F^{\mu \nu}F_{\mu \nu}
    \notag \\
  & \quad 
    -
    g_{\sigma} \sigma \overline{\psi} \psi
    -
    g_{\omega} \omega_{\mu} \overline{\psi} \gamma^{\mu} \psi
    -
    g_{\rho} \vec{\rho}_{\mu} \cdot \overline{\psi} \gamma^{\mu} \vec{\tau} \psi
    -
    e A_{\mu} j_{\urm{em}}^{\mu},
    \label{eq:Lagrangian}
\end{align}
where $ m_0 $ and $ m_1 $ are isoscalar and isovector nucleon masses
defined by 
$ m_0 = \left( m_ n + m_p \right) / 2 $ and $ m_1 = \left( m_n - m_p \right) / 2 $, respectively,
and 
$ \Omega^{\mu \nu} $, $ \vec{R}^{\mu \nu} $, and $ F^{\mu \nu} $ are the field strength tensors 
of $ \omega $ and $ \rho $ mesons and photon, respectively.
The arrows denote the vectors in the isospin space. 
The vector $ \left( \omega, \vec{\rho} \right)_{\mu}^T $ in the mass term of the vector mesons in Eq.~\eqref{eq:Lagrangian} is defined by
\begin{equation}
  \begin{pmatrix}
    \omega \\
    \vec{\rho}
  \end{pmatrix}_{\mu}
  =
  \begin{pmatrix}
    \omega_{\mu} \\
    \rho^{+}_{\mu} \\
    \rho^0_{\mu} \\
    \rho^{-}_{\mu} 
  \end{pmatrix}
\end{equation}
and the $ 4 \times 4 $ mass matrix $ M_{\urm{v}}^2 $ is given by
\begin{equation}
  M_{\urm{v}}^2
  =
  \begin{pmatrix}
    m_{\omega}^2   & 0          & \Delta m_{\urm{v}}^2   & 0 \\
    0              & m_{\rho}^2 & 0              & 0 \\
    \Delta m_{\urm{v}}^2   & 0          & m_{\rho}^2     & 0 \\
    0              & 0          & 0              & m_{\rho}^2 
  \end{pmatrix}.
\end{equation}
The off-diagonal element $ \Delta m_{\urm{v}}^2 $,
which couples the $ \omega $ and the neutral $ \rho $ mesons, represents the $ \omega $-$ \rho^0 $ mixing. 
\par
The nucleon masses are taken from Ref.~\cite{pdg24} as
$ m_n = 939.565420 $ 
and 
$ m_p = 938.272088 \, \mathrm{MeV} $.
The meson-nucleon coupling constants are density dependent as in DD-ME2, 
\begin{equation}
  g_i \left( \rho \right)
  =
  f_i
  \left( \rho / \rho_{\urm{sat}} \right)
  g_i \left( \rho_{\urm{sat}} \right)
  \qquad
  \text{($ i = \sigma $, $ \omega $, and $ \rho $)},
\end{equation}
where $ \rho_{\urm{sat}} $ is the saturation density. 
For $ i = \sigma $ and $ \omega $, 
the function $ f_i $ is parameterized as
\begin{equation}
  f_i \left( x \right)
  =
  a_i
  \frac{1 + b_i \left( x + d_i \right)^2}{1 + c_i \left( x + d_i \right)^2}
\end{equation}
with the constraints
$ f_i \left( 1 \right) = 1 $,
$ f''_{\sigma} \left( 1 \right) = f''_{\omega} \left( 1 \right) $,
and $ f''_i \left( 0 \right) = 0 $~\cite{ddme2} 
to reduce the number of independent parameters in $ f_{\sigma} $ and $ f_{\omega} $ to three. 
The density dependence for $ \rho $ meson is given by 
\begin{equation}
  f_{\rho}
  \left( \rho / \rho_{\urm{sat}} \right)
  =
  e^{- a_{\rho} \left( \rho / \rho_{\uurm{sat}} - 1 \right)}.
\end{equation}
\subsubsection{Equations of motion for meson fields}
\par
We impose the time-reversal symmetry of the ground states of even-even nuclei. 
Then, the space-like components of the meson and electromagnetic fields all vanish. 
In addition, due to the charge conservation, the charged $ \rho $ meson fields also vanish. 
Thus, only the time-even nucleon densities act as sources for the time-like fields.
The scalar, vector-isoscalar, and vector-isovector densities of nucleon are given 
in terms of the canonical single-particle wave functions $ \psi_k \left( \ve{r} \right) $ as
\begin{subequations}
  \begin{align}
    \rho_S \left( \ve{r} \right)
    & =
      \avr{\overline{\psi} \psi}
      =
      \sum_k
      v_k^2
      \overline{\psi}_k \left( \ve{r} \right)
      \psi_k \left( \ve{r} \right), \\
    \rho \left( \ve{r} \right)
    & =
      \avr{\overline{\psi} \gamma^0 \psi}
      =
      \sum_k
      v_k^2
      \psi_k^{\dagger} \left( \ve{r} \right)
      \psi_k \left( \ve{r} \right), \\
    \rho_3 \left( \ve{r} \right)
    & =
      \avr{\overline{\psi} \gamma^0 \tau_3 \psi}
      =
      \sum_k
      v_k^2
      \psi_k^{\dagger} \tau_3 \psi_k \left( \ve{r} \right),
  \end{align}
\end{subequations}
respectively, where $ v_k^2 $ is the occupation probability of the state $ k $.
\par
The equations of motion for the meson fields are given by
\begin{subequations}
  \begin{align}
    \left( - \Nabla^2 + m_{\sigma}^2 \right)
    \sigma \left( \ve{r} \right) 
    & =
      - g_{\sigma} \rho_S \left( \ve{r} \right), \\
    \left( - \Nabla^2 + m_{\omega}^2 \right)
    \omega \left( \ve{r} \right)
    & =
      g_{\omega} \rho \left( \ve{r} \right)
      -
      \Delta m_{\urm{v}}^2 \, \rho^0 \left( \ve{r} \right), 
      \label{eq:KG-omg} \\
    \left( - \Nabla^2 + m_{\rho}^2 \right)
    \rho^0 \left( \ve{r} \right)
    & =
      g_{\rho} \rho_3 \left( \ve{r} \right)
      -
      \Delta m_{\urm{v}}^2 \, \omega \left( \ve{r} \right).
      \label{eq:KG-rho}
  \end{align}
\end{subequations}
Notice that $ \omega $ and $ \rho^0 $ fields are coupled due to the $ \omega $-$ \rho^0 $ mixing. 
\subsubsection{Electromagnetic field}
\par
In calculating the charge density and electromagnetic field, 
we take into account both the finite size of nucleons and the vacuum polarization effect. 
The nuclear charge form factor is given by~\cite{KuSu00,PPV07,Horowitz2012Phys.Rev.C86_045503,KuSu19,ReNa21,Naito2021Phys.Rev.C104_024316}
\begin{equation}
  \label{eq:rhoq}
  \tilde{\rho}_{\urm{ch}} \left( \ve{q} \right) 
  =
  \sum_{\tau = p, \, n} 
  \left[
    F_{1 \tau} \left( \ve{q}^2 \right)
    \tilde{\rho}_{\tau} \left( \ve{q} \right)
    +
    F_{2 \tau} \left( \ve{q}^2 \right)
    \tilde{\rho}_{\kappa \tau} \left( \ve{q} \right)
  \right], 
\end{equation}
where the Fourier transforms of the point nucleon densities are as
\begin{subequations}
  \begin{align}
    \tilde{\rho}_{\tau} \left( \ve{q} \right)
    & =
      \int 
      d^3 r \,
      e^{i \ve{q} \cdot \ve{r}}
      \rho_{\tau} \left( \ve{r} \right), \\
    \tilde{\rho}_{\kappa \tau} \left( \ve{q} \right)
    & =
      \int
      d^3 r \,
      e^{i \ve{q} \cdot \ve{r}}
      \rho_{\kappa \tau} \left( \ve{r} \right). 
  \end{align}
\end{subequations}
The coordinate-space densities here are defined by
\begin{subequations}
  \begin{align}
    \rho_{\tau} \left( \ve{r} \right)
    & =
      \sum_{k \in \tau}
      v_k^2
      \psi_k^{\dagger} \left( \ve{r} \right)
      \psi_k \left( \ve{r} \right), \\
    \rho_{\kappa \tau} \left( \ve{r} \right)
    & =
      \frac{\kappa_{\tau}}{2m}
      \Nabla
      \cdot 
      \sum_{k \in \tau}
      v_k^2
      \overline{\psi}_k \left( \ve{r} \right)
      i
      \ve{\alpha}
      \psi_k \left( \ve{r} \right), 
      \label{eq:rho_k}
  \end{align}
\end{subequations}
where $ m $ is taken to be the average nucleon mass,
$ \kappa_p = 1.793$ and $ \kappa_n = -1.913 $ are the anomalous magnetic moments of proton and neutron, respectively, 
and $ \ve{\alpha} = \gamma^0 \ve{\gamma} $ is the usual Dirac matrix. 
The nucleon form factors $ F_{1 \tau} \left( \ve{q}^2 \right) $ and $ F_{2 \tau} \left( \ve{q}^2 \right) $
contain the information of the internal electromagnetic structure of a nucleon. 
They are related to the Sachs form factors, $ G_{\urm{E} \tau} $ and $ G_{\urm{M} \tau} $, as \cite{PPV07}
\begin{subequations}
  \begin{align}
    G_{\urm{E} \tau} \left( \ve{q}^2 \right)
    & =
      F_{1 \tau} \left( \ve{q}^2 \right)
      -
      \frac{\ve{q}^2}{4m}
      \kappa_{\tau} F_{2 \tau} \left( \ve{q}^2 \right), \\
    G_{\urm{M} \tau} \left( \ve{q}^2 \right)
    & =
      F_{1 \tau} \left( \ve{q}^2 \right)
      +
      \kappa_{\tau} F_{2 \tau} \left( \ve{q}^2 \right).
  \end{align}
\end{subequations}
The empirical relation, 
\begin{align}
  G_{\urm{E} p} \left( \ve{q}^2 \right)
  & \approx
    \frac{G_{\urm{M} p} \left( \ve{q}^2 \right)}{1 + \kappa_p}
    \approx
    \frac{G_{\urm{M} n} \left( \ve{q}^2 \right)}{\kappa_n}
    \approx
    G_{\urm{D}} \left( \ve{q}^2 \right)
    \notag \\
  & =
    \frac{1}{\left( 1 + \ve{q}^2 / \Lambda^2 \right)^2}
\end{align}
with $ 1 / \Lambda^2 \approx 5.48 \times 10^{-2} \, \mathrm{fm}^2 $~\cite{PPV07}, 
leads to
\begin{subequations}
  \begin{align}
    F_{1 p} \left( \ve{q}^2 \right)
    & \approx
      \frac{1}{1 + \ve{q}^2/4m^2}
      \left[
      1
      +
      \left( 1 + \kappa_p \right)
      \frac{\ve{q}^2}{4m^2}
      \right]
      G_{\urm{E} p} \left( \ve{q}^2 \right), \\
    F_{1 n}
    \left( \ve{q}^2 \right)
    & =
      \frac{1}{1 + \ve{q}^2/4m^2}
      \left[
      G_{\urm{E} n} \left( \ve{q}^2 \right)
      +
      \frac{\ve{q}^2}{4m^2}
      \kappa_n
      G_{\urm{E} p} \left( \ve{q}^2 \right)
      \right], \\
    \kappa_p
    F_{2 p} \left( \ve{q}^2 \right)
    & \approx
      \frac{\kappa_p}{1 + \ve{q}^2/4m^2}
      G_{\urm{E} p} \left( \ve{q}^2 \right), \\
    \kappa_n
    F_{2 n} \left( \ve{q}^2 \right)
    & \approx
      \frac{1}{1 + \ve{q}^2/4m^2}
      \left[
      -
      G_{\urm{E} n} \left( \ve{q}^2 \right)
      +
      \kappa_n G_{\urm{E} p} \left( \ve{q}^2 \right)
      \right],
  \end{align}
\end{subequations}
where we adopt the Sachs form factors from Ref.~\cite{KuSu19} as
\begin{subequations}
  \begin{align}
    G_{\urm{E} p} \left( \ve{q}^2 \right)
    & =
      \frac{1}{\left( 1 + r_p^2 \ve{q}^2/12 \right)^2}, \\
    G_{\urm{E} n} \left( \ve{q}^2 \right)
    & =
      \frac{a}{\left( 1 + r_{+}^2 \ve{q}^2/12 \right)^2}
      -
      \frac{a}{\left( 1 + r_{-}^2 \ve{q}^2/12 \right)^2}, 
  \end{align}
\end{subequations}
where $ r_p = 0.841 \, \mathrm{fm} $, taken from Ref.~\cite{pdg24},
and $ a = 0.115 $, $ r_{\pm}^2 = r_{\urm{av}}^2 \pm r_n^2/2a $ with
$ r_{\urm{av}} = 0.856 \, \mathrm{fm} $ and $ r_n^2 = -0.1147 \, \mathrm{fm}^2 $, taken from Ref.~\cite{Ge11}.
Note that the nucleon form factors are normalized as
$ F_{1 p} \left( 0 \right) = F_{2 p} \left( 0 \right) = F_{2 n} \left( 0 \right) = 1 $
and $ F_{1 n} \left( 0 \right) = 0 $.
\par
To compute the electrostatic potential $ A_0 $ from the aforementioned nuclear charge density, 
we take into account the vacuum polarization effect due to an electron loop, i.e., 
to the first order of fine structure constant $ \alpha \simeq 1/137 $.
The electrostatic potential with the VP correction is given by~\cite{GreinerQED}
\begin{equation}
  \label{eq:A0_VP}
  e A_0 \left( \ve{r} \right)
  =
  e^2
  \int \frac{d^3q}{\left( 2 \pi \right)^3}
  e^{i \ve{q} \cdot \ve{r}}
  \left[
    1
    +
    \Pi^R \left( \ve{q}^2 \right)
  \right]
  \frac{4\pi}{\ve{q}^2}
  \tilde{\rho}_{\urm{ch}} \left( \ve{q} \right),
\end{equation}
where the polarization function $ \Pi^R $ is
\begin{equation}
  \Pi^R \left( \ve{q}^2 \right)
  =
  \frac{\alpha}{\pi}
  \left[
    -
    \frac{5}{9}
    +
    \frac{1}{3}
    \frac{4 m_e^2}{\ve{q}^2}
    +
    \frac{1}{3}
    \left(
      1
      -
      \frac{2 m_e^2}{\ve{q}^2}
    \right)
    f \left( \ve{q}^2 \right)
  \right]
\end{equation}
with
\begin{equation}
  f \left( \ve{q}^2 \right)
  =
  \sqrt{1 + \frac{4 m_e^2}{\ve{q}^2}}
  \log
  \frac{\sqrt{1 + \frac{4m_e^2}{\ve{q}^2}} + 1}{\sqrt{1 + \frac{4 m_e^2}{\ve{q}^2}} - 1}
\end{equation}
and $ m_e $ being the electron mass.
If the polarization function $ \Pi^R $ is neglected, 
the potential in Eq.~\eqref{eq:A0_VP} reduces to the one without the VP correction.  
\par
Using the Hartree fields in the $ ph $ channel as described above,
and the Gogny D1S interaction in the $ pp $ channel, we solve the Hartree-Bogoliubov equations self-consistently.
In calculating the total energy,
the exact Coulomb exchange and the pairing energies are included perturbatively (see Appendix \ref{app:coulomb}).
The vacuum polarization and finite nucleon size effects are neglected in these exchange and pairing contributions.
\subsection{Parameter fitting}
\par
The parameters in the Lagrangian as given in Eq.~\eqref{eq:Lagrangian}, 
$ \Delta m_{\urm{v}}^2 $, $ m_{\sigma} $, $ g_{\sigma} \left( \rho_{\urm{sat}} \right) $,
$ g_{\omega} \left( \rho_{\urm{sat}} \right) $, $ g_{\rho} \left( \rho_{\urm{sat}} \right) $,
and the four independent parameters in the density dependences 
$ f_i $ (three from $ i = \sigma $ and $ \omega $, and one from $ i = \rho $) of the coupling constants, 
are optimized to the binding energies and charge radii of 
$ \nuc{O}{16}{} $, $ \nuc{Ca}{40}{} $, $ \nuc{Ca}{48}{} $, $ \nuc{Ni}{56}{} $, $ \nuc{Sn}{132}{} $, and $ \nuc{Pb}{208}{} $,
and 
the mass difference of $ T = 1/2 $ mirror pairs
$ \nuc{N}{15}{} $-$ \nuc{O}{15}{} $,
$ \nuc{O}{17}{} $-$ \nuc{F}{17}{} $,
$ \nuc{K}{39}{} $-$ \nuc{Ca}{39}{} $,
$ \nuc{Ca}{41}{} $-$ \nuc{Sc}{41}{} $,
$ \nuc{Co}{55}{} $-$ \nuc{Ni}{55}{} $, and
$ \nuc{Ni}{57}{} $-$ \nuc{Cu}{57}{} $
defined by
\begin{equation}
  \Delta B \left( N, Z \right)
  \equiv
  B \left( N, Z \right)
  -
  B \left( Z, N \right)
  \qquad
  \text{($ N > Z $)},
\end{equation}
where $ B \left( N, Z \right) $ is the binding energy of nucleus
with the given neutron and proton numbers $ \left( N, Z \right) $.
The vector meson masses are fixed to be $ m_{\omega} = 783 $ and $ m_{\rho} = 763 \, \mathrm{MeV} $
as is done in DD-ME2.
The parameter set thus obtained will be called DD-ME-CSB. 
Experimental data of binding energies are taken from Refs.~\cite{AME2020-1, AME2020-2},
and those of charge radii are taken from Refs.~\cite{Ang13,Li21,Sommer22}. 
In calculating the charge radius, we take into account the center-of-mass correction~\cite{TaCh24}. 
See Appendix~\ref{app:cmcorr} for the details. 
\par
In order to investigate the $ \Delta m_{\urm{v}}^2 $ dependence of the observables, we have also performed 
optimizations using $ \Delta m_{\urm{v}}^2 = 0.00 $, $ 0.05 $, $ 0.10 $, and $ 0.15 \, \mathrm{fm}^{-2} $, 
excluding the mass differences $\Delta B$ from the dataset.
This is  
because the
mixing strength $ \Delta m_{\urm{v}}^2 $ is intimately correlated with $\Delta B$, 
  and fixing its value could lead to unnatural or potentially undesirable readjustment of the other parameters in the charge-symmetry-conserving  part of the interaction. 
These parameter sets will be labelled according to the $ \Delta m_{\urm{v}}^2 $ values as 
DD-ME-CSB00, DD-ME-CSB05, DD-ME-CSB10, and DD-ME-CSB15, respectively.
\section{Results and Discussion}
\label{sec:results}
\par
The parameters of the DD-ME-CSB model are listed in Table~\ref{tb:best_fit_param}, 
in comparison with those of the DD-ME2 parameter set
[see Appendix~\ref{app:CSB(00-15)} for parameter values of the other sets, DD-ME-CSB (00--15)].
The values of the two models are similar except for the parameters $ g_{\rho} \left( \rho_{\urm{sat}} \right) $ 
and $ a_{\rho} $ of $ \rho $ meson. 
In the present model, the nucleon-$ \rho $ meson ($ N $-$ \rho $) coupling is stronger.
This difference may be understood as follows.
The contributions of $ \omega $-$ \rho^0 $ mixing and $ N $-$ \rho $ coupling to the binding energy 
for $ N > Z $ nuclei have opposite signs:
The former is attractive while the latter is repulsive.
The additional attraction due to the $ \omega $-$ \rho^0 $ mixing has been compensated 
by the $ N $-$ \rho $ coupling via fitting to $ N > Z $ nuclei such as $ \nuc{Pb}{208}{} $.
Notice that, at $ N < Z $ side, the both contributions become positive, reducing the binding energy. 
\par
The value of $ \Delta m_{\urm{v}}^2 \approx 0.07 \, \mathrm{fm}^{-2} $ is comparable with the empirical estimates, 
$ 0.087 \pm 0.01 \, \mathrm{fm}^{-2} $~\cite{Coon1977Nucl.Phys.A287_381} and 
$ 0.116 \pm 0.02 \, \mathrm{fm}^{-2} $~\cite{Coon1987Phys.Rev.C36_2189}. 
The ratio $ \Delta m_{\urm{v}}^2 / m_{\omega}^2 \approx \Delta m_{\urm{v}}^2 / m_{\rho}^2 \approx 1/220 $, 
which is similar to the fine-structure constant, is also consistent with one of the possible 
mechanisms for $ \omega $-$ \rho^0 $ conversion by an electromagnetic process.
\begin{table}
  \caption{DD-ME-CSB and DD-ME2 parameter sets.
    Note that the vecror-meson masses $ m_{\omega} $ and $ m_{\rho} $ are fixed.
    The other parameter sets with fixed values of $\Delta m_{\urm{v}}^2$ are summarized in Table~\ref{tb:CSB00-15}.}
  \label{tb:best_fit_param}
  \begin{ruledtabular}
    \begin{tabular}{ldd}
      Parameter & \multicolumn{1}{c}{DD-ME-CSB} & \multicolumn{1}{c}{DD-ME2~\cite{ddme2}} \\
      \hline
      $ m_{\sigma} $ ($ \mathrm{MeV} $)                 & 549.9985 & 550.1238 \\
      $ m_{\omega} $ ($ \mathrm{MeV} $)                 & 783.0000 & 783.0000 \\
      $ m_{\rho} $   ($ \mathrm{MeV} $)                 & 763.0000 & 763.0000 \\
      $ g_{\sigma} \left( \rho_{\urm{sat}} \right) $    & 10.78917 &  10.5396 \\     
      $ g_{\omega} \left( \rho_{\urm{sat}} \right) $    & 13.38742 &  13.0189 \\     
      $ g_{\rho}   \left( \rho_{\urm{sat}} \right) $    & 4.230683 &   3.6836 \\     
      $ a_{\sigma} $                                    & 1.356091 &   1.3881 \\     
      $ b_{\sigma} $                                    & 1.260987 &   1.0943 \\     
      $ c_{\sigma} $                                    & 1.886524 &   1.7057 \\     
      $ d_{\sigma} $                                    & 0.420347 &   0.4421 \\     
      $ a_{\omega} $                                    & 1.355118 &   1.3892 \\     
      $ b_{\omega} $                                    & 1.054689 &   0.9240 \\     
      $ c_{\omega} $                                    & 1.596527 &   1.4620 \\     
      $ d_{\omega} $                                    & 0.456932 &   0.4775 \\     
      $ a_{\rho} $                                      & 0.242512 &   0.5647 \\     
      $ \Delta m_{\urm{v}}^2$ ($ 10^{-2} \, \mathrm{fm}^{-2} $) & 7.206445 &   0      \\
    \end{tabular}
  \end{ruledtabular}
\end{table}
\subsection{Matter properties}
\par
The equation of state (EoS) of nuclear matter 
of the present model is given as a function of the total nucleon density $ \rho = \rho_n + \rho_p $ 
and the relative asymmetry $ \delta = \left( \rho_n - \rho_p \right) / \rho$. 
The EoS can be expanded around $ \left( \rho, \delta \right) = \left( \rho_{\urm{sat}}, 0 \right) $ as 
\begin{equation}
  \label{exp1}
  \frac{E}{A} \left( \rho, \delta \right)
  =
  \sum_{m = 0}^{\infty}
  \sum_{n = 0}^{\infty}
  e_{m, n}
  \frac{1}{m!}
  \left( \frac{\rho - \rho_{\urm{sat}}}{3 \rho_{\urm{sat}}} \right)^m
  \delta^n   
\end{equation}
with
\begin{equation}
  \label{exp2}
  e_{m, n}
  =
  \frac{1}{n!}
  \left( 3 \rho_{\urm{sat}} \right)^m
  \left.
    \frac{\partial^{n+m} \left( E / A \right)}{\partial \rho^m \, \partial \delta^n}
  \right|_{\left( \rho, \delta \right) = \left( \rho_{\uurm{sat}}, 0 \right)}.
\end{equation}
Note that $ e_{0, 0} $ corresponds to the energy per nucleon of symmetric nuclear matter at saturation. 
By definition, $ e_{1, 0} = 0 $ holds, and $ e_{2, 0} = K_{\infty} $ represents the incompressibility of symmetric matter.
Only the terms with even $ n $ appear in the expansions \eqref{exp1} and \eqref{exp2} if the CSB interaction is not considered.
The parameters characterizing the symmetry energy are $ e_{0, 2} = J_{\urm{sym}} $ and $ e_{1, 2} = L_{\urm{sym}} $. 
The present model also includes the term linear in $ \delta $ due to CSB,
with parameters defined as $ e_{0, 1} \equiv J_{\urm{CSB}} $ and $ e_{1, 1} \equiv L_{\urm{CSB}} $.
The energy per nucleon (relative to $ e_{0, 0} $) of pure neutron matter at saturation density, 
$ \left( \rho, \delta \right) = \left( \rho_{\urm{sat}}, 1 \right) $,
is given approximately by $ J_{\urm{sym}} + J_{\urm{CSB}} $, 
and the corresponding slope parameter by $ L_{\urm{sym}} + L_{\urm{CSB}} $.
\par
Table~\ref{tb:MatterProperties} summarizes the nuclear matter properties of DD-ME-CSB 
in comparison with those of DD-ME2.
The two models yield similar results, while DD-ME-CSB gives larger values
for the incompressibility $ K_{\infty} $ and the slope parameter $ L $ of symmetry energy. 
The larger value of $ L $ may arise from the significant difference of
$ N $-$ \rho $ coupling in DD-ME-CSB compared to DD-ME2. 
The incompressibility of $ K_{\infty} = 277 \, \mathrm{MeV} $ obtained with DD-ME-CSB is comparable 
with that of NL3 parameter set ($ K_{\infty} = 272 \, \mathrm{MeV} $)~\cite{NL3}. 
Unlike DD-ME2, where $ K_{\infty} $ was constrained around $ 250 \, \mathrm{MeV} $
in order to reproduce experimental giant monopole resonance energies~\cite{ddme2}, 
no such constraint was imposed in the present fit.
\par
\begin{figure}
  \includegraphics[width=1.0\linewidth]{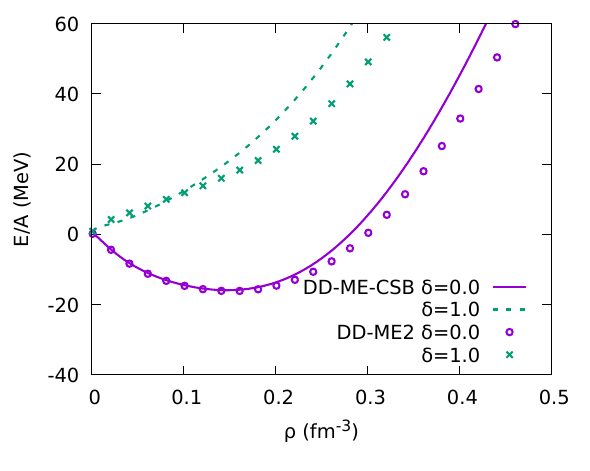}
  \caption{Comparison of nuclear matter equations of state (EoS) for DD-ME-CSB and DD-ME2. 
    The solid and dashed curves show the EoS of symmetric and pure neutron matters, 
    respectively, for DD-ME-CSB while open circles and crosses show those for DD-ME2.}
  \label{fig:eos}
\end{figure}

Figure~\ref{fig:eos} shows a comparison of the EoSs of the two models. 
While they are similar at sub-saturation densities, the differences become increasingly 
pronounced at densities above saturation.
\begin{table}
  \caption{Comparison of the nuclear matter properties of DD-ME-CSB and DD-ME2 parameter sets. 
    The properties of the other sets with fixed values of $\Delta m_{\urm{v}}^2$ are summarized in Table \ref{tb:MatterProperties00-15}.}
  \begin{ruledtabular}
    \label{tb:MatterProperties}
    \begin{tabular}{ldd}
      & \multicolumn{1}{c}{DD-ME-CSB} & \multicolumn{1}{c}{DD-ME2} \\
      \hline
      $ \rho_{\urm{sat}} $              ($ \mathrm{fm}^{-3} $) & 0.147  & 0.152  \\
      $ e_{0, 0} $                      ($ \mathrm{MeV} $)     & -15.88 & -16.14 \\
      $ m_n^* / m_n $                                          & 0.566  & 0.572  \\
      $ m_p^* / m_p $                                          & 0.565  & 0.572  \\
      $ K_{\infty} $                    ($ \mathrm{MeV} $)     & 276.92 & 250.89 \\
      $ J_{\urm{sym}} $                 ($ \mathrm{MeV} $)     & 35.9   & 32.2   \\
      $ J_{\urm{CSB}} $                 ($ \mathrm{MeV} $)     & -0.568 & 0      \\
      $ J_{\urm{CSB}} + J_{\urm{sym}} $ ($ \mathrm{MeV} $)     & 35.34  & 32.2   \\
      $ L_{\urm{sym}} $                 ($ \mathrm{MeV} $)     & 84.3   & 51.3   \\
      $ L_{\urm{CSB}} $                 ($ \mathrm{MeV} $)     & -1.26  & 0      \\
      $ L_{\urm{CSB}} + L_{\urm{sym}} $ ($ \mathrm{MeV} $)     & 83.04  & 51.3   \\
    \end{tabular}
  \end{ruledtabular}
\end{table}
\subsection{Binding energies and charge radii}
\par
\begin{figure}
  \includegraphics[width=\linewidth]{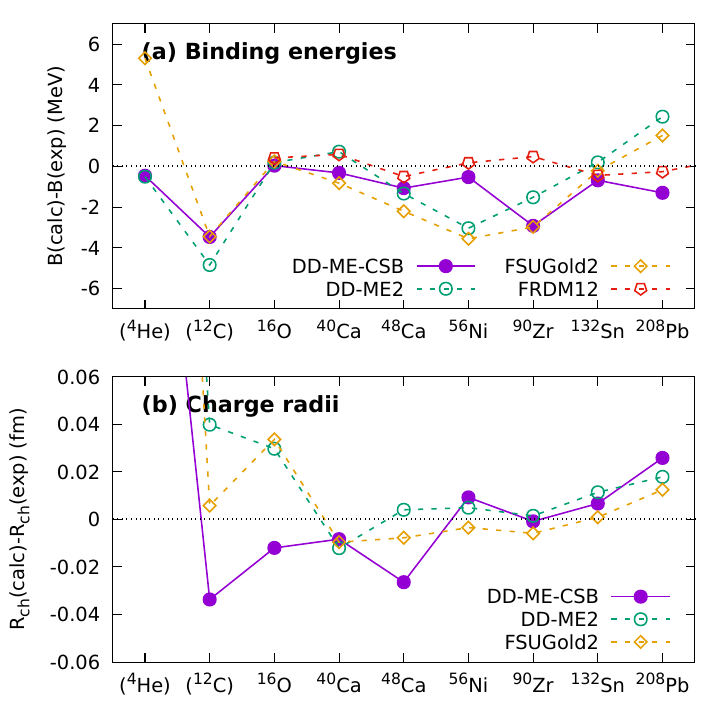}
  \caption{Deviations from the experimental data of
    (a) binding energies and (b) charge radii 
    for doubly- and semi-magic nuclei.
    The results of DD-ME-CSB parameter set are compared to 
    those obtained with DD-ME2~\cite{ddme2}, FSUGold2~\cite{fsugold2} and FRDM~\cite{frdm12}.
    Note that $ \nuc{He}{4}{} $ and $ \nuc{C}{12}{} $ are not included in the fit.}
  \label{fig:fit1}
\end{figure}

Figure~\ref{fig:fit1} shows the deviations from the experimental data of 
(a) binding energies and (b) charge radii of doubly- or semi-magic nuclei. 
The results of DD-ME-CSB parameter set are compared to those obtained with the RMF models DD-ME2~\cite{ddme2},
FSUGold2~\cite{fsugold2}, and the mass model FRDM~\cite{frdm12}.
The overall performance of DD-ME-CSB is comparable with the other mean-field models, 
DD-ME2 and FSUGold2, but somewhat worse than FRDM, which is specially designed for the mass prediction.
In particular, the agreement of binding energies 
for $ \nuc{O}{16}{} $, $ \nuc{Ca}{40}{} $, and $ \nuc{Ni}{56}{} $ is better than DD-ME2 and FSUGold2 models. 
The deviation of charge radii for DD-ME-CSB appears to increase from a negative value 
for $ \nuc{C}{12}{} $ towards the heavier nuclei. 
An anomalous decrease at $ \nuc{Ca}{48}{} $ for DD-ME-CSB is due to the negative spin-orbit contribution 
(from $ f_{7/2} $ neutrons) to the charge radius, which is not taken into account in DD-ME2 and FSUGold2. 
This contribution tends to be significant in relativistic mean-field models~\cite{KuSu00,KuSu19,Naito2023Phys.Rev.C107_054307,TaCh24}.
\subsection{Mass differencs of mirror nuclei}
\par
\begin{figure}
  \includegraphics[width=\linewidth]{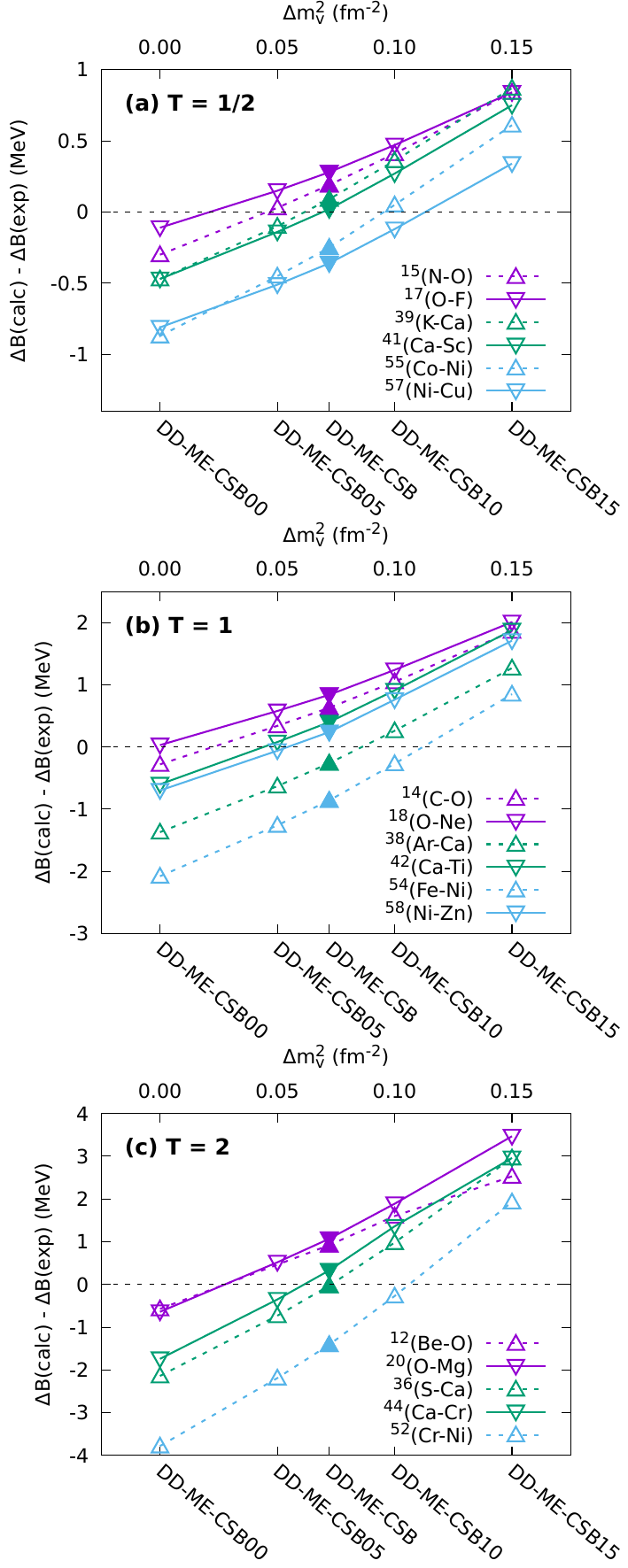}
  \caption{Deviations of the mass differences of mirror nuclei from the measured values 
    of (a) $ T = 1/2 $, (b) $ T = 1 $, and (c) $ T = 2 $ systems for parameter sets 
    DD-ME-CSB and DD-ME-CSB (00--15),
    plotted as functions of the $ \Delta m_{\urm{v}}^2 $ values.}
  \label{fig:DelB}
\end{figure}

Figure~\ref{fig:DelB} shows the deviations of the mass differences of mirror nuclei $ \Delta B $
from the measured values for for nuclei with isospin $ T = 1/2 $, $ 1 $, and $ 2 $,
as functions of the value of the $ \omega $-$ \rho^0 $ mixing parameter $ \Delta m_{\urm{v}}^2 $. 
It is evident that the DD-ME-CSB parameter set yields the best agreement with experimental data 
not only for the $ T = 1/2 $ nuclei,
which were included in the parameter fitting,
but also for the $ T = 1 $ and $ 2 $ systems.
The results also show that the deviations tend to be smaller for heavier nuclei within each 
isospin group, suggesting that the mixing parameter $ \Delta m_{\urm{v}}^2 $ could,
in general, be density dependent due to many-body effects.
\par
\begin{figure}
  \includegraphics[width=\linewidth]{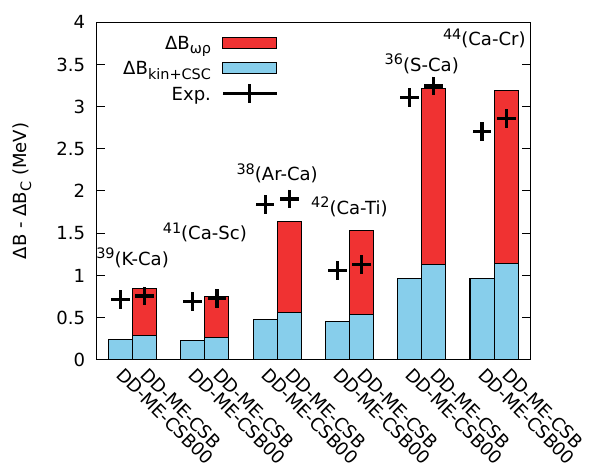}
  \caption{Decomposition of the mass difference of mirror nuclei subtracted by its Coulomb part,
    $ \Delta B - \Delta B_{\urm{C}} $,
    into the contributions from the $ \omega $-$ \rho^0 $ mixing 
    and the sum of kinetic and charge-symmetry-conserving (CSC) potential energies 
    for DD-ME-CSB00 ($ \Delta m_{\urm{v}}^2 = 0 $) and DD-ME-CSB models. 
    For comparison, the experimental values of $ \Delta B $ subtracted by the calculated Coulomb 
    contributions are also plotted by crosses.
    The experimental uncertainties are smaller than the vertical size of the crosses for all the cases.}
  \label{fig:DelB_dcmp}
\end{figure}

The calculated mass difference of mirror nuclei in the present model can be decomposed into three components as
\begin{equation}
  \Delta B
  =
  \Delta B_{\urm{C}}
  +
  \Delta B_{\omega \rho}
  +
  \Delta B_{\urm{kin} + \urm{CSC}}, 
\end{equation}
where $ \Delta B_{\urm{C}} $, $ \Delta B_{\omega \rho} $, and $ \Delta B_{\urm{kin} + \urm{CSC}} $
denote contributions from the Coulomb interaction,
$ \omega $-$ \rho^0 $ mixing,
and the sum of the kinetic energy and the charge-symmetry-conserving (CSC) part of the potential energy, respectively.
The Coulomb term $ \Delta B_{\urm{C}} $ includes the direct, exchange, pairing, and VP contributions. 
In general, $ \Delta B $ is dominated 
by the Coulomb interaction, which violates the charge symmetry in a trivial way. 
To examine the significance of the $ \omega $-$ \rho^0 $ mixing in $ \Delta B $, 
we present in Fig.~\ref{fig:DelB_dcmp} the decomposition of the mass difference of mirror nuclei, 
after subtracting the Coulomb contribution ($ \Delta B - \Delta B_{\urm{C}} $),
into the contributions from the $ \omega $-$ \rho^0 $ mixing $ \Delta B_{\omega \rho} $
and the sum of kinetic and CSC potential energies $ \Delta B_{\urm{kin} + \urm{CSC}} $. 
The results are shown for both the DD-ME-CSB00 model (without $ \omega $-$ \rho^0 $ mixing) 
and the DD-ME-CSB model. For comparison, the experimental values of $ \Delta B $ with 
the calculated Coulomb contribution subtracted are also shown by crosses.
In $ \nuc{K}{39}{} $-$ \nuc{Ca}{39}{} $, $ \nuc{Ca}{41}{} $-$ \nuc{Sc}{41}{} $,
$ \nuc{Ar}{38}{} $-$ \nuc{Ca}{38}{} $, $ \nuc{Ca}{42}{} $-$ \nuc{Ti}{42}{} $,
$ \nuc{S}{36}{} $-$ \nuc{Ca}{36}{} $, and $ \nuc{Ca}{44}{} $-$ \nuc{Cr}{44}{} $ pairs shown here,
the $ \omega $-$ \rho^0 $ mixing $ \Delta B_{\omega \rho} $ gives the major contribution to $ \Delta B - \Delta B_{\urm{C}} $.
This is nothing but the well-known Okamoto-Nolen-Schiffer anomaly~\cite{
  Okamoto1964Phys.Lett.11_150,
  Nolen1969Annu.Rev.Nucl.Sci.19_471}.
The term $ \Delta B_{\urm{kin} + \urm{CSC}} $, arising mainly from the 
nuclear rearrangement effects by the Coulomb and CSB forces, also play a nonnegligible role.
\subsection{Connection with Skyrme-type functional}
\par
Finally, we present a connection between the $ \omega $-$ \rho^0 $ mixing 
model and a Skyrme-type CSB interaction. 
An approximate correspondence between the two models can be established 
through a gradient expansion of the energy density functional. 
\par
The CSB part of the Lagrangian density of the present model is given by
\begin{equation}
  \ca{L}_{\urm{CSB}}
  =
  \Delta m_{\urm{v}}^2 \, \omega^{\mu} \rho^0_{\mu}. 
\end{equation}
Applying a gradient expansion~\cite{TaHa12,BMMR02} of the meson propagators
$\left( -\Nabla^2 + m_{\omega,\rho}^2\right)^{-1}$
[See Eqs.~\eqref{eq:KG-omg} and \eqref{eq:KG-rho}],
we obtain the following expressions 
for $ \omega $ and $ \rho^0 $ fields to the first order in $ \Nabla^2 / m_{\omega, \, \rho}^2$ as
\begin{subequations}
  \begin{align}
    \omega
    & \approx 
      \frac{g_{\omega}}{m_{\omega}^2}
      \left( 1 + \frac{\Nabla^2}{m_{\omega}^2} \right)
      \left( \rho - \Delta m_{\urm{v}}^2 \, \rho^0 \right), \\
    \rho^0
    & \approx 
      \frac{g_{\rho}}{m_{\rho}^2}
      \left( 1 + \frac{\Nabla^2}{m_{\rho}^2} \right)
      \left( \rho_3 - \Delta m_{\urm{v}}^2 \, \omega \right).
  \end{align}
\end{subequations}
The corresponding CSB contribution of EDF,
retaining terms only to the first order in $ \Delta m_{\urm{v}}^2 $, becomes
\begin{widetext}
  \begin{align}
    E_{\urm{CSB}}
    & =
      -
      \Delta m_{\urm{v}}^2
      \int
      d^3 r \, \omega \rho^0
      \notag \\
    & =
      -
      \frac{\Delta m_{\urm{v}}^2}{m_{\omega}^2 m_{\rho}^2}
      \int
      d^3 r
      \,
      g_{\omega}
      g_{\rho}
      \left[
      \left( \rho_n^2 - \rho_p^2 \right)
      +
      \left(\frac{1}{m_{\omega}^2}+\frac{1}{m_{\rho}^2}\right)
      \left( \rho_n \Nabla^2 \rho_n - \rho_p \Nabla^2 \rho_p \right)
      +
      \left(\frac{1}{m_{\omega}^2}-\frac{1}{m_{\rho}^2}\right)
      \left( \rho_n \Nabla^2 \rho_p - \rho_p \Nabla^2 \rho_n \right)
      \right]
      \notag \\
    & \approx
      \int
      d^3 r \, 
      \left[
      a_0 \left( \rho_n^2 - \rho_p^2 \right) 
      +
      a_1 \left( \rho_n \Nabla^2 \rho_n - \rho_p \Nabla^2 \rho_p \right)
      \right],
      \label{eq:ECSB_Sk}
  \end{align}
\end{widetext}
where we have defined
\begin{subequations}
  \begin{align}
    a_0
    & =
      -\Delta m_{\urm{v}}^2
      \frac{g_{\omega} g_{\rho}}{m_{\omega}^2 m_{\rho}^2},
      \label{eq:Ska0} \\
    a_1
    & =
      \left(
      \frac{1}{m_{\omega}^2}
      +
      \frac{1}{m_{\rho}^2}
      \right)
      a_0.
      \label{eq:Ska1}
  \end{align}
\end{subequations}
The cross term $ \rho_n \Nabla^2 \rho_p - \rho_p \Nabla^2 \rho_n $ cancels out 
upon integration by parts, where for simplicity we have neglected 
the spatial dependence of the coupling constants $ g_{\omega} $ and $g_{\rho} $ 
via their density dependence.
\par
Table~\ref{tb:Skyrme} lists the values of $ a_0 $ and $ a_1 $ extracted 
from the parameter sets in the present work.
In calculating these parameters, we used the coupling constants at the saturation density, i.e., 
$ g_{\omega} \left( \rho_{\urm{sat}} \right) $ and $ g_{\rho} \left( \rho_{\urm{sat}} \right) $.
For comparison, the corresponding values of Skyrme models~\cite{
  Roca-Maza2018Phys.Rev.Lett.120_202501,
  Baczyk2019J.Phys.G46_03LT01,SNRH24}
are also shown. 
\par
The parameter $ a_0 $ of DD-ME-CSB model is 
comparable to those of the Skyrme-type CSB interactions, 
except for $ \text{SV}_{\text{T}} $-ISB (next-to-leading order)~\cite{
  Baczyk2019J.Phys.G46_03LT01}, in which the parameters of
the leading and next-to-leading order in the gradient expansion 
are fitted independently. 
\par
The values of the parameter $ a_1 $ vary significantly among the models 
listed in Table \ref{tb:Skyrme}. 
It is important to note that the Skyrme interaction of the first order
in $ \Nabla^2 $ yields not only terms of the form 
$ \rho \Nabla^2 \rho $ in the EDF like the $a_1$ term in Eq.~\eqref{eq:ECSB_Sk},
but also $ \rho \tau $ type one~\cite{
  Naito2023Phys.Rev.C107_064302}, where $\tau$ is the kinetic energy density. 
Furthermore, the interactions in Refs.~\cite{
  Roca-Maza2018Phys.Rev.Lett.120_202501,
  Baczyk2019J.Phys.G46_03LT01} 
include also a charge-independence-breaking component as well. 
However, it is worth noting that $ a_1 $ here is given simultaneously with $ a_0 $
via the relation \eqref{eq:Ska1}. 
This is in contrast with the Skyrme model, where the parameters of the leading- 
and the next-to-leading-order gradient terms are typically fitted independently, 
as is discussed in Refs.~\cite{NaGi24,SNRH24}. 
In this regard, it is also notable that the QCD-based approach in Ref.~\cite{SNRH24} 
determines the Skyrme parameters simultaneously up to the first order in $ \Nabla^2 $.
\begin{table*}
  \caption{The parameters $ a_0 $ and $ a_1 $ as defined in Eqs.~\eqref{eq:Ska0} and \eqref{eq:Ska1}. The corresponding values of 
    Skyrme models are also shown for comparison. 
    The Skyrme parameters in Refs.~\cite{
      Roca-Maza2018Phys.Rev.Lett.120_202501,
      SNRH24}
    are related to $ a_0 $ and $ a_1 $ as 
    $ a_0 = \tilde{s}_0 / 8 $ and 
    $ a_1 = \frac{3}{64} \left( - \tilde{s}_1 + \tilde{s}_2 \right) $
    with 
    $ \tilde{s}_0 = s_0 \left( 1 - y_0 \right) $,
    $ \tilde{s}_1 = s_1 \left( 1 - y_1 \right) $, and 
    $ \tilde{s}_2 = s_2 \left( 1 + y_2 \right) $,
    while those in Ref.~\cite{Baczyk2019J.Phys.G46_03LT01} as 
    $ a_0 = t_0^{\urm{III}}/2 $ and 
    $ a_1 = \frac{3}{16} \left( - t_1^{\urm{III}} + t_2^{\urm{III}} \right) $.}
  \label{tb:Skyrme}
  \begin{ruledtabular}
    \begin{tabular}{ldd}
      Parameter set & \multicolumn{1}{c}{$ a_0 $ ($ \mathrm{MeV} \, \mathrm{fm}^3 $)} & \multicolumn{1}{c}{$ a_1 $ ($ \mathrm{MeV} \, \mathrm{fm}^5 $)} \\
      \hline
      DD-ME-CSB   & -3.421 & -0.4461 \\
      DD-ME-CSB00 &  0     &  0      \\
      DD-ME-CSB05 & -2.364 & -0.3083 \\
      DD-ME-CSB10 & -4.774 & -0.6226 \\
      DD-ME-CSB15 & -7.104 & -0.9264 \\
      \hline
      SAMi-ISB \cite{Roca-Maza2018Phys.Rev.Lett.120_202501}
                    & -6.6 \pm 0.2 & 0 \\
      $ \text{SV}_{\text{T}} $-ISB (leading order) \cite{Baczyk2019J.Phys.G46_03LT01}
                    & -3.65 \pm 0.15 & 0 \\
      $ \text{SV}_{\text{T}} $-ISB (next-to-leading order) \cite{Baczyk2019J.Phys.G46_03LT01}
                    & 5.5 \pm 1.0 & 1.16\pm 0.77 \\
      QCD-based (case I) \cite{SNRH24}
                    & -1.9^{+1.1}_{-1.6} & -0.024^{+0.020}_{-0.023} \\
      QCD-based (case II) \cite{SNRH24}
                    & -1.9^{+1.1}_{-1.6} & 0.0084^{+0.0066}_{-0.0047} \\
    \end{tabular}
  \end{ruledtabular}
\end{table*}
\section{Summary and perspectives}
\label{sec:summary}
\par
In summary, we have developed a relativistic mean-field model that incorporates the CSB of the nuclear force through
$ \omega $-$ \rho^0 $ mixing, and we have investigated its performance in the nuclear matter and the mirror nuclei with the isospin $T$=1/2, 1 and 2.
To better constrain the $ \omega $-$ \rho^0 $ mixing effect, we have also included corrections to 
the electromagnetic interaction: namely, the nucleon electromagnetic form factors, the first-order vacuum polarization (VP) effect, and the exact Coulomb exchange and the pairing terms.
The parameters of the meson-exchange interaction in the $ ph $ channel,
including the $ \omega $-$ \rho^0 $ mixing strength $ \Delta m_{\urm{v}}^2 $,
were optimized to reproduce the mass differences of mirror nuclei with the isospin $ T = 1/2 $,
as well as the binding energies and charge radii 
of doubly and semi-magic nuclei.
The resulting parameter set is named DD-ME-CSB.
In addition, we introduced four other parameter sets: DD-ME-CSB00, DD-ME-CSB05, DD-ME-CSB10, and DD-ME-CSB15,
in which $ \Delta m_{\urm{v}}^2 $ is fixed at $ 0 $, $ 0.05 $, $ 0.10 $, and $ 0.15 \, \mathrm{fm}^{-2} $, respectively.
\par
We evaluated the performance of these five parameter sets and found that all perform 
comparably to existing models in reproducing the binding energies and charge radii of 
doubly- and semi-magic nuclei (See also Appendix~\ref{app:CSB(00-15)}).
Regarding the mass differences of the mirror nuclei,
the DD-ME-CSB set yields the best agreement with experimental data not only for the $ T = 1/2 $ nuclei included
in the optimization but also for the $ T = 1 $ and $ 2 $ systems.
The $ \omega $-$ \rho^0 $ mixing term gives the major contribution to the mirror binding 
energy differences,
although the effects of the nuclear 
rearrangement induced by the Coulomb and CSB forces are also non-negligible.
\par
Furthermore, a connection between the $ \omega $-$ \rho^0 $ mixing model
and a Skyrme-type CSB interaction was established by the gradient expansion method.
The corresponding leading-order Skyrme CSB parameter extracted from the DD-ME-CSB 
model is found to be consistent with those of existing phenomenological Skyrme energy density functionals.
\par
One possible extension of the present model is to incorporate CSB in the 
$ N $-$ \Lambda $ interaction. Experimental evidence for such effects has been observed 
in the binding energy differences of mirror $ \Lambda $ hypernuclei 
in the $ s $- and $ p $-shell regions~\cite{Botta17,STSH25}. 
While mechanisms other than $ \omega $-$ \rho^0 $ mixing, 
such as $ \Lambda $-$ \Sigma^0 $ mixing~\cite{Dalitz64},
are expected to play roles,
the $ \omega $-$ \rho^0 $ mixing may have significant contribution  to the CSB of $ N $-$ \Lambda $ 
interaction as well as to the $ N $-$ N $ interaction,
and thus provides a global starting point for developing a CSB model for hypernuclei as well as non-hyper nuclei based on the same RMF model.

\begin{acknowledgments}
  Y.~T.~acknowledges supports from the Basis Science Research Program of the National 
  Research Foundation of Korea under Grant Nos.~RS-2024-00361003, RS-2024-00460031, and RS-2021-NR60129.
  T.~N.~acknolwedges
  the JSPS Grant-in-Aid for Transformative Research Areas (A) under Grant No.~JP25H01558,
  the JSPS Grant-in-Aid for Scientific Research (S) under Grant No.~JP25H00402,
  the JSPS Grant-in-Aid for Scientific Research (B) under Grant Nos.~JP23K26538 and JP25K01003,
  the JSPS Grant-in-Aid for Scientific Research (C) under Grant No.~JP23K03426,
  the JSPS Grant-in-Aid for Early-Career Scientists under Grant No.~JP24K17057,
  and
  the JSPS Grant-in-Aid for JSPS Fellows under Grant No.~JP25KJ0405.
\end{acknowledgments}
\appendix
\begin{widetext}
  \section{Coulomb-exchange and -pairing energies}
  \label{app:coulomb}
  \par
  In this appendix, we present the exchange and pairing contributions of the Coulomb interaction energy. 
  In the present calculations, we neglect the finite size of the nucleons and vacuum polarization (VP) effects. 
  In the second-quantization formalism, the charge density operator is expressed as
  \begin{equation}
    \hat{\rho}_{\urm{ch}} \left( \ve{r} \right)
    =
    \sum_{\tau = p, \, n} 
    \sum_{\alpha \beta}
    \left(
      \delta_{\tau p}
      \delta_{\alpha \beta}
      +
      \frac{\kappa_{\tau}}{2m_{\tau}}
      \Nabla \cdot i \ve{\gamma}_{\alpha \beta}
    \right)
    \hat{\psi}_{\tau}^{\dagger} \left(\ve{r} \alpha \right)
    \hat{\psi}_{\tau} \left( \ve{r} \beta \right)
  \end{equation}
  where $ \hat{\psi}_{\tau} $ ($ \tau = n $ or $ p $) denotes the neutron or proton field operator, 
  $ \alpha $ and $ \beta $ are Dirac indices, and $ \kappa_{\tau} $ is the anomalous magnetic moment of the nucleon of type $ \tau $.
  The exchange ($ E_{\urm{Cx}} $) and pairing ($ E_{\urm{Cpair}} $) contributions to the Coulomb energy are given by
  \begin{align}
    E_{\urm{Cx}} + E_{\urm{Cpair}}
    & =
      -
      \frac{e^2}{2}
      \sum_{\tau = p, \, n} 
      \sum_{\alpha \beta \gamma \delta}
      \int d^3r \, d^3r' \, 
      \frac{1}{\left| \ve{r} -\ve{r}' \right|}
      \left( 
      \delta_{\tau p} \delta_{\alpha \beta}
      +
      \frac{\kappa_{\tau}}{2m_{\tau}}
      i \Nabla \cdot \ve{\gamma}_{\alpha \beta}
      \right)
      \left(
      \delta_{\tau p} \delta_{\gamma \delta}
      +
      \frac{\kappa_{\tau}}{2m_{\tau}}
      i \Nabla' \cdot \ve{\gamma}_{\gamma \delta}
      \right)
      \notag \\
    & \qquad
      \times
      \left[ 
      -
      \avr{\hat{\psi}_{\tau}^{\dagger} \left( \ve{r} \alpha \right) \hat{\psi}_{\tau} \left( \ve{r}' \delta \right)}
      \avr{\hat{\psi}_{\tau}^{\dagger} \left( \ve{r}' \gamma \right) \hat{\psi}_{\tau} \left( \ve{r} \beta \right)}
      +
      \avr{\hat{\psi}_{\tau}^{\dagger} \left( \ve{r} \alpha \right) \hat{\psi}_{\tau}^{\dagger} \left( \ve{r}' \gamma \right)}
      \avr{\hat{\psi}_{\tau} \left( \ve{r}' \delta \right) \hat{\psi}_{\tau} \left( \ve{r} \beta \right)}
      \right], 
  \end{align}
  where the first term inside the square brackets represents the exchange contribution, 
  and the second term corresponds to the pairing contribution.
  \par
  With spherical symmetry, the quasiparticle wave functions are eigenstates of total angular momentum 
  and are written as
  \begin{equation}
    \psi_{a m} \left( \ve{r} \right)
    =
    \frac{1}{r}
    \begin{pmatrix}
      G_a^{\urm{U}}    \left( r \right) \\
      - iF_a^{\urm{U}} \left( r \right) \ve{\sigma} \cdot \hat{\ve{r}} \\
      G_a^{\urm{V}}    \left( r \right) \\
      - iF_a^{\urm{V}} \left( r \right) \ve{\sigma} \cdot \hat{\ve{r}}
    \end{pmatrix}
    \ca{Y}_{l_a j_a m} \left( \theta, \phi \right),
  \end{equation}
  where $ \ca{Y}_{ljm} \left( \theta, \phi \right) $ is the spinor spherical harmonic with 
  orbital angular momentum $ l $, total angular momentum $ j $, and magnetic quantum number $ m $, 
  and $ \hat{\ve{r}} = \ve{r}/r $ is the unit radial vector. 
  The label $ a \equiv \left( n_a, \kappa_a \right) $ denotes the radial quantum number $ n_a $ and 
  the angular quantum number $ \kappa_a $ specifying $ j_a $ and $ l_a $ as 
  $ \kappa_a = \mp \left( j_a + 1/2 \right) $ for $ j_a = l_a \pm 1/2 $.
  To evaluate the exchange and pairing contributions in the spherical case, 
  we make use of the multipole expansion of the Coulomb potential: 
  \begin{equation}
    \frac{1}{\left| \ve{r} - \ve{r}' \right|}
    =
    \sum_{\lambda}
    f_{\lambda} \left( r, r' \right)
    \frac{4 \pi}{2 \lambda + 1}
    \sum_{\mu}
    Y_{\lambda \mu} \left( \ve{r} \right)^*
    Y_{\lambda \mu} \left( \ve{r}' \right), 
  \end{equation}
  with $ f_{\lambda} \left(r, r' \right) = r_{<}^{\lambda} / r_{>}^{\lambda + 1} $,
  $ r_{<} \equiv \min \left( r, r' \right) $ and $ r_{>} \equiv \max \left(  r, r' \right) $.
  \par 
  The exchange energy is then expressed as 
  \begin{align}
    E_{\urm{Cx}} 
    & = 
      -
      \frac{e^2}{2}
      \sum_{\tau = p, \, n} 
      \sum_{ab \in \tau}
      \left( 2 j_a + 1 \right)
      \left( 2 j_b + 1 \right)
      \sum_{\lambda}
      \frac{1 + \left( -1 \right)^{l_a + \lambda + l_b}}{2}
      \begin{pmatrix}
        j_a & \lambda & j_b \\
        1/2 & 0 & -1/2
      \end{pmatrix}^2
                  \notag \\
    & \qquad
      \times
      \int dr \, dr'\, 
      f_{\lambda} \left( r, r' \right)
      \left[
      \delta_{\tau p}
      R_{ab}^{\urm{(1)}} \left( r \right)
      +
      \frac{\kappa_{\tau}}{2m_{\tau}}
      R_{ab}^{\urm{(2)}} \left( r \right)
      \right]
      \left[
      \delta_{\tau p}
      R_{ba}^{\urm{(1)}} \left( r' \right)
      +
      \frac{\kappa_{\tau}}{2m_{\tau}}
      R_{ba}^{\urm{(2)}} \left( r' \right) 
      \right], 
      \label{eq:ECx}
  \end{align}
  where 
  \begin{subequations}
    \begin{align}
      R_{ab}^{\urm{(1)}} \left( r \right) 
      & \equiv
        G_a^{\urm{V} *} \left( r \right) 
        G_b^{\urm{V}} \left( r \right)
        +
        F_a^{\urm{V} *} \left( r \right)
        F_b^{\urm{V}} \left( r \right), \\
      R_{ab}^{\urm{(2)}} \left( r \right) 
      & \equiv
        G_a^{\urm{V} *} \left( r \right) 
        \tilde{F}_b^{\urm{V}} \left( r \right)
        +
        F_a^{\urm{V} *} \left( r \right)
        \tilde{G}_b^{\urm{V}} \left( r \right)
        + 
        \tilde{G}_a^{\urm{V} *} \left( r \right)
        F_b^{\urm{V}} \left( r \right)
        +
        \tilde{F}_a^{\urm{V} *} \left( r \right)
        G_b^{\urm{V}} \left( r \right), 
    \end{align}
  \end{subequations}
  and
  \begin{equation}
    \tilde{G}_a \left( r \right) 
    = 
    \left( \frac{d}{dr} + \frac{\kappa_a}{r} \right) G_a \left( r \right),
    \qquad    
    \tilde{F}_a \left( r \right) = \left( \frac{d}{dr} - \frac{\kappa_a}{r}\right) F_a \left( r \right).
  \end{equation}
  \par
  The pairing energy is calculated by neglecting the small components $ F_a^{\urm{U}} $ and $ F_a^{\urm{V}} $ of the quasiparticle wave functions:
  \begin{align}
    E_{\urm{Cpair}}
    & = 
      \frac{e^2}{2}
      \sum_{ab \in p}
      \left( 2 j_a + 1 \right)
      \left( 2 j_b + 1 \right)
      \sum_{\lambda}
      \frac{1 + \left( -1 \right)^{l_a + \lambda + l_b}}{2}
      \begin{pmatrix}
        j_a & \lambda & j_b \\
        1/2 & 0 & -1/2
      \end{pmatrix}^2
                  \notag \\
    & \qquad
      \times
      \int dr \, dr'\, 
      G^{\urm{V} *}_{b} \left( r \right)
      G_a^{\urm{V}} \left( r \right) 
      f_{\lambda} \left( r, r' \right)
      G_a^{\urm{U} *} \left( r' \right)
      G_b^{\urm{U}} \left( r' \right).
      \label{eq:ECp}
  \end{align}
  \par
  The Coulomb exchange and pairing energies as given in Eqs.~\eqref{eq:ECx} and \eqref{eq:ECp} are 
  added perturbatively after we obtain a self consistent solution. 
  \section{Charge radius with center-of-mass correction}
  \label{app:cmcorr}
  \par
  The mean-square charge radius is defined as 
  \begin{equation}
    \avr{r^2}_{\urm{ch}}
    =
    -
    \frac{\left. \Nabla^2 \tilde{\rho}_{\urm{ch}} \left( \ve{q} \right) \right|_{\ve{q} = \ve{0}}}{Z}.
  \end{equation}
  \par
  To account for the center-of-mass (CM) correction to the mean-square charge radius~\cite{TaCh24},
  the nuclear charge form factor is modified using the corrected charge-density operator as 
  \begin{equation}
    \tilde{\rho}'_{\urm{ch}} \left( \ve{q} \right)
    =
    \avr{
      \sum_{k = 1}^A
      e^{i \ve{q} \cdot \left( \hat{\ve{r}}_k - \hat{\ve{R}}_G \right)}
      \left[
        F_{1k} \left( \ve{q}^2 \right)
        +
        \frac{\kappa_k}{2m}
        F_{2k} \left( \ve{q}^2 \right)
        \ve{q} \cdot \hat{\ve{\gamma}}_k
      \right]},
  \end{equation}
  where $ \hat{\ve{r}}_k $ is the position operator of the $ k $th nucleon, 
  $ \hat{\ve{R}}_G = \left( 1 / A \right) \sum_{k = 1}^A \hat{\ve{r}}_k $ is the CM operator, 
  and $ \hat{\ve{\gamma}}_k $ is the Dirac matrix acting on the $ k $th nucleon,
  Here, $ F_{1k} $, $ F_{2k} $, and $ \kappa_k $ are
  the nucleon form factors and anomalous magnetic moment, respectively, for the isospin of nucleon $ k $. 
  This correction leads to the following expression for the mean-square charge radius:
  \begin{align}
    \avr{r^2}'_{\urm{ch}}
    & =
      - \frac{\left. \Nabla^2 \tilde{\rho}'_{\urm{ch}} \left( \ve{q} \right) \right|_{\ve{q} = \ve{0}}}{Z}
      \notag \\
    & = 
      \frac{1}{Z}
      \avr{\sum_{k \in p} \left( \hat{\ve{r}}_k - \hat{\ve{R}}_G \right)^2}
      -
      \frac{2}{Z}
      \sum_{\tau = p, \, n}
      \frac{\kappa_{\tau}}{2m}
      \avr{\sum_{k \in \tau} \left( \hat{\ve{r}}_k - \hat{\ve{R}}_G \right) \cdot i \hat{\ve{\gamma}}_k}
      +
      C_p
      +
      \frac{N}{Z}
      C_n
  \end{align}
  with
  \begin{equation}
    C_{\tau}
    =
    - 6
    \left.
      \frac{d F_{1 \tau} \left( 0 \right)}{d \ve{q}^2}
    \right|_{\ve{q}^2 = 0}.
  \end{equation}
  The root-mean-square charge radius $ R_{\urm{ch}} $ corrected for the CM motion is then given by 
  \begin{equation}
    R_{\urm{ch}}
    =
    \sqrt{\avr{r^2}'_{\urm{ch}}}.
  \end{equation}
  \section{Parameter sets DD-ME-CSB (00--15)}
  \label{app:CSB(00-15)}
  \par
  In this appendix, we summarize and compare the performance of four parameter sets
  DD-ME-CSB00, DD-ME-CSB05, DD-ME-CSB10, and DD-ME-CSB15,
  in which the $ \omega $-$ \rho^0 $ mixing parameter is fixed at
  $ \Delta m_{\urm{v}}^2 = 0 $, $ 0.05 $, $ 0.10 $, and $ 0.15 \, \mathrm{fm}^{-2} $, respectively.
  The corresponding parameter values are listed in Table~\ref{tb:CSB00-15}.
  \par
  Table~\ref{tb:MatterProperties00-15} presents the nuclear matter properties of these sets 
  as those shown in Table~\ref{tb:MatterProperties} for DD-ME-CSB. 
  The matter properties are generally similar
  among the DD-ME-CSB (00--15) sets and comparable with those of DD-ME-CSB,
  except that the slope parameter $ L $ of the symmetry energy tends to decrease
  as $ \Delta m_{\urm{v}}^2 $ increases.
  As discussed in Sec.~\ref{sec:results},
  this behavior arises from the correlation between the isovector and CSB components of the interaction.
  \par
  Figure~\ref{fig:fit2} shows a comparison, analogous to Fig.~\ref{fig:fit1},
  among the DD-ME-CSB and DD-ME-CSB (00--15). 
  The performance of all five parameter sets is found to be quite similar.
  \begin{table}
    \caption{Parameter sets DD-ME-CSB (00--15).}
    \label{tb:CSB00-15}
    \begin{ruledtabular}
      \begin{tabular}{ldddd}
        Parameter & \multicolumn{1}{c}{DD-ME-CSB00} & \multicolumn{1}{c}{DD-ME-CSB05} & \multicolumn{1}{c}{DD-ME-CSB10} & \multicolumn{1}{c}{DD-ME-CSB15}  \\
        \hline
        $ m_{\sigma} $ ($ \mathrm{MeV} $)              & 550.0379 & 549.9987 & 550.0015 & 550.0417 \\
        $ m_{\omega} $ ($ \mathrm{MeV} $)              & 783.0000 & 783.0000 & 783.0000 & 783.0000 \\
        $ m_{\rho} $   ($ \mathrm{MeV} $)              & 763.0000 & 763.0000 & 763.0000 & 763.0000 \\
        $ g_{\sigma} \left( \rho_{\urm{sat}} \right) $ & 10.78006 & 10.79196 & 10.79436 & 10.80112 \\     
        $ g_{\omega} \left( \rho_{\urm{sat}} \right) $ & 13.37023 & 13.38988 & 13.39438 & 13.40461 \\     
        $ g_{\rho}   \left( \rho_{\urm{sat}} \right) $ & 4.051779 & 4.213082 & 4.252404 & 4.215265 \\     
        $ a_{\sigma} $                                 & 1.342274 & 1.347672 & 1.352526 & 1.360665 \\     
        $ b_{\sigma} $                                 & 1.236565 & 1.261488 & 1.262817 & 1.246289 \\     
        $ c_{\sigma} $                                 & 1.827884 & 1.872017 & 1.882629 & 1.874215 \\     
        $ d_{\sigma} $                                 & 0.427037 & 0.421973 & 0.420782 & 0.421725 \\     
        $ a_{\omega} $                                 & 1.341773 & 1.346804 & 1.351604 & 1.360025 \\     
        $ b_{\omega} $                                 & 1.073411 & 1.074346 & 1.056895 & 1.014033 \\     
        $ c_{\omega} $                                 & 1.601441 & 1.610772 & 1.594069 & 1.547045 \\     
        $ d_{\omega} $                                 & 0.456230 & 0.454907 & 0.457284 & 0.464182 \\     
        $ a_{\rho} $                                   & 0.242800 & 0.224797 & 0.329958 & 0.475520 \\     
        $ \Delta m_{\urm{v}}^2 $ ($ \mathrm{fm}^{-2} $)        & 0 & 0.05 & 0.10 & 0.15\\
      \end{tabular}
    \end{ruledtabular}
  \end{table}
  \begin{table}
    \caption{Matter properties of the parameter sets DD-ME-CSB (00--15).}
    \label{tb:MatterProperties00-15}
    \begin{ruledtabular}
      \begin{tabular}{ldddd}
        & \multicolumn{1}{c}{DD-ME-CSB00} & \multicolumn{1}{c}{DD-ME-CSB05} & \multicolumn{1}{c}{DD-ME-CSB10} & \multicolumn{1}{c}{DD-ME-CSB15}  \\
        \hline
        $ \rho_{\urm{sat}} $              ($ \mathrm{fm}^{-3} $) & 0.147  & 0.147  & 0.147  & 0.148  \\
        $ e_{0, 0} $                      ($ \mathrm{MeV} $)     & -15.94 & -15.90 & -15.89 & -15.86 \\
        $ m_n^* / m_n $                                          & 0.568  & 0.567  & 0.565  & 0.563  \\
        $ m_p^* / m_p $                                          & 0.567  & 0.566  & 0.565  & 0.562  \\
        $ K_{\infty} $                    ($ \mathrm{MeV} $)     & 284.36 & 281.39 & 279.43 & 274.69 \\
        $ J_{\urm{sym}} $                 ($ \mathrm{MeV} $)     & 34.3   & 35.6   & 36.1   & 36.0   \\
        $ J_{\urm{CSB}} $                 ($ \mathrm{MeV} $)     & -0.064 & -0.411 & -0.767 & -1.116 \\
        $ J_{\urm{CSB}} + J_{\urm{sym}} $ ($ \mathrm{MeV} $)     & 34.25  & 35.23  & 35.33  & 34.87  \\
        $ L_{\urm{sym}} $                 ($ \mathrm{MeV} $)     & 81.7   & 85.7   & 75.5   & 60.4   \\
        $ L_{\urm{CSB}} $                 ($ \mathrm{MeV} $)     & -0.283 & -0.978 & -1.466 & -1.582 \\
        $ L_{\urm{CSB}} + L_{\urm{sym}} $ ($ \mathrm{MeV} $)     & 81.45  & 84.73  & 74.01  & 58.79  \\
      \end{tabular}
    \end{ruledtabular}
  \end{table}
  \begin{figure}
    \includegraphics[width=9cm]{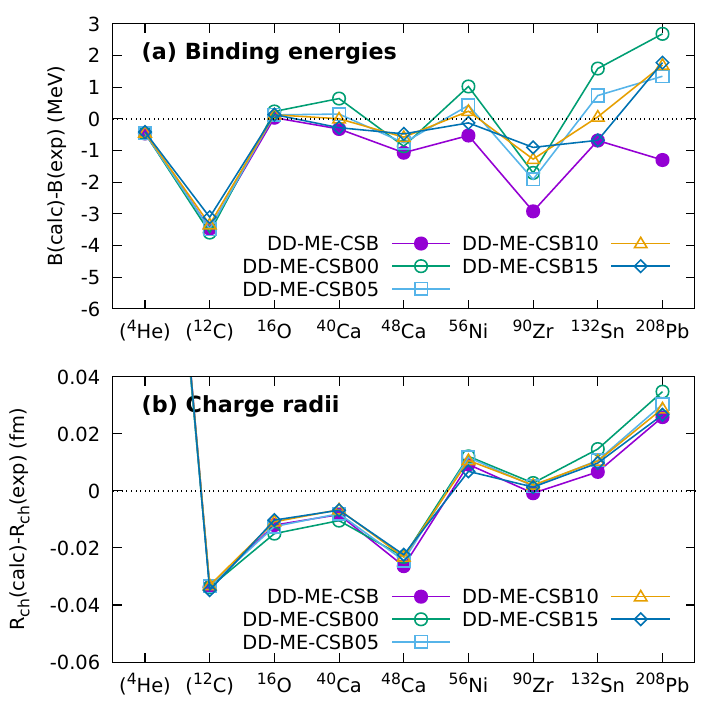}
    \caption{Deviations from the experimental data of
      (a) binding energies and (b) charge radii 
      for doubly- or semi-magic nuclei.
      The results of the DD-ME-CSB parameter set are compared to 
      those obtained with DD-ME-CSB (00--15).
      Note that $ \nuc{He}{4}{} $ and $ \nuc{C}{12}{} $ are not included in the fit.}
    \label{fig:fit2}
  \end{figure}
  \section{Neutron-skin thicknesses}
  \label{app:nskin}
  \par
  Here, we examine the correlation between the neutron-skin thicknesses and the slope parameter of 
  the pure neutron matter EoS.
  Figure~\ref{fig:nskin} shows the neutron-skin thicknesses $ \Delta r_{np} $ of
  $ \nuc{Ca}{48}{} $ and $ \nuc{Pb}{208}{} $ nuclei plotted against the slope parameter
  $ L_{\urm{sym}} + L_{\urm{CSB}} $ for the five models developed in this work. 
  Clear correlations between $ \Delta r_{np} $ and the slope parameter are observed, 
  consistent with trends found in other mean-field models. 
  The present results lie close to the established $ \Delta r_{np} $-$ L $ correlation reported in the literature 
  (see, e.g., Refs.~\cite{crex,RCVW11}).
  \begin{figure}
    \includegraphics[width=9cm]{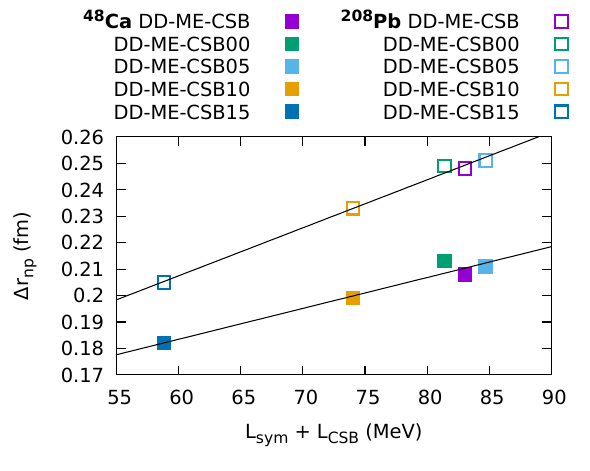}
    \caption{Neutron skin thicknesses of $ \nuc{Ca}{48}{} $ (solid squares) and $ \nuc{Pb}{208}{} $ (open squares) 
      obtained using DD-ME-CSB and DD-ME-CSB (00--15). 
      The black lines represent linear fits of the form $ aL + b $ for each nucleus:
      $ a = 0.0018 \, \mathrm{fm} / \mathrm{MeV} $ and $ b = 0.099 \, \mathrm{fm} $ for $ \nuc{Ca}{48}{} $
      and $ a = 0.0012 \, \mathrm{fm} / \mathrm{MeV} $ and $ b = 0.11 \, \mathrm{fm} $ for $ \nuc{Pb}{208}{} $.}
    \label{fig:nskin}
  \end{figure}
  \section{Decompositions of mass differences of mirror nuclei}
  \par
  In this appendix, we present a decomposition of the mass difference of mirror nuclei
  $\Delta B$ into the following contributions:  
  the Coulomb direct energy      $ \Delta B_{\urm{Cd}} $, 
  the Coulomb exchange energy    $ \Delta B_{\urm{Cx}} $, 
  the Coulomb pairing energy     $ \Delta B_{\urm{Cpair}} $, 
  the vacuum-polarization energy $ \Delta B_{\urm{CVP}} $, 
  the kinetic energy             $ \Delta B_{\urm{kin}} $, 
  the charge-symmetry-conserving part of nuclear potential energy $ \Delta B_{\urm{CSC}} $,
  and the $ \omega $-$ \rho^0 $ mixing energy $ \Delta B_{\omega \rho} $. 
  These are listed in Table~\ref{tb:DelB-decomposition} for 
  $ \nuc{K}{39}{} $-$ \nuc{Ca}{39}{} $, $ \nuc{Ca}{41}{} $-$ \nuc{Sc}{41}{} $,
  $ \nuc{Ar}{38}{} $-$ \nuc{Ca}{38}{} $, $ \nuc{Ca}{42}{} $-$ \nuc{Ti}{42}{} $,
  $ \nuc{S}{36}{} $-$ \nuc{Ca}{36}{} $, and $ \nuc{Ca}{44}{} $-$ \nuc{Cr}{44}{} $ pairs.
  \par
  As also seen in Fig.~\ref{fig:DelB_dcmp}, in addition to the dominant Coulomb terms,
  the kinetic energy ($ \Delta B_{\urm{kin}} $),
  charge-symmetry-conserving nuclear potential ($ \Delta B_{\urm{CSC}} $),
  and $ \omega $-$ \rho^0 $ mixing energy ($ \Delta B_{\omega \rho} $)
  make significant contributions to the total $ \Delta B $.
  Since the mirror nuclei listed in the table are close to $ \nuc{Ca}{40}{} $
  and have similar masses,
  the value of $ \Delta B_{\omega \rho}$ is approximately proportional to $ \left| T_z \right| = T $.
  \begin{table}
    \caption{Decomposition of the mirror binding energy difference $ \Delta B $,
      obtained with DD-ME-CSB parameter set,
      into differences of 
      the Coulomb direct energy      $ \Delta B_{\urm{Cd}} $, 
      the Coulomb exchange energy    $ \Delta B_{\urm{Cx}} $, 
      the Coulomb pairing energy     $ \Delta B_{\urm{Cpair}} $, 
      the vacuum-polarization energy $ \Delta B_{\urm{CVP}} $, 
      the kinetic energy             $ \Delta B_{\urm{kin}} $, 
      the charge-symmetry-conserving part of nuclear potential energy $ \Delta B_{\urm{CSC}} $,
      and the $ \omega $-$ \rho^0 $ mixing energy $ \Delta B_{\omega \rho} $. 
      The measured values of $ \Delta B $ are also listed in the last row.
      All values are given in unit of MeV.}
    \label{tb:DelB-decomposition}
    \begin{ruledtabular}
      \begin{tabular}{ldddddd}
        & \multicolumn{2}{c}{$ T = 1/2 $}  
        & \multicolumn{2}{c}{$ T = 1 $}  
        & \multicolumn{2}{c}{$ T = 2 $}  \\
        & \multicolumn{1}{c}{$ \nuc{K}{39}{} $-$ \nuc{Ca}{39}{} $}
        & \multicolumn{1}{c}{$ \nuc{Ca}{41}{} $-$ \nuc{Sc}{41}{} $}
        & \multicolumn{1}{c}{$ \nuc{Ar}{38}{} $-$ \nuc{Ca}{38}{} $}
        & \multicolumn{1}{c}{$ \nuc{Ca}{42}{} $-$ \nuc{Ti}{42}{} $}
        & \multicolumn{1}{c}{$ \nuc{S}{36}{} $-$ \nuc{Ca}{36}{} $}
        & \multicolumn{1}{c}{$ \nuc{Ca}{44}{} $-$ \nuc{Cr}{44}{} $} \\
        \hline
        $ \Delta B_{\urm{Cd}} $     &  7.00 &  6.73 & 13.60 &  13.76 &  25.90 &  28.62 \\
        $ \Delta B_{\urm{Cx}} $     & -0.52 & -0.23 & -1.00 &  -0.46 &  -1.68 &  -1.14 \\
        $ \Delta B_{\urm{Cpair}} $  &  0.00 &  0.00 & -0.40 &   0.48 &  -0.37 &   0.63 \\
        $ \Delta B_{\urm{CVP}} $    &  0.07 &  0.05 &  0.13 &   0.11 &   0.25 &   0.22 \\
        $ \Delta B_{\urm{kin}} $    & -0.30 & -0.70 & -0.50 &  -1.12 &  -1.54 &  -2.57 \\
        $ \Delta B_{\urm{CSC}} $    &  0.58 &  0.96 &  1.06 &   1.66 &   2.66 &   3.72 \\
        $ \Delta B_{\omega \rho} $  &  0.56 &  0.49 &  1.08 &   0.99 &   2.09 &   2.04 \\
        Total $ \Delta B$           &  7.40 &  7.30 & 13.97 &  15.41 &  27.32 &  31.52 \\
        \hline
        Exp.~$ \Delta B$           &  7.31 &  7.28 & 14.23 &  15.01 &  27.35 &  31.19 \\
      \end{tabular}
    \end{ruledtabular}
  \end{table}
\end{widetext}
\bibliography{DD-ME-CSB}

\begin{thebibliography}{88}%
\makeatletter
\providecommand \@ifxundefined [1]{%
 \@ifx{#1\undefined}
}%
\providecommand \@ifnum [1]{%
 \ifnum #1\expandafter \@firstoftwo
 \else \expandafter \@secondoftwo
 \fi
}%
\providecommand \@ifx [1]{%
 \ifx #1\expandafter \@firstoftwo
 \else \expandafter \@secondoftwo
 \fi
}%
\providecommand \natexlab [1]{#1}%
\providecommand \enquote  [1]{``#1''}%
\providecommand \bibnamefont  [1]{#1}%
\providecommand \bibfnamefont [1]{#1}%
\providecommand \citenamefont [1]{#1}%
\providecommand \href@noop [0]{\@secondoftwo}%
\providecommand \href [0]{\begingroup \@sanitize@url \@href}%
\providecommand \@href[1]{\@@startlink{#1}\@@href}%
\providecommand \@@href[1]{\endgroup#1\@@endlink}%
\providecommand \@sanitize@url [0]{\catcode `\\12\catcode `\$12\catcode `\&12\catcode `\#12\catcode `\^12\catcode `\_12\catcode `\%12\relax}%
\providecommand \@@startlink[1]{}%
\providecommand \@@endlink[0]{}%
\providecommand \url  [0]{\begingroup\@sanitize@url \@url }%
\providecommand \@url [1]{\endgroup\@href {#1}{\urlprefix }}%
\providecommand \urlprefix  [0]{URL }%
\providecommand \Eprint [0]{\href }%
\providecommand \doibase [0]{https://doi.org/}%
\providecommand \selectlanguage [0]{\@gobble}%
\providecommand \bibinfo  [0]{\@secondoftwo}%
\providecommand \bibfield  [0]{\@secondoftwo}%
\providecommand \translation [1]{[#1]}%
\providecommand \BibitemOpen [0]{}%
\providecommand \bibitemStop [0]{}%
\providecommand \bibitemNoStop [0]{.\EOS\space}%
\providecommand \EOS [0]{\spacefactor3000\relax}%
\providecommand \BibitemShut  [1]{\csname bibitem#1\endcsname}%
\let\auto@bib@innerbib\@empty
\bibitem [{\citenamefont {Okamoto}(1964)}]{Okamoto1964Phys.Lett.11_150}%
  \BibitemOpen
  \bibfield  {author} {\bibinfo {author} {\bibfnamefont {K.}~\bibnamefont {Okamoto}},\ }\bibfield  {title} {\bibinfo {title} {{Coulomb energy of $ \mathrm{He}^3 $ and possible charge asymmetry of nuclear forces}},\ }\href {https://doi.org/10.1016/0031-9163(64)90650-X} {\bibfield  {journal} {\bibinfo  {journal} {Phys. Lett.}\ }\textbf {\bibinfo {volume} {11}},\ \bibinfo {pages} {150} (\bibinfo {year} {1964})}\BibitemShut {NoStop}%
\bibitem [{\citenamefont {Nolen}\ and\ \citenamefont {Schiffer}(1969)}]{Nolen1969Annu.Rev.Nucl.Sci.19_471}%
  \BibitemOpen
  \bibfield  {author} {\bibinfo {author} {\bibfnamefont {J.~A.}\ \bibnamefont {Nolen}, \bibfnamefont {Jr.}}\ and\ \bibinfo {author} {\bibfnamefont {J.~P.}\ \bibnamefont {Schiffer}},\ }\bibfield  {title} {\bibinfo {title} {{Coulomb Energies}},\ }\href {https://doi.org/10.1146/annurev.ns.19.120169.002351} {\bibfield  {journal} {\bibinfo  {journal} {Annu. Rev. Nucl. Sci.}\ }\textbf {\bibinfo {volume} {19}},\ \bibinfo {pages} {471} (\bibinfo {year} {1969})}\BibitemShut {NoStop}%
\bibitem [{\citenamefont {Hatsuda}\ \emph {et~al.}(1990)\citenamefont {Hatsuda}, \citenamefont {Hogaasen},\ and\ \citenamefont {Prakash}}]{Hatsuda1990Phys.Rev.C42_2212}%
  \BibitemOpen
  \bibfield  {author} {\bibinfo {author} {\bibfnamefont {T.}~\bibnamefont {Hatsuda}}, \bibinfo {author} {\bibfnamefont {H.}~\bibnamefont {Hogaasen}},\ and\ \bibinfo {author} {\bibfnamefont {M.}~\bibnamefont {Prakash}},\ }\bibfield  {title} {\bibinfo {title} {{Neutron-proton mass difference in nuclei and the Okamoto-Nolen-Schiffer anomaly}},\ }\href {https://doi.org/10.1103/PhysRevC.42.2212} {\bibfield  {journal} {\bibinfo  {journal} {Phys. Rev. C}\ }\textbf {\bibinfo {volume} {42}},\ \bibinfo {pages} {2212} (\bibinfo {year} {1990})}\BibitemShut {NoStop}%
\bibitem [{\citenamefont {Hatsuda}\ \emph {et~al.}(1991)\citenamefont {Hatsuda}, \citenamefont {H\o{}gaasen},\ and\ \citenamefont {Prakash}}]{Hatsuda1991Phys.Rev.Lett.66_2851}%
  \BibitemOpen
  \bibfield  {author} {\bibinfo {author} {\bibfnamefont {T.}~\bibnamefont {Hatsuda}}, \bibinfo {author} {\bibfnamefont {H.}~\bibnamefont {H\o{}gaasen}},\ and\ \bibinfo {author} {\bibfnamefont {M.}~\bibnamefont {Prakash}},\ }\bibfield  {title} {\bibinfo {title} {{QCD sum rules in medium and the Okamoto-Nolen-Schiffer anomaly}},\ }\href {https://doi.org/10.1103/PhysRevLett.66.2851} {\bibfield  {journal} {\bibinfo  {journal} {Phys. Rev. Lett.}\ }\textbf {\bibinfo {volume} {66}},\ \bibinfo {pages} {2851} (\bibinfo {year} {1991})}\BibitemShut {NoStop}%
\bibitem [{\citenamefont {Auerbach}(1992)}]{Auerbach1992Phys.Lett.B282_263}%
  \BibitemOpen
  \bibfield  {author} {\bibinfo {author} {\bibfnamefont {N.}~\bibnamefont {Auerbach}},\ }\bibfield  {title} {\bibinfo {title} {{Comment on QCD effects in the nuclear medium, the effective nucleon mass and the Nolen-Schiffer anomaly}},\ }\href {https://doi.org/10.1016/0370-2693(92)90635-H} {\bibfield  {journal} {\bibinfo  {journal} {Phys. Lett. B}\ }\textbf {\bibinfo {volume} {282}},\ \bibinfo {pages} {263} (\bibinfo {year} {1992})}\BibitemShut {NoStop}%
\bibitem [{\citenamefont {Shahnas}(1994)}]{Shahnas1994Phys.Rev.C50_2346}%
  \BibitemOpen
  \bibfield  {author} {\bibinfo {author} {\bibfnamefont {M.~H.}\ \bibnamefont {Shahnas}},\ }\bibfield  {title} {\bibinfo {title} {{Nolen-Schiffer anomaly of mirror nuclei and charge symmetry breaking in nuclear interactions}},\ }\href {https://doi.org/10.1103/PhysRevC.50.2346} {\bibfield  {journal} {\bibinfo  {journal} {Phys. Rev. C}\ }\textbf {\bibinfo {volume} {50}},\ \bibinfo {pages} {2346} (\bibinfo {year} {1994})}\BibitemShut {NoStop}%
\bibitem [{\citenamefont {Saito}\ and\ \citenamefont {Thomas}(1994)}]{Saito1994Phys.Lett.B335_17}%
  \BibitemOpen
  \bibfield  {author} {\bibinfo {author} {\bibfnamefont {K.}~\bibnamefont {Saito}}\ and\ \bibinfo {author} {\bibfnamefont {A.~W.}\ \bibnamefont {Thomas}},\ }\bibfield  {title} {\bibinfo {title} {{The Nolen-Schiffer anomaly and isospin symmetry breaking in nuclear matter}},\ }\href {https://doi.org/10.1016/0370-2693(94)91551-2} {\bibfield  {journal} {\bibinfo  {journal} {Phys. Lett. B}\ }\textbf {\bibinfo {volume} {335}},\ \bibinfo {pages} {17} (\bibinfo {year} {1994})}\BibitemShut {NoStop}%
\bibitem [{\citenamefont {Brown}\ \emph {et~al.}(2000)\citenamefont {Brown}, \citenamefont {Richter},\ and\ \citenamefont {Lindsay}}]{Brown00}%
  \BibitemOpen
  \bibfield  {author} {\bibinfo {author} {\bibfnamefont {B.~A.}\ \bibnamefont {Brown}}, \bibinfo {author} {\bibfnamefont {W.~A.}\ \bibnamefont {Richter}},\ and\ \bibinfo {author} {\bibfnamefont {R.}~\bibnamefont {Lindsay}},\ }\bibfield  {title} {\bibinfo {title} {{Displacement energies with the Skyrme Hartree--Fock method}},\ }\href {https://doi.org/10.1016/S0370-2693(00)00589-X} {\bibfield  {journal} {\bibinfo  {journal} {Phys. Lett. B}\ }\textbf {\bibinfo {volume} {483}},\ \bibinfo {pages} {49} (\bibinfo {year} {2000})}\BibitemShut {NoStop}%
\bibitem [{\citenamefont {B\k{a}czyk}\ \emph {et~al.}(2018)\citenamefont {B\k{a}czyk}, \citenamefont {Dobaczewski}, \citenamefont {Konieczka}, \citenamefont {Satu\l{}a}, \citenamefont {Nakatsukasa},\ and\ \citenamefont {Sato}}]{Baczyk2018Phys.Lett.B778_178}%
  \BibitemOpen
  \bibfield  {author} {\bibinfo {author} {\bibfnamefont {P.}~\bibnamefont {B\k{a}czyk}}, \bibinfo {author} {\bibfnamefont {J.}~\bibnamefont {Dobaczewski}}, \bibinfo {author} {\bibfnamefont {M.}~\bibnamefont {Konieczka}}, \bibinfo {author} {\bibfnamefont {W.}~\bibnamefont {Satu\l{}a}}, \bibinfo {author} {\bibfnamefont {T.}~\bibnamefont {Nakatsukasa}},\ and\ \bibinfo {author} {\bibfnamefont {K.}~\bibnamefont {Sato}},\ }\bibfield  {title} {\bibinfo {title} {{Isospin-symmetry breaking in masses of $ N \simeq Z $ nuclei}},\ }\href {https://doi.org/10.1016/j.physletb.2017.12.068} {\bibfield  {journal} {\bibinfo  {journal} {Phys. Lett. B}\ }\textbf {\bibinfo {volume} {778}},\ \bibinfo {pages} {178} (\bibinfo {year} {2018})}\BibitemShut {NoStop}%
\bibitem [{\citenamefont {Dong}\ \emph {et~al.}(2018)\citenamefont {Dong}, \citenamefont {Zhang}, \citenamefont {Zuo}, \citenamefont {Gu}, \citenamefont {Wang},\ and\ \citenamefont {Sun}}]{Dong2018Phys.Rev.C97_021301}%
  \BibitemOpen
  \bibfield  {author} {\bibinfo {author} {\bibfnamefont {J.~M.}\ \bibnamefont {Dong}}, \bibinfo {author} {\bibfnamefont {Y.~H.}\ \bibnamefont {Zhang}}, \bibinfo {author} {\bibfnamefont {W.}~\bibnamefont {Zuo}}, \bibinfo {author} {\bibfnamefont {J.~Z.}\ \bibnamefont {Gu}}, \bibinfo {author} {\bibfnamefont {L.~J.}\ \bibnamefont {Wang}},\ and\ \bibinfo {author} {\bibfnamefont {Y.}~\bibnamefont {Sun}},\ }\bibfield  {title} {\bibinfo {title} {{Generalized isobaric multiplet mass equation and its application to the Nolen-Schiffer anomaly}},\ }\href {https://doi.org/10.1103/PhysRevC.97.021301} {\bibfield  {journal} {\bibinfo  {journal} {Phys. Rev. C}\ }\textbf {\bibinfo {volume} {97}},\ \bibinfo {pages} {021301} (\bibinfo {year} {2018})}\BibitemShut {NoStop}%
\bibitem [{\citenamefont {B{\k{a}}czyk}\ \emph {et~al.}(2019)\citenamefont {B{\k{a}}czyk}, \citenamefont {Satu{\l}a}, \citenamefont {Dobaczewski},\ and\ \citenamefont {Konieczka}}]{Baczyk2019J.Phys.G46_03LT01}%
  \BibitemOpen
  \bibfield  {author} {\bibinfo {author} {\bibfnamefont {P.}~\bibnamefont {B{\k{a}}czyk}}, \bibinfo {author} {\bibfnamefont {W.}~\bibnamefont {Satu{\l}a}}, \bibinfo {author} {\bibfnamefont {J.}~\bibnamefont {Dobaczewski}},\ and\ \bibinfo {author} {\bibfnamefont {M.}~\bibnamefont {Konieczka}},\ }\bibfield  {title} {\bibinfo {title} {{Isobaric multiplet mass equation within nuclear density functional theory}},\ }\href {https://doi.org/10.1088/1361-6471/aaffe4} {\bibfield  {journal} {\bibinfo  {journal} {J. Phys. G}\ }\textbf {\bibinfo {volume} {46}},\ \bibinfo {pages} {03LT01} (\bibinfo {year} {2019})}\BibitemShut {NoStop}%
\bibitem [{\citenamefont {Sagawa}\ \emph {et~al.}(2022)\citenamefont {Sagawa}, \citenamefont {Yoshida}, \citenamefont {Naito}, \citenamefont {Uesaka}, \citenamefont {Zenihiro}, \citenamefont {Tanaka},\ and\ \citenamefont {Suzuki}}]{Sagawa2022Phys.Lett.B829_137072}%
  \BibitemOpen
  \bibfield  {author} {\bibinfo {author} {\bibfnamefont {H.}~\bibnamefont {Sagawa}}, \bibinfo {author} {\bibfnamefont {S.}~\bibnamefont {Yoshida}}, \bibinfo {author} {\bibfnamefont {T.}~\bibnamefont {Naito}}, \bibinfo {author} {\bibfnamefont {T.}~\bibnamefont {Uesaka}}, \bibinfo {author} {\bibfnamefont {J.}~\bibnamefont {Zenihiro}}, \bibinfo {author} {\bibfnamefont {J.}~\bibnamefont {Tanaka}},\ and\ \bibinfo {author} {\bibfnamefont {T.}~\bibnamefont {Suzuki}},\ }\bibfield  {title} {\bibinfo {title} {{Isovector density and isospin impurity in $ {}^{40} \mathrm{Ca} $}},\ }\href {https://doi.org/10.1016/j.physletb.2022.137072} {\bibfield  {journal} {\bibinfo  {journal} {Phys. Lett. B}\ }\textbf {\bibinfo {volume} {829}},\ \bibinfo {pages} {137072} (\bibinfo {year} {2022})}\BibitemShut {NoStop}%
\bibitem [{\citenamefont {Naito}\ \emph {et~al.}(2022{\natexlab{a}})\citenamefont {Naito}, \citenamefont {Col\`o}, \citenamefont {Liang}, \citenamefont {Roca-Maza},\ and\ \citenamefont {Sagawa}}]{Naito2022Phys.Rev.C105_L021304}%
  \BibitemOpen
  \bibfield  {author} {\bibinfo {author} {\bibfnamefont {T.}~\bibnamefont {Naito}}, \bibinfo {author} {\bibfnamefont {G.}~\bibnamefont {Col\`o}}, \bibinfo {author} {\bibfnamefont {H.}~\bibnamefont {Liang}}, \bibinfo {author} {\bibfnamefont {X.}~\bibnamefont {Roca-Maza}},\ and\ \bibinfo {author} {\bibfnamefont {H.}~\bibnamefont {Sagawa}},\ }\bibfield  {title} {\bibinfo {title} {{Toward \textit{ab initio} charge symmetry breaking in nuclear energy density functionals}},\ }\href {https://doi.org/10.1103/PhysRevC.105.L021304} {\bibfield  {journal} {\bibinfo  {journal} {Phys. Rev. C}\ }\textbf {\bibinfo {volume} {105}},\ \bibinfo {pages} {L021304} (\bibinfo {year} {2022}{\natexlab{a}})}\BibitemShut {NoStop}%
\bibitem [{\citenamefont {Naito}\ \emph {et~al.}(2022{\natexlab{b}})\citenamefont {Naito}, \citenamefont {Roca-Maza}, \citenamefont {Col\`o}, \citenamefont {Liang},\ and\ \citenamefont {Sagawa}}]{Naito2022Phys.Rev.C106_L061306}%
  \BibitemOpen
  \bibfield  {author} {\bibinfo {author} {\bibfnamefont {T.}~\bibnamefont {Naito}}, \bibinfo {author} {\bibfnamefont {X.}~\bibnamefont {Roca-Maza}}, \bibinfo {author} {\bibfnamefont {G.}~\bibnamefont {Col\`o}}, \bibinfo {author} {\bibfnamefont {H.}~\bibnamefont {Liang}},\ and\ \bibinfo {author} {\bibfnamefont {H.}~\bibnamefont {Sagawa}},\ }\bibfield  {title} {\bibinfo {title} {{Isospin symmetry breaking in the charge radius difference of mirror nuclei}},\ }\href {https://doi.org/10.1103/PhysRevC.106.L061306} {\bibfield  {journal} {\bibinfo  {journal} {Phys. Rev. C}\ }\textbf {\bibinfo {volume} {106}},\ \bibinfo {pages} {L061306} (\bibinfo {year} {2022}{\natexlab{b}})}\BibitemShut {NoStop}%
\bibitem [{\citenamefont {Naito}\ \emph {et~al.}(2023{\natexlab{a}})\citenamefont {Naito}, \citenamefont {Col\`o}, \citenamefont {Liang}, \citenamefont {Roca-Maza},\ and\ \citenamefont {Sagawa}}]{Naito2023Phys.Rev.C107_064302}%
  \BibitemOpen
  \bibfield  {author} {\bibinfo {author} {\bibfnamefont {T.}~\bibnamefont {Naito}}, \bibinfo {author} {\bibfnamefont {G.}~\bibnamefont {Col\`o}}, \bibinfo {author} {\bibfnamefont {H.}~\bibnamefont {Liang}}, \bibinfo {author} {\bibfnamefont {X.}~\bibnamefont {Roca-Maza}},\ and\ \bibinfo {author} {\bibfnamefont {H.}~\bibnamefont {Sagawa}},\ }\bibfield  {title} {\bibinfo {title} {{Effects of Coulomb and isospin symmetry breaking interactions on neutron-skin thickness}},\ }\href {https://doi.org/10.1103/PhysRevC.107.064302} {\bibfield  {journal} {\bibinfo  {journal} {Phys. Rev. C}\ }\textbf {\bibinfo {volume} {107}},\ \bibinfo {pages} {064302} (\bibinfo {year} {2023}{\natexlab{a}})}\BibitemShut {NoStop}%
\bibitem [{\citenamefont {Sagawa}\ \emph {et~al.}(2024)\citenamefont {Sagawa}, \citenamefont {Naito}, \citenamefont {Roca-Maza},\ and\ \citenamefont {Hatsuda}}]{SNRH24}%
  \BibitemOpen
  \bibfield  {author} {\bibinfo {author} {\bibfnamefont {H.}~\bibnamefont {Sagawa}}, \bibinfo {author} {\bibfnamefont {T.}~\bibnamefont {Naito}}, \bibinfo {author} {\bibfnamefont {X.}~\bibnamefont {Roca-Maza}},\ and\ \bibinfo {author} {\bibfnamefont {T.}~\bibnamefont {Hatsuda}},\ }\bibfield  {title} {\bibinfo {title} {{QCD-based charge symmetry breaking interaction and the Okamoto-Nolen-Schiffer anomaly}},\ }\href {https://doi.org/10.1103/PhysRevC.109.L011302} {\bibfield  {journal} {\bibinfo  {journal} {Phys. Rev. C}\ }\textbf {\bibinfo {volume} {109}},\ \bibinfo {pages} {L011302} (\bibinfo {year} {2024})}\BibitemShut {NoStop}%
\bibitem [{\citenamefont {Naito}\ \emph {et~al.}(2025)\citenamefont {Naito}, \citenamefont {Hijikata}, \citenamefont {Zenihiro}, \citenamefont {Col\`o},\ and\ \citenamefont {Sagawa}}]{Naito:2025qub}%
  \BibitemOpen
  \bibfield  {author} {\bibinfo {author} {\bibfnamefont {T.}~\bibnamefont {Naito}}, \bibinfo {author} {\bibfnamefont {Y.}~\bibnamefont {Hijikata}}, \bibinfo {author} {\bibfnamefont {J.}~\bibnamefont {Zenihiro}}, \bibinfo {author} {\bibfnamefont {G.}~\bibnamefont {Col\`o}},\ and\ \bibinfo {author} {\bibfnamefont {H.}~\bibnamefont {Sagawa}},\ }\bibfield  {title} {\bibinfo {title} {{Mirror-skin thickness: A possible observable sensitive to the charge symmetry breaking energy density functional}},\ }\Eprint {https://arxiv.org/abs/2503.05147} {arXiv:2503.05147 [nucl-th]}  (\bibinfo {year} {2025})\BibitemShut {NoStop}%
\bibitem [{\citenamefont {Suzuki}\ \emph {et~al.}(1993)\citenamefont {Suzuki}, \citenamefont {Sagawa},\ and\ \citenamefont {Van~Giai}}]{Suzuki1993Phys.Rev.C47_R1360}%
  \BibitemOpen
  \bibfield  {author} {\bibinfo {author} {\bibfnamefont {T.}~\bibnamefont {Suzuki}}, \bibinfo {author} {\bibfnamefont {H.}~\bibnamefont {Sagawa}},\ and\ \bibinfo {author} {\bibfnamefont {N.}~\bibnamefont {Van~Giai}},\ }\bibfield  {title} {\bibinfo {title} {{Charge independence and charge symmetry breaking interactions and the Coulomb energy anomaly in isobaric analog states}},\ }\href {https://doi.org/10.1103/PhysRevC.47.R1360} {\bibfield  {journal} {\bibinfo  {journal} {Phys. Rev. C}\ }\textbf {\bibinfo {volume} {47}},\ \bibinfo {pages} {R1360} (\bibinfo {year} {1993})}\BibitemShut {NoStop}%
\bibitem [{\citenamefont {Roca-Maza}\ \emph {et~al.}(2018)\citenamefont {Roca-Maza}, \citenamefont {Col\`{o}},\ and\ \citenamefont {Sagawa}}]{Roca-Maza2018Phys.Rev.Lett.120_202501}%
  \BibitemOpen
  \bibfield  {author} {\bibinfo {author} {\bibfnamefont {X.}~\bibnamefont {Roca-Maza}}, \bibinfo {author} {\bibfnamefont {G.}~\bibnamefont {Col\`{o}}},\ and\ \bibinfo {author} {\bibfnamefont {H.}~\bibnamefont {Sagawa}},\ }\bibfield  {title} {\bibinfo {title} {{Nuclear Symmetry Energy and the Breaking of the Isospin Symmetry: How Do They Reconcile with Each Other?}},\ }\href {https://doi.org/10.1103/PhysRevLett.120.202501} {\bibfield  {journal} {\bibinfo  {journal} {Phys. Rev. Lett.}\ }\textbf {\bibinfo {volume} {120}},\ \bibinfo {pages} {202501} (\bibinfo {year} {2018})}\BibitemShut {NoStop}%
\bibitem [{\citenamefont {Sagawa}\ \emph {et~al.}(1996)\citenamefont {Sagawa}, \citenamefont {Giai},\ and\ \citenamefont {Suzuki}}]{Sagawa1996Phys.Rev.C53_2163}%
  \BibitemOpen
  \bibfield  {author} {\bibinfo {author} {\bibfnamefont {H.}~\bibnamefont {Sagawa}}, \bibinfo {author} {\bibfnamefont {N.~V.}\ \bibnamefont {Giai}},\ and\ \bibinfo {author} {\bibfnamefont {T.}~\bibnamefont {Suzuki}},\ }\bibfield  {title} {\bibinfo {title} {{Effect of isospin mixing on superallowed Fermi $ \beta $ decay}},\ }\href {https://doi.org/10.1103/PhysRevC.53.2163} {\bibfield  {journal} {\bibinfo  {journal} {Phys. Rev. C}\ }\textbf {\bibinfo {volume} {53}},\ \bibinfo {pages} {2163} (\bibinfo {year} {1996})}\BibitemShut {NoStop}%
\bibitem [{\citenamefont {Liang}\ \emph {et~al.}(2009)\citenamefont {Liang}, \citenamefont {Van~Giai},\ and\ \citenamefont {Meng}}]{Liang2009Phys.Rev.C79_064316}%
  \BibitemOpen
  \bibfield  {author} {\bibinfo {author} {\bibfnamefont {H.}~\bibnamefont {Liang}}, \bibinfo {author} {\bibfnamefont {N.}~\bibnamefont {Van~Giai}},\ and\ \bibinfo {author} {\bibfnamefont {J.}~\bibnamefont {Meng}},\ }\bibfield  {title} {\bibinfo {title} {{Isospin corrections for superallowed Fermi $ \beta $ decay in self-consistent relativistic random-phase approximation approaches}},\ }\href {https://doi.org/10.1103/PhysRevC.79.064316} {\bibfield  {journal} {\bibinfo  {journal} {Phys. Rev. C}\ }\textbf {\bibinfo {volume} {79}},\ \bibinfo {pages} {064316} (\bibinfo {year} {2009})}\BibitemShut {NoStop}%
\bibitem [{\citenamefont {Satu\l{}a}\ \emph {et~al.}(2011)\citenamefont {Satu\l{}a}, \citenamefont {Dobaczewski}, \citenamefont {Nazarewicz},\ and\ \citenamefont {Rafalski}}]{Satula2011Phys.Rev.Lett.106_132502}%
  \BibitemOpen
  \bibfield  {author} {\bibinfo {author} {\bibfnamefont {W.}~\bibnamefont {Satu\l{}a}}, \bibinfo {author} {\bibfnamefont {J.}~\bibnamefont {Dobaczewski}}, \bibinfo {author} {\bibfnamefont {W.}~\bibnamefont {Nazarewicz}},\ and\ \bibinfo {author} {\bibfnamefont {M.}~\bibnamefont {Rafalski}},\ }\bibfield  {title} {\bibinfo {title} {{Microscopic Calculations of Isospin-Breaking Corrections to Superallowed Beta Decay}},\ }\href {https://doi.org/10.1103/PhysRevLett.106.132502} {\bibfield  {journal} {\bibinfo  {journal} {Phys. Rev. Lett.}\ }\textbf {\bibinfo {volume} {106}},\ \bibinfo {pages} {132502} (\bibinfo {year} {2011})}\BibitemShut {NoStop}%
\bibitem [{\citenamefont {Satu\l{}a}\ \emph {et~al.}(2012)\citenamefont {Satu\l{}a}, \citenamefont {Dobaczewski}, \citenamefont {Nazarewicz},\ and\ \citenamefont {Werner}}]{Satula2012Phys.Rev.C86_054316}%
  \BibitemOpen
  \bibfield  {author} {\bibinfo {author} {\bibfnamefont {W.}~\bibnamefont {Satu\l{}a}}, \bibinfo {author} {\bibfnamefont {J.}~\bibnamefont {Dobaczewski}}, \bibinfo {author} {\bibfnamefont {W.}~\bibnamefont {Nazarewicz}},\ and\ \bibinfo {author} {\bibfnamefont {T.~R.}\ \bibnamefont {Werner}},\ }\bibfield  {title} {\bibinfo {title} {{Isospin-breaking corrections to superallowed Fermi $ \beta $ decay in isospin- and angular-momentum-projected nuclear density functional theory}},\ }\href {https://doi.org/10.1103/PhysRevC.86.054316} {\bibfield  {journal} {\bibinfo  {journal} {Phys. Rev. C}\ }\textbf {\bibinfo {volume} {86}},\ \bibinfo {pages} {054316} (\bibinfo {year} {2012})}\BibitemShut {NoStop}%
\bibitem [{\citenamefont {Rafalski}\ and\ \citenamefont {Satu{\l}a}(2012)}]{Rafalski2012Phys.Scr.T150_014032}%
  \BibitemOpen
  \bibfield  {author} {\bibinfo {author} {\bibfnamefont {M.}~\bibnamefont {Rafalski}}\ and\ \bibinfo {author} {\bibfnamefont {W.}~\bibnamefont {Satu{\l}a}},\ }\bibfield  {title} {\bibinfo {title} {{Microscopic calculations of isospin mixing in $N \approx Z$ nuclei and isospin-symmetry-breaking corrections to the superallowed $\beta$-decay}},\ }\href {https://doi.org/10.1088/0031-8949/2012/t150/014032} {\bibfield  {journal} {\bibinfo  {journal} {Phys. Scr.}\ }\textbf {\bibinfo {volume} {T150}},\ \bibinfo {pages} {014032} (\bibinfo {year} {2012})}\BibitemShut {NoStop}%
\bibitem [{\citenamefont {Kaneko}\ \emph {et~al.}(2017)\citenamefont {Kaneko}, \citenamefont {Sun}, \citenamefont {Mizusaki}, \citenamefont {Tazaki},\ and\ \citenamefont {Ghorui}}]{Kaneko2017Phys.Lett.B773_521}%
  \BibitemOpen
  \bibfield  {author} {\bibinfo {author} {\bibfnamefont {K.}~\bibnamefont {Kaneko}}, \bibinfo {author} {\bibfnamefont {Y.}~\bibnamefont {Sun}}, \bibinfo {author} {\bibfnamefont {T.}~\bibnamefont {Mizusaki}}, \bibinfo {author} {\bibfnamefont {S.}~\bibnamefont {Tazaki}},\ and\ \bibinfo {author} {\bibfnamefont {S.~K.}\ \bibnamefont {Ghorui}},\ }\bibfield  {title} {\bibinfo {title} {{Isospin-symmetry breaking in superallowed Fermi $ \beta $-decay due to isospin-nonconserving forces}},\ }\href {https://doi.org/10.1016/j.physletb.2017.08.056} {\bibfield  {journal} {\bibinfo  {journal} {Phys. Lett. B}\ }\textbf {\bibinfo {volume} {773}},\ \bibinfo {pages} {521} (\bibinfo {year} {2017})}\BibitemShut {NoStop}%
\bibitem [{\citenamefont {Hardy}\ and\ \citenamefont {Towner}(2020)}]{Hardy2020Phys.Rev.C102_045501}%
  \BibitemOpen
  \bibfield  {author} {\bibinfo {author} {\bibfnamefont {J.~C.}\ \bibnamefont {Hardy}}\ and\ \bibinfo {author} {\bibfnamefont {I.~S.}\ \bibnamefont {Towner}},\ }\bibfield  {title} {\bibinfo {title} {{Superallowed ${0}^{+}\rightarrow{0}^{+}$ nuclear $\beta$ decays: 2020 critical survey, with implications for ${V}_{\mathit{ud}}$ and CKM unitarity}},\ }\href {https://doi.org/10.1103/PhysRevC.102.045501} {\bibfield  {journal} {\bibinfo  {journal} {Phys. Rev. C}\ }\textbf {\bibinfo {volume} {102}},\ \bibinfo {pages} {045501} (\bibinfo {year} {2020})}\BibitemShut {NoStop}%
\bibitem [{\citenamefont {Xayavong}\ and\ \citenamefont {Smirnova}(2022)}]{Xayavong2022Phys.Rev.C105_044308}%
  \BibitemOpen
  \bibfield  {author} {\bibinfo {author} {\bibfnamefont {L.}~\bibnamefont {Xayavong}}\ and\ \bibinfo {author} {\bibfnamefont {N.~A.}\ \bibnamefont {Smirnova}},\ }\bibfield  {title} {\bibinfo {title} {{Radial overlap correction to superallowed ${0}^{+}\rightarrow{0}^{+}$ nuclear $\beta$ decays using the shell model with Hartree-Fock radial wave functions}},\ }\href {https://doi.org/10.1103/PhysRevC.105.044308} {\bibfield  {journal} {\bibinfo  {journal} {Phys. Rev. C}\ }\textbf {\bibinfo {volume} {105}},\ \bibinfo {pages} {044308} (\bibinfo {year} {2022})}\BibitemShut {NoStop}%
\bibitem [{\citenamefont {Hardy}\ and\ \citenamefont {Towner}(2013)}]{Hardy2013Ann.Phys.525_443}%
  \BibitemOpen
  \bibfield  {author} {\bibinfo {author} {\bibfnamefont {J.~C.}\ \bibnamefont {Hardy}}\ and\ \bibinfo {author} {\bibfnamefont {I.~S.}\ \bibnamefont {Towner}},\ }\bibfield  {title} {\bibinfo {title} {{CKM unitarity normalization tests, present and future}},\ }\href {https://doi.org/10.1002/andp.201300004} {\bibfield  {journal} {\bibinfo  {journal} {Ann. Phys.}\ }\textbf {\bibinfo {volume} {525}},\ \bibinfo {pages} {443} (\bibinfo {year} {2013})}\BibitemShut {NoStop}%
\bibitem [{\citenamefont {Porter}(2016)}]{Porter2016Prog.Part.Nucl.Phys.91_101}%
  \BibitemOpen
  \bibfield  {author} {\bibinfo {author} {\bibfnamefont {F.~C.}\ \bibnamefont {Porter}},\ }\bibfield  {title} {\bibinfo {title} {{Experimental status of the CKM matrix}},\ }\href {https://doi.org/10.1016/j.ppnp.2016.06.003} {\bibfield  {journal} {\bibinfo  {journal} {Prog. Part. Nucl. Phys.}\ }\textbf {\bibinfo {volume} {91}},\ \bibinfo {pages} {101} (\bibinfo {year} {2016})}\BibitemShut {NoStop}%
\bibitem [{\citenamefont {Navas}\ \emph {et~al.}(2024)\citenamefont {Navas}, \citenamefont {Amsler}, \citenamefont {Gutsche}, \citenamefont {Hanhart}, \citenamefont {Hern\'andez-Rey}, \citenamefont {Louren\c{c}o}, \citenamefont {Masoni}, \citenamefont {Mikhasenko}, \citenamefont {Mitchell}, \citenamefont {Patrignani}, \citenamefont {Schwanda}, \citenamefont {Spanier}, \citenamefont {Venanzoni}, \citenamefont {Yuan}, \citenamefont {Agashe}, \citenamefont {Aielli}, \citenamefont {Allanach}, \citenamefont {Alvarez-Mu\~niz}, \citenamefont {Antonelli}, \citenamefont {Aschenauer}, \citenamefont {Asner}, \citenamefont {Assamagan}, \citenamefont {Baer}, \citenamefont {Banerjee}, \citenamefont {Barnett}, \citenamefont {Baudis}, \citenamefont {Bauer}, \citenamefont {Beatty}, \citenamefont {Beringer}, \citenamefont {Bettini}, \citenamefont {Biebel}, \citenamefont {Black}, \citenamefont {Blucher}, \citenamefont {Bonventre}, \citenamefont {Briere}, \citenamefont {Buckley}, \citenamefont {Burkert}, \citenamefont {Bychkov}, \citenamefont {Cahn}, \citenamefont {Cao}, \citenamefont {Carena}, \citenamefont {Casarosa}, \citenamefont {Ceccucci}, \citenamefont {Cerri}, \citenamefont {Chivukula}, \citenamefont {Cowan}, \citenamefont {Cranmer}, \citenamefont {Crede}, \citenamefont {Cremonesi}, \citenamefont {D'Ambrosio}, \citenamefont {Damour}, \citenamefont {de~Florian}, \citenamefont {de~Gouv\^ea}, \citenamefont {DeGrand}, \citenamefont {Demers}, \citenamefont {Demiragli}, \citenamefont {Dobrescu}, \citenamefont {D'Onofrio}, \citenamefont {Doser}, \citenamefont {Dreiner}, \citenamefont {Eerola}, \citenamefont {Egede}, \citenamefont {Eidelman}, \citenamefont {El-Khadra}, \citenamefont {Ellis}, \citenamefont {Eno}, \citenamefont {Erler}, \citenamefont {Ezhela}, \citenamefont {Fava}, \citenamefont {Fetscher}, \citenamefont {Fields}, \citenamefont {Freitas}, \citenamefont {Gallagher}, \citenamefont {Gershon}, \citenamefont {Gershtein}, \citenamefont {Gherghetta}, \citenamefont {Gonzalez-Garcia}, \citenamefont {Goodman}, \citenamefont {Grab}, \citenamefont {Gritsan}, \citenamefont {Grojean}, \citenamefont {Groom}, \citenamefont {Gr\"unewald}, \citenamefont {Gurtu}, \citenamefont {Haber}, \citenamefont {Hamel}, \citenamefont {Hashimoto}, \citenamefont {Hayato}, \citenamefont {Hebecker}, \citenamefont {Heinemeyer}, \citenamefont {Hikasa}, \citenamefont {Hisano}, \citenamefont {H\"ocker}, \citenamefont {Holder}, \citenamefont {Hsu}, \citenamefont {Huston}, \citenamefont {Hyodo}, \citenamefont {Ianni}, \citenamefont {Kado}, \citenamefont {Karliner}, \citenamefont {Katz}, \citenamefont {Kenzie}, \citenamefont {Khoze}, \citenamefont {Klein}, \citenamefont {Krauss}, \citenamefont {Kreps}, \citenamefont {Kri\v{z}an}, \citenamefont {Krusche}, \citenamefont {Kwon}, \citenamefont {Lahav}, \citenamefont {Lellouch}, \citenamefont {Lesgourgues}, \citenamefont {Liddle}, \citenamefont {Ligeti}, \citenamefont {Lin}, \citenamefont {Lippmann}, \citenamefont {Liss}, \citenamefont {Lister}, \citenamefont {Littenberg}, \citenamefont {Lugovsky}, \citenamefont {Lugovsky}, \citenamefont {Lusiani}, \citenamefont {Makida}, \citenamefont {Maltoni}, \citenamefont {Manohar}, \citenamefont {Marciano}, \citenamefont {Matthews}, \citenamefont {Mei\ss{}ner}, \citenamefont {Melzer-Pellmann}, \citenamefont {Mertsch}, \citenamefont {Miller}, \citenamefont {Milstead}, \citenamefont {M\"onig}, \citenamefont {Molaro}, \citenamefont {Moortgat}, \citenamefont {Moskovic}, \citenamefont {Nagata}, \citenamefont {Nakamura}, \citenamefont {Narain}, \citenamefont {Nason}, \citenamefont {Nelles}, \citenamefont {Neubert}, \citenamefont {Nir}, \citenamefont {O'Connell}, \citenamefont {O'Hare}, \citenamefont {Olive}, \citenamefont {Peacock}, \citenamefont {Pianori}, \citenamefont {Pich}, \citenamefont {Piepke}, \citenamefont {Pietropaolo}, \citenamefont {Pomarol}, \citenamefont {Pordes}, \citenamefont {Profumo}, \citenamefont {Quadt}, \citenamefont {Rabbertz}, \citenamefont {Rademacker}, \citenamefont {Raffelt}, \citenamefont {Ramsey-Musolf}, \citenamefont {Richardson}, \citenamefont {Ringwald}, \citenamefont {Robinson}, \citenamefont {Roesler}, \citenamefont {Rolli}, \citenamefont {Romaniouk}, \citenamefont {Rosenberg}, \citenamefont {Rosner}, \citenamefont {Rybka}, \citenamefont {Ryskin}, \citenamefont {Ryutin}, \citenamefont {Safdi}, \citenamefont {Sakai}, \citenamefont {Sarkar}, \citenamefont {Sauli}, \citenamefont {Schneider}, \citenamefont {Sch\"onert}, \citenamefont {Scholberg}, \citenamefont {Schwartz}, \citenamefont {Schwiening}, \citenamefont {Scott}, \citenamefont {Sefkow}, \citenamefont {Seljak}, \citenamefont {Sharma}, \citenamefont {Sharpe}, \citenamefont {Shiltsev}, \citenamefont {Signorelli}, \citenamefont {Silari}, \citenamefont {Simon}, \citenamefont {Sj\"ostrand}, \citenamefont {Skands}, \citenamefont {Skwarnicki}, \citenamefont {Smoot}, \citenamefont {Soffer}, \citenamefont {Sozzi}, \citenamefont {Spiering}, \citenamefont {Stahl}, \citenamefont {Sumino}, \citenamefont {Takahashi}, \citenamefont {Tanabashi}, \citenamefont {Tanaka}, \citenamefont {Ta\v{s}evsk\'y}, \citenamefont {Terao}, \citenamefont {Terashi}, \citenamefont {Terning}, \citenamefont {Thoma}, \citenamefont {Thorne}, \citenamefont {Tiator}, \citenamefont {Titov}, \citenamefont {Tovey}, \citenamefont {Trabelsi}, \citenamefont {Urquijo}, \citenamefont {Valencia}, \citenamefont {Van~de Water}, \citenamefont {Varelas}, \citenamefont {Verde}, \citenamefont {Vivarelli}, \citenamefont {Vogel}, \citenamefont {Vogelsang}, \citenamefont {Vorobyev}, \citenamefont {Wakely}, \citenamefont {Walkowiak}, \citenamefont {Walter}, \citenamefont {Wands}, \citenamefont {Weinberg}, \citenamefont {Weinberg}, \citenamefont {Wermes}, \citenamefont {White}, \citenamefont {Wiencke}, \citenamefont {Willocq}, \citenamefont {Woody}, \citenamefont {Workman}, \citenamefont {Yao}, \citenamefont {Yokoyama}, \citenamefont {Yoshida}, \citenamefont {Zanderighi}, \citenamefont {Zeller}, \citenamefont {Zhu}, \citenamefont {Zhu}, \citenamefont {Zimmermann}, \citenamefont {Zyla}, \citenamefont {Anderson}, \citenamefont {Kramer}, \citenamefont {Schaffner},\ and\ \citenamefont {Zheng}}]{pdg24}%
  \BibitemOpen
  \bibfield  {author} {\bibinfo {author} {\bibfnamefont {S.}~\bibnamefont {Navas}}, \bibinfo {author} {\bibfnamefont {C.}~\bibnamefont {Amsler}}, \bibinfo {author} {\bibfnamefont {T.}~\bibnamefont {Gutsche}}, \bibinfo {author} {\bibfnamefont {C.}~\bibnamefont {Hanhart}}, \bibinfo {author} {\bibfnamefont {J.~J.}\ \bibnamefont {Hern\'andez-Rey}}, \bibinfo {author} {\bibfnamefont {C.}~\bibnamefont {Louren\c{c}o}}, \bibinfo {author} {\bibfnamefont {A.}~\bibnamefont {Masoni}}, \bibinfo {author} {\bibfnamefont {M.}~\bibnamefont {Mikhasenko}}, \bibinfo {author} {\bibfnamefont {R.~E.}\ \bibnamefont {Mitchell}}, \bibinfo {author} {\bibfnamefont {C.}~\bibnamefont {Patrignani}}, \bibinfo {author} {\bibfnamefont {C.}~\bibnamefont {Schwanda}}, \bibinfo {author} {\bibfnamefont {S.}~\bibnamefont {Spanier}}, \bibinfo {author} {\bibfnamefont {G.}~\bibnamefont {Venanzoni}}, \bibinfo {author} {\bibfnamefont {C.~Z.}\ \bibnamefont {Yuan}}, \bibinfo {author} {\bibfnamefont {K.}~\bibnamefont {Agashe}}, \bibinfo {author} {\bibfnamefont {G.}~\bibnamefont {Aielli}}, \bibinfo {author} {\bibfnamefont {B.~C.}\ \bibnamefont {Allanach}}, \bibinfo {author} {\bibfnamefont {J.}~\bibnamefont {Alvarez-Mu\~niz}}, \bibinfo {author} {\bibfnamefont {M.}~\bibnamefont {Antonelli}}, \bibinfo {author} {\bibfnamefont {E.~C.}\ \bibnamefont {Aschenauer}}, \bibinfo {author} {\bibfnamefont {D.~M.}\ \bibnamefont {Asner}}, \bibinfo {author} {\bibfnamefont {K.}~\bibnamefont {Assamagan}}, \bibinfo {author} {\bibfnamefont {H.}~\bibnamefont {Baer}}, \bibinfo {author} {\bibfnamefont {S.}~\bibnamefont {Banerjee}}, \bibinfo {author} {\bibfnamefont {R.~M.}\ \bibnamefont {Barnett}}, \bibinfo {author} {\bibfnamefont {L.}~\bibnamefont {Baudis}}, \bibinfo {author} {\bibfnamefont {C.~W.}\ \bibnamefont {Bauer}}, \bibinfo {author} {\bibfnamefont {J.~J.}\ \bibnamefont {Beatty}}, \bibinfo {author} {\bibfnamefont {J.}~\bibnamefont {Beringer}}, \bibinfo {author} {\bibfnamefont {A.}~\bibnamefont {Bettini}}, \bibinfo {author} {\bibfnamefont {O.}~\bibnamefont {Biebel}}, \bibinfo {author} {\bibfnamefont {K.~M.}\ \bibnamefont {Black}}, \bibinfo {author} {\bibfnamefont {E.}~\bibnamefont {Blucher}}, \bibinfo {author} {\bibfnamefont {R.}~\bibnamefont {Bonventre}}, \bibinfo {author} {\bibfnamefont {R.~A.}\ \bibnamefont {Briere}}, \bibinfo {author} {\bibfnamefont {A.}~\bibnamefont {Buckley}}, \bibinfo {author} {\bibfnamefont {V.~D.}\ \bibnamefont {Burkert}}, \bibinfo {author} {\bibfnamefont {M.~A.}\ \bibnamefont {Bychkov}}, \bibinfo {author} {\bibfnamefont {R.~N.}\ \bibnamefont {Cahn}}, \bibinfo {author} {\bibfnamefont {Z.}~\bibnamefont {Cao}}, \bibinfo {author} {\bibfnamefont {M.}~\bibnamefont {Carena}}, \bibinfo {author} {\bibfnamefont {G.}~\bibnamefont {Casarosa}}, \bibinfo {author} {\bibfnamefont {A.}~\bibnamefont {Ceccucci}}, \bibinfo {author} {\bibfnamefont {A.}~\bibnamefont {Cerri}}, \bibinfo {author} {\bibfnamefont {R.~S.}\ \bibnamefont {Chivukula}}, \bibinfo {author} {\bibfnamefont {G.}~\bibnamefont {Cowan}}, \bibinfo {author} {\bibfnamefont {K.}~\bibnamefont {Cranmer}}, \bibinfo {author} {\bibfnamefont {V.}~\bibnamefont {Crede}}, \bibinfo {author} {\bibfnamefont {O.}~\bibnamefont {Cremonesi}}, \bibinfo {author} {\bibfnamefont {G.}~\bibnamefont {D'Ambrosio}}, \bibinfo {author} {\bibfnamefont {T.}~\bibnamefont {Damour}}, \bibinfo {author} {\bibfnamefont {D.}~\bibnamefont {de~Florian}}, \bibinfo {author} {\bibfnamefont {A.}~\bibnamefont {de~Gouv\^ea}}, \bibinfo {author} {\bibfnamefont {T.}~\bibnamefont {DeGrand}}, \bibinfo {author} {\bibfnamefont {S.}~\bibnamefont {Demers}}, \bibinfo {author} {\bibfnamefont {Z.}~\bibnamefont {Demiragli}}, \bibinfo {author} {\bibfnamefont {B.~A.}\ \bibnamefont {Dobrescu}}, \bibinfo {author} {\bibfnamefont {M.}~\bibnamefont {D'Onofrio}}, \bibinfo {author} {\bibfnamefont {M.}~\bibnamefont {Doser}}, \bibinfo {author} {\bibfnamefont {H.~K.}\ \bibnamefont {Dreiner}}, \bibinfo {author} {\bibfnamefont {P.}~\bibnamefont {Eerola}}, \bibinfo {author} {\bibfnamefont {U.}~\bibnamefont {Egede}}, \bibinfo {author} {\bibfnamefont {S.}~\bibnamefont {Eidelman}}, \bibinfo {author} {\bibfnamefont {A.~X.}\ \bibnamefont {El-Khadra}}, \bibinfo {author} {\bibfnamefont {J.}~\bibnamefont {Ellis}}, \bibinfo {author} {\bibfnamefont {S.~C.}\ \bibnamefont {Eno}}, \bibinfo {author} {\bibfnamefont {J.}~\bibnamefont {Erler}}, \bibinfo {author} {\bibfnamefont {V.~V.}\ \bibnamefont {Ezhela}}, \bibinfo {author} {\bibfnamefont {A.}~\bibnamefont {Fava}}, \bibinfo {author} {\bibfnamefont {W.}~\bibnamefont {Fetscher}}, \bibinfo {author} {\bibfnamefont {B.~D.}\ \bibnamefont {Fields}}, \bibinfo {author} {\bibfnamefont {A.}~\bibnamefont {Freitas}}, \bibinfo {author} {\bibfnamefont {H.}~\bibnamefont {Gallagher}}, \bibinfo {author} {\bibfnamefont {T.}~\bibnamefont {Gershon}}, \bibinfo {author} {\bibfnamefont {Y.}~\bibnamefont {Gershtein}}, \bibinfo {author} {\bibfnamefont {T.}~\bibnamefont {Gherghetta}}, \bibinfo {author} {\bibfnamefont {M.~C.}\ \bibnamefont {Gonzalez-Garcia}}, \bibinfo {author} {\bibfnamefont {M.}~\bibnamefont {Goodman}}, \bibinfo {author} {\bibfnamefont {C.}~\bibnamefont {Grab}}, \bibinfo {author} {\bibfnamefont {A.~V.}\ \bibnamefont {Gritsan}}, \bibinfo {author} {\bibfnamefont {C.}~\bibnamefont {Grojean}}, \bibinfo {author} {\bibfnamefont {D.~E.}\ \bibnamefont {Groom}}, \bibinfo {author} {\bibfnamefont {M.}~\bibnamefont {Gr\"unewald}}, \bibinfo {author} {\bibfnamefont {A.}~\bibnamefont {Gurtu}}, \bibinfo {author} {\bibfnamefont {H.~E.}\ \bibnamefont {Haber}}, \bibinfo {author} {\bibfnamefont {M.}~\bibnamefont {Hamel}}, \bibinfo {author} {\bibfnamefont {S.}~\bibnamefont {Hashimoto}}, \bibinfo {author} {\bibfnamefont {Y.}~\bibnamefont {Hayato}}, \bibinfo {author} {\bibfnamefont {A.}~\bibnamefont {Hebecker}}, \bibinfo {author} {\bibfnamefont {S.}~\bibnamefont {Heinemeyer}}, \bibinfo {author} {\bibfnamefont {K.}~\bibnamefont {Hikasa}}, \bibinfo {author} {\bibfnamefont {J.}~\bibnamefont {Hisano}}, \bibinfo {author} {\bibfnamefont {A.}~\bibnamefont {H\"ocker}}, \bibinfo {author} {\bibfnamefont {J.}~\bibnamefont {Holder}}, \bibinfo {author} {\bibfnamefont {L.}~\bibnamefont {Hsu}}, \bibinfo {author} {\bibfnamefont {J.}~\bibnamefont {Huston}}, \bibinfo {author} {\bibfnamefont {T.}~\bibnamefont {Hyodo}}, \bibinfo {author} {\bibfnamefont {A.}~\bibnamefont {Ianni}}, \bibinfo {author} {\bibfnamefont {M.}~\bibnamefont {Kado}}, \bibinfo {author} {\bibfnamefont {M.}~\bibnamefont {Karliner}}, \bibinfo {author} {\bibfnamefont {U.~F.}\ \bibnamefont {Katz}}, \bibinfo {author} {\bibfnamefont {M.}~\bibnamefont {Kenzie}}, \bibinfo {author} {\bibfnamefont {V.~A.}\ \bibnamefont {Khoze}}, \bibinfo {author} {\bibfnamefont {S.~R.}\ \bibnamefont {Klein}}, \bibinfo {author} {\bibfnamefont {F.}~\bibnamefont {Krauss}}, \bibinfo {author} {\bibfnamefont {M.}~\bibnamefont {Kreps}}, \bibinfo {author} {\bibfnamefont {P.}~\bibnamefont {Kri\v{z}an}}, \bibinfo {author} {\bibfnamefont {B.}~\bibnamefont {Krusche}}, \bibinfo {author} {\bibfnamefont {Y.}~\bibnamefont {Kwon}}, \bibinfo {author} {\bibfnamefont {O.}~\bibnamefont {Lahav}}, \bibinfo {author} {\bibfnamefont {L.~P.}\ \bibnamefont {Lellouch}}, \bibinfo {author} {\bibfnamefont {J.}~\bibnamefont {Lesgourgues}}, \bibinfo {author} {\bibfnamefont {A.~R.}\ \bibnamefont {Liddle}}, \bibinfo {author} {\bibfnamefont {Z.}~\bibnamefont {Ligeti}}, \bibinfo {author} {\bibfnamefont {C.-J.}\ \bibnamefont {Lin}}, \bibinfo {author} {\bibfnamefont {C.}~\bibnamefont {Lippmann}}, \bibinfo {author} {\bibfnamefont {T.~M.}\ \bibnamefont {Liss}}, \bibinfo {author} {\bibfnamefont {A.}~\bibnamefont {Lister}}, \bibinfo {author} {\bibfnamefont {L.}~\bibnamefont {Littenberg}}, \bibinfo {author} {\bibfnamefont {K.~S.}\ \bibnamefont {Lugovsky}}, \bibinfo {author} {\bibfnamefont {S.~B.}\ \bibnamefont {Lugovsky}}, \bibinfo {author} {\bibfnamefont {A.}~\bibnamefont {Lusiani}}, \bibinfo {author} {\bibfnamefont {Y.}~\bibnamefont {Makida}}, \bibinfo {author} {\bibfnamefont {F.}~\bibnamefont {Maltoni}}, \bibinfo {author} {\bibfnamefont {A.~V.}\ \bibnamefont {Manohar}}, \bibinfo {author} {\bibfnamefont {W.~J.}\ \bibnamefont {Marciano}}, \bibinfo {author} {\bibfnamefont {J.}~\bibnamefont {Matthews}}, \bibinfo {author} {\bibfnamefont {U.-G.}\ \bibnamefont {Mei\ss{}ner}}, \bibinfo {author} {\bibfnamefont {I.-A.}\ \bibnamefont {Melzer-Pellmann}}, \bibinfo {author} {\bibfnamefont {P.}~\bibnamefont {Mertsch}}, \bibinfo {author} {\bibfnamefont {D.~J.}\ \bibnamefont {Miller}}, \bibinfo {author} {\bibfnamefont {D.}~\bibnamefont {Milstead}}, \bibinfo {author} {\bibfnamefont {K.}~\bibnamefont {M\"onig}}, \bibinfo {author} {\bibfnamefont {P.}~\bibnamefont {Molaro}}, \bibinfo {author} {\bibfnamefont {F.}~\bibnamefont {Moortgat}}, \bibinfo {author} {\bibfnamefont {M.}~\bibnamefont {Moskovic}}, \bibinfo {author} {\bibfnamefont {N.}~\bibnamefont {Nagata}}, \bibinfo {author} {\bibfnamefont {K.}~\bibnamefont {Nakamura}}, \bibinfo {author} {\bibfnamefont {M.}~\bibnamefont {Narain}}, \bibinfo {author} {\bibfnamefont {P.}~\bibnamefont {Nason}}, \bibinfo {author} {\bibfnamefont {A.}~\bibnamefont {Nelles}}, \bibinfo {author} {\bibfnamefont {M.}~\bibnamefont {Neubert}}, \bibinfo {author} {\bibfnamefont {Y.}~\bibnamefont {Nir}}, \bibinfo {author} {\bibfnamefont {H.~B.}\ \bibnamefont {O'Connell}}, \bibinfo {author} {\bibfnamefont {C.~A.~J.}\ \bibnamefont {O'Hare}}, \bibinfo {author} {\bibfnamefont {K.~A.}\ \bibnamefont {Olive}}, \bibinfo {author} {\bibfnamefont {J.~A.}\ \bibnamefont {Peacock}}, \bibinfo {author} {\bibfnamefont {E.}~\bibnamefont {Pianori}}, \bibinfo {author} {\bibfnamefont {A.}~\bibnamefont {Pich}}, \bibinfo {author} {\bibfnamefont {A.}~\bibnamefont {Piepke}}, \bibinfo {author} {\bibfnamefont {F.}~\bibnamefont {Pietropaolo}}, \bibinfo {author} {\bibfnamefont {A.}~\bibnamefont {Pomarol}}, \bibinfo {author} {\bibfnamefont {S.}~\bibnamefont {Pordes}}, \bibinfo {author} {\bibfnamefont {S.}~\bibnamefont {Profumo}}, \bibinfo {author} {\bibfnamefont {A.}~\bibnamefont {Quadt}}, \bibinfo {author} {\bibfnamefont
  {K.}~\bibnamefont {Rabbertz}}, \bibinfo {author} {\bibfnamefont {J.}~\bibnamefont {Rademacker}}, \bibinfo {author} {\bibfnamefont {G.}~\bibnamefont {Raffelt}}, \bibinfo {author} {\bibfnamefont {M.}~\bibnamefont {Ramsey-Musolf}}, \bibinfo {author} {\bibfnamefont {P.}~\bibnamefont {Richardson}}, \bibinfo {author} {\bibfnamefont {A.}~\bibnamefont {Ringwald}}, \bibinfo {author} {\bibfnamefont {D.~J.}\ \bibnamefont {Robinson}}, \bibinfo {author} {\bibfnamefont {S.}~\bibnamefont {Roesler}}, \bibinfo {author} {\bibfnamefont {S.}~\bibnamefont {Rolli}}, \bibinfo {author} {\bibfnamefont {A.}~\bibnamefont {Romaniouk}}, \bibinfo {author} {\bibfnamefont {L.~J.}\ \bibnamefont {Rosenberg}}, \bibinfo {author} {\bibfnamefont {J.~L.}\ \bibnamefont {Rosner}}, \bibinfo {author} {\bibfnamefont {G.}~\bibnamefont {Rybka}}, \bibinfo {author} {\bibfnamefont {M.~G.}\ \bibnamefont {Ryskin}}, \bibinfo {author} {\bibfnamefont {R.~A.}\ \bibnamefont {Ryutin}}, \bibinfo {author} {\bibfnamefont {B.}~\bibnamefont {Safdi}}, \bibinfo {author} {\bibfnamefont {Y.}~\bibnamefont {Sakai}}, \bibinfo {author} {\bibfnamefont {S.}~\bibnamefont {Sarkar}}, \bibinfo {author} {\bibfnamefont {F.}~\bibnamefont {Sauli}}, \bibinfo {author} {\bibfnamefont {O.}~\bibnamefont {Schneider}}, \bibinfo {author} {\bibfnamefont {S.}~\bibnamefont {Sch\"onert}}, \bibinfo {author} {\bibfnamefont {K.}~\bibnamefont {Scholberg}}, \bibinfo {author} {\bibfnamefont {A.~J.}\ \bibnamefont {Schwartz}}, \bibinfo {author} {\bibfnamefont {J.}~\bibnamefont {Schwiening}}, \bibinfo {author} {\bibfnamefont {D.}~\bibnamefont {Scott}}, \bibinfo {author} {\bibfnamefont {F.}~\bibnamefont {Sefkow}}, \bibinfo {author} {\bibfnamefont {U.}~\bibnamefont {Seljak}}, \bibinfo {author} {\bibfnamefont {V.}~\bibnamefont {Sharma}}, \bibinfo {author} {\bibfnamefont {S.~R.}\ \bibnamefont {Sharpe}}, \bibinfo {author} {\bibfnamefont {V.}~\bibnamefont {Shiltsev}}, \bibinfo {author} {\bibfnamefont {G.}~\bibnamefont {Signorelli}}, \bibinfo {author} {\bibfnamefont {M.}~\bibnamefont {Silari}}, \bibinfo {author} {\bibfnamefont {F.}~\bibnamefont {Simon}}, \bibinfo {author} {\bibfnamefont {T.}~\bibnamefont {Sj\"ostrand}}, \bibinfo {author} {\bibfnamefont {P.}~\bibnamefont {Skands}}, \bibinfo {author} {\bibfnamefont {T.}~\bibnamefont {Skwarnicki}}, \bibinfo {author} {\bibfnamefont {G.~F.}\ \bibnamefont {Smoot}}, \bibinfo {author} {\bibfnamefont {A.}~\bibnamefont {Soffer}}, \bibinfo {author} {\bibfnamefont {M.~S.}\ \bibnamefont {Sozzi}}, \bibinfo {author} {\bibfnamefont {C.}~\bibnamefont {Spiering}}, \bibinfo {author} {\bibfnamefont {A.}~\bibnamefont {Stahl}}, \bibinfo {author} {\bibfnamefont {Y.}~\bibnamefont {Sumino}}, \bibinfo {author} {\bibfnamefont {F.}~\bibnamefont {Takahashi}}, \bibinfo {author} {\bibfnamefont {M.}~\bibnamefont {Tanabashi}}, \bibinfo {author} {\bibfnamefont {J.}~\bibnamefont {Tanaka}}, \bibinfo {author} {\bibfnamefont {M.}~\bibnamefont {Ta\v{s}evsk\'y}}, \bibinfo {author} {\bibfnamefont {K.}~\bibnamefont {Terao}}, \bibinfo {author} {\bibfnamefont {K.}~\bibnamefont {Terashi}}, \bibinfo {author} {\bibfnamefont {J.}~\bibnamefont {Terning}}, \bibinfo {author} {\bibfnamefont {U.}~\bibnamefont {Thoma}}, \bibinfo {author} {\bibfnamefont {R.~S.}\ \bibnamefont {Thorne}}, \bibinfo {author} {\bibfnamefont {L.}~\bibnamefont {Tiator}}, \bibinfo {author} {\bibfnamefont {M.}~\bibnamefont {Titov}}, \bibinfo {author} {\bibfnamefont {D.~R.}\ \bibnamefont {Tovey}}, \bibinfo {author} {\bibfnamefont {K.}~\bibnamefont {Trabelsi}}, \bibinfo {author} {\bibfnamefont {P.}~\bibnamefont {Urquijo}}, \bibinfo {author} {\bibfnamefont {G.}~\bibnamefont {Valencia}}, \bibinfo {author} {\bibfnamefont {R.}~\bibnamefont {Van~de Water}}, \bibinfo {author} {\bibfnamefont {N.}~\bibnamefont {Varelas}}, \bibinfo {author} {\bibfnamefont {L.}~\bibnamefont {Verde}}, \bibinfo {author} {\bibfnamefont {I.}~\bibnamefont {Vivarelli}}, \bibinfo {author} {\bibfnamefont {P.}~\bibnamefont {Vogel}}, \bibinfo {author} {\bibfnamefont {W.}~\bibnamefont {Vogelsang}}, \bibinfo {author} {\bibfnamefont {V.}~\bibnamefont {Vorobyev}}, \bibinfo {author} {\bibfnamefont {S.~P.}\ \bibnamefont {Wakely}}, \bibinfo {author} {\bibfnamefont {W.}~\bibnamefont {Walkowiak}}, \bibinfo {author} {\bibfnamefont {C.~W.}\ \bibnamefont {Walter}}, \bibinfo {author} {\bibfnamefont {D.}~\bibnamefont {Wands}}, \bibinfo {author} {\bibfnamefont {D.~H.}\ \bibnamefont {Weinberg}}, \bibinfo {author} {\bibfnamefont {E.~J.}\ \bibnamefont {Weinberg}}, \bibinfo {author} {\bibfnamefont {N.}~\bibnamefont {Wermes}}, \bibinfo {author} {\bibfnamefont {M.}~\bibnamefont {White}}, \bibinfo {author} {\bibfnamefont {L.~R.}\ \bibnamefont {Wiencke}}, \bibinfo {author} {\bibfnamefont {S.}~\bibnamefont {Willocq}}, \bibinfo {author} {\bibfnamefont {C.~L.}\ \bibnamefont {Woody}}, \bibinfo {author} {\bibfnamefont {R.~L.}\ \bibnamefont {Workman}}, \bibinfo {author} {\bibfnamefont {W.-M.}\ \bibnamefont {Yao}}, \bibinfo {author} {\bibfnamefont {M.}~\bibnamefont {Yokoyama}}, \bibinfo {author} {\bibfnamefont {R.}~\bibnamefont {Yoshida}}, \bibinfo {author} {\bibfnamefont {G.}~\bibnamefont {Zanderighi}}, \bibinfo {author} {\bibfnamefont {G.~P.}\ \bibnamefont {Zeller}}, \bibinfo {author} {\bibfnamefont {R.-Y.}\ \bibnamefont {Zhu}}, \bibinfo {author} {\bibfnamefont {S.-L.}\ \bibnamefont {Zhu}}, \bibinfo {author} {\bibfnamefont {F.}~\bibnamefont {Zimmermann}}, \bibinfo {author} {\bibfnamefont {P.~A.}\ \bibnamefont {Zyla}}, \bibinfo {author} {\bibfnamefont {J.}~\bibnamefont {Anderson}}, \bibinfo {author} {\bibfnamefont {M.}~\bibnamefont {Kramer}}, \bibinfo {author} {\bibfnamefont {P.}~\bibnamefont {Schaffner}},\ and\ \bibinfo {author} {\bibfnamefont {W.}~\bibnamefont {Zheng}} (\bibinfo {collaboration} {Particle Data Group Collaboration}),\ }\bibfield  {title} {\bibinfo {title} {{Review of particle physics}},\ }\href {https://doi.org/10.1103/PhysRevD.110.030001} {\bibfield  {journal} {\bibinfo  {journal} {Phys. Rev. D}\ }\textbf {\bibinfo {volume} {110}},\ \bibinfo {pages} {030001} (\bibinfo {year} {2024})}\BibitemShut {NoStop}%
\bibitem [{\citenamefont {Henley}(1969)}]{Henley1969IsospininNuclearPhysics_15}%
  \BibitemOpen
  \bibfield  {author} {\bibinfo {author} {\bibfnamefont {E.~M.}\ \bibnamefont {Henley}},\ }\bibfield  {title} {\bibinfo {title} {{Charge independece and charge symmetry of nuclear forces}},\ }in\ \href@noop {} {\emph {\bibinfo {booktitle} {{Isospin in Nuclear Physics}}}},\ \bibinfo {editor} {edited by\ \bibinfo {editor} {\bibfnamefont {D.~H.}\ \bibnamefont {Wilkinson}}}\ (\bibinfo  {publisher} {North-Holland},\ \bibinfo {address} {Amsterdam},\ \bibinfo {year} {1969})\ Chap.~\bibinfo {chapter} {2}, p.~\bibinfo {pages} {15}\BibitemShut {NoStop}%
\bibitem [{\citenamefont {Miller}\ \emph {et~al.}(2006)\citenamefont {Miller}, \citenamefont {Opper},\ and\ \citenamefont {Stephenson}}]{Miller2006Annu.Rev.Nucl.Part.Sci.56_253}%
  \BibitemOpen
  \bibfield  {author} {\bibinfo {author} {\bibfnamefont {G.~A.}\ \bibnamefont {Miller}}, \bibinfo {author} {\bibfnamefont {A.~K.}\ \bibnamefont {Opper}},\ and\ \bibinfo {author} {\bibfnamefont {E.~J.}\ \bibnamefont {Stephenson}},\ }\bibfield  {title} {\bibinfo {title} {{Charge Symmetry Breaking and QCD}},\ }\href {https://doi.org/10.1146/annurev.nucl.56.080805.140446} {\bibfield  {journal} {\bibinfo  {journal} {Annu. Rev. Nucl. Part. Sci.}\ }\textbf {\bibinfo {volume} {56}},\ \bibinfo {pages} {253} (\bibinfo {year} {2006})}\BibitemShut {NoStop}%
\bibitem [{\citenamefont {Coon}\ and\ \citenamefont {Scadron}(1982{\natexlab{a}})}]{Coon1982Phys.Rev.C26_2402}%
  \BibitemOpen
  \bibfield  {author} {\bibinfo {author} {\bibfnamefont {S.~A.}\ \bibnamefont {Coon}}\ and\ \bibinfo {author} {\bibfnamefont {M.~D.}\ \bibnamefont {Scadron}},\ }\bibfield  {title} {\bibinfo {title} {{Two-pion exchange contributions to charge asymmetric and charge dependent nuclear forces}},\ }\href {https://doi.org/10.1103/PhysRevC.26.2402} {\bibfield  {journal} {\bibinfo  {journal} {Phys. Rev. C}\ }\textbf {\bibinfo {volume} {26}},\ \bibinfo {pages} {2402} (\bibinfo {year} {1982}{\natexlab{a}})}\BibitemShut {NoStop}%
\bibitem [{\citenamefont {Coon}\ \emph {et~al.}(1977)\citenamefont {Coon}, \citenamefont {Scadron},\ and\ \citenamefont {McNamee}}]{Coon1977Nucl.Phys.A287_381}%
  \BibitemOpen
  \bibfield  {author} {\bibinfo {author} {\bibfnamefont {S.~A.}\ \bibnamefont {Coon}}, \bibinfo {author} {\bibfnamefont {M.~D.}\ \bibnamefont {Scadron}},\ and\ \bibinfo {author} {\bibfnamefont {P.~C.}\ \bibnamefont {McNamee}},\ }\bibfield  {title} {\bibinfo {title} {{On the sign of the $ \rho $-$ \omega $ mixing charge asymmetric $ nn $ potential}},\ }\href {https://doi.org/10.1016/0375-9474(77)90052-5} {\bibfield  {journal} {\bibinfo  {journal} {Nucl. Phys. A}\ }\textbf {\bibinfo {volume} {287}},\ \bibinfo {pages} {381} (\bibinfo {year} {1977})}\BibitemShut {NoStop}%
\bibitem [{\citenamefont {Coon}\ and\ \citenamefont {Scadron}(1982{\natexlab{b}})}]{Coon1982Phys.Rev.C26_562}%
  \BibitemOpen
  \bibfield  {author} {\bibinfo {author} {\bibfnamefont {S.~A.}\ \bibnamefont {Coon}}\ and\ \bibinfo {author} {\bibfnamefont {M.~D.}\ \bibnamefont {Scadron}},\ }\bibfield  {title} {\bibinfo {title} {{Role of $ \pi^0 \eta' $ mixing in nuclear charge asymmetry}},\ }\href {https://doi.org/10.1103/PhysRevC.26.562} {\bibfield  {journal} {\bibinfo  {journal} {Phys. Rev. C}\ }\textbf {\bibinfo {volume} {26}},\ \bibinfo {pages} {562} (\bibinfo {year} {1982}{\natexlab{b}})}\BibitemShut {NoStop}%
\bibitem [{\citenamefont {Coon}\ and\ \citenamefont {Barrett}(1987)}]{Coon1987Phys.Rev.C36_2189}%
  \BibitemOpen
  \bibfield  {author} {\bibinfo {author} {\bibfnamefont {S.~A.}\ \bibnamefont {Coon}}\ and\ \bibinfo {author} {\bibfnamefont {R.~C.}\ \bibnamefont {Barrett}},\ }\bibfield  {title} {\bibinfo {title} {{$ \rho $-$ \omega $ mixing in nuclear charge asymmetry}},\ }\href {https://doi.org/10.1103/PhysRevC.36.2189} {\bibfield  {journal} {\bibinfo  {journal} {Phys. Rev. C}\ }\textbf {\bibinfo {volume} {36}},\ \bibinfo {pages} {2189} (\bibinfo {year} {1987})}\BibitemShut {NoStop}%
\bibitem [{\citenamefont {van Kolck}(1995)}]{Kolck1995Few-BodySyst.Suppl.9_444}%
  \BibitemOpen
  \bibfield  {author} {\bibinfo {author} {\bibfnamefont {U.}~\bibnamefont {van Kolck}},\ }\bibfield  {title} {\bibinfo {title} {{Isospin Violation in Low-energy Hadronic Physics}},\ }\href {https://doi.org/10.1007/978-3-7091-9453-9_64} {\bibfield  {journal} {\bibinfo  {journal} {Few-Body Syst. Suppl.}\ }\textbf {\bibinfo {volume} {9}},\ \bibinfo {pages} {444} (\bibinfo {year} {1995})}\BibitemShut {NoStop}%
\bibitem [{\citenamefont {van Kolck}\ \emph {et~al.}(1996)\citenamefont {van Kolck}, \citenamefont {Friar},\ and\ \citenamefont {Goldman}}]{Kolck1996Phys.Lett.B371_169}%
  \BibitemOpen
  \bibfield  {author} {\bibinfo {author} {\bibfnamefont {U.}~\bibnamefont {van Kolck}}, \bibinfo {author} {\bibfnamefont {J.~L.}\ \bibnamefont {Friar}},\ and\ \bibinfo {author} {\bibfnamefont {T.}~\bibnamefont {Goldman}},\ }\bibfield  {title} {\bibinfo {title} {{Phenomenological aspects of isospin violation in the nuclear force}},\ }\href {https://doi.org/10.1016/0370-2693(96)00009-3} {\bibfield  {journal} {\bibinfo  {journal} {Phys. Lett. B}\ }\textbf {\bibinfo {volume} {371}},\ \bibinfo {pages} {169} (\bibinfo {year} {1996})}\BibitemShut {NoStop}%
\bibitem [{\citenamefont {Epelbaum}\ and\ \citenamefont {Mei\ss{}ner}(2005)}]{Epelbaum2005Phys.Rev.C72_044001}%
  \BibitemOpen
  \bibfield  {author} {\bibinfo {author} {\bibfnamefont {E.}~\bibnamefont {Epelbaum}}\ and\ \bibinfo {author} {\bibfnamefont {U.-G.}\ \bibnamefont {Mei\ss{}ner}},\ }\bibfield  {title} {\bibinfo {title} {{Isospin-violating nucleon-nucleon forces using the method of unitary transformation}},\ }\href {https://doi.org/10.1103/PhysRevC.72.044001} {\bibfield  {journal} {\bibinfo  {journal} {Phys. Rev. C}\ }\textbf {\bibinfo {volume} {72}},\ \bibinfo {pages} {044001} (\bibinfo {year} {2005})}\BibitemShut {NoStop}%
\bibitem [{\citenamefont {Mei\ss{}ner}\ \emph {et~al.}(2008)\citenamefont {Mei\ss{}ner}, \citenamefont {Rakhimov}, \citenamefont {Wirzba},\ and\ \citenamefont {Yakhshiev}}]{Meissner2008Eur.Phys.J.A36_37}%
  \BibitemOpen
  \bibfield  {author} {\bibinfo {author} {\bibfnamefont {U.~G.}\ \bibnamefont {Mei\ss{}ner}}, \bibinfo {author} {\bibfnamefont {A.~M.}\ \bibnamefont {Rakhimov}}, \bibinfo {author} {\bibfnamefont {A.}~\bibnamefont {Wirzba}},\ and\ \bibinfo {author} {\bibfnamefont {U.~T.}\ \bibnamefont {Yakhshiev}},\ }\bibfield  {title} {\bibinfo {title} {{Neutron-proton mass difference in finite nuclei and the Nolen-Schiffer anomaly}},\ }\href {https://doi.org/10.1140/epja/i2008-10571-0} {\bibfield  {journal} {\bibinfo  {journal} {Eur. Phys. J. A}\ }\textbf {\bibinfo {volume} {36}},\ \bibinfo {pages} {37} (\bibinfo {year} {2008})}\BibitemShut {NoStop}%
\bibitem [{\citenamefont {Hohenberg}\ and\ \citenamefont {Kohn}(1964)}]{Hohenberg1964Phys.Rev.136_B864}%
  \BibitemOpen
  \bibfield  {author} {\bibinfo {author} {\bibfnamefont {P.}~\bibnamefont {Hohenberg}}\ and\ \bibinfo {author} {\bibfnamefont {W.}~\bibnamefont {Kohn}},\ }\bibfield  {title} {\bibinfo {title} {{Inhomogeneous Electron Gas}},\ }\href {https://doi.org/10.1103/PhysRev.136.B864} {\bibfield  {journal} {\bibinfo  {journal} {Phys. Rev.}\ }\textbf {\bibinfo {volume} {136}},\ \bibinfo {pages} {B864} (\bibinfo {year} {1964})}\BibitemShut {NoStop}%
\bibitem [{\citenamefont {Kohn}\ and\ \citenamefont {Sham}(1965)}]{Kohn1965Phys.Rev.140_A1133}%
  \BibitemOpen
  \bibfield  {author} {\bibinfo {author} {\bibfnamefont {W.}~\bibnamefont {Kohn}}\ and\ \bibinfo {author} {\bibfnamefont {L.~J.}\ \bibnamefont {Sham}},\ }\bibfield  {title} {\bibinfo {title} {{Self-Consistent Equations Including Exchange and Correlation Effects}},\ }\href {https://doi.org/10.1103/PhysRev.140.A1133} {\bibfield  {journal} {\bibinfo  {journal} {Phys. Rev.}\ }\textbf {\bibinfo {volume} {140}},\ \bibinfo {pages} {A1133} (\bibinfo {year} {1965})}\BibitemShut {NoStop}%
\bibitem [{\citenamefont {Vautherin}\ and\ \citenamefont {Brink}(1972)}]{Vautherin1972Phys.Rev.C5_626}%
  \BibitemOpen
  \bibfield  {author} {\bibinfo {author} {\bibfnamefont {D.}~\bibnamefont {Vautherin}}\ and\ \bibinfo {author} {\bibfnamefont {D.~M.}\ \bibnamefont {Brink}},\ }\bibfield  {title} {\bibinfo {title} {{Hartree-Fock Calculations with Skyrme's Interaction. I. Spherical Nuclei}},\ }\href {https://doi.org/10.1103/PhysRevC.5.626} {\bibfield  {journal} {\bibinfo  {journal} {Phys. Rev. C}\ }\textbf {\bibinfo {volume} {5}},\ \bibinfo {pages} {626} (\bibinfo {year} {1972})}\BibitemShut {NoStop}%
\bibitem [{\citenamefont {Kohn}(1999)}]{Kohn1999Rev.Mod.Phys.71_1253}%
  \BibitemOpen
  \bibfield  {author} {\bibinfo {author} {\bibfnamefont {W.}~\bibnamefont {Kohn}},\ }\bibfield  {title} {\bibinfo {title} {{Nobel Lecture: Electronic structure of matter---wave functions and density functionals}},\ }\href {https://doi.org/10.1103/RevModPhys.71.1253} {\bibfield  {journal} {\bibinfo  {journal} {Rev. Mod. Phys.}\ }\textbf {\bibinfo {volume} {71}},\ \bibinfo {pages} {1253} (\bibinfo {year} {1999})}\BibitemShut {NoStop}%
\bibitem [{\citenamefont {Bender}\ \emph {et~al.}(2003)\citenamefont {Bender}, \citenamefont {Heenen},\ and\ \citenamefont {Reinhard}}]{Bender2003Rev.Mod.Phys.75_121}%
  \BibitemOpen
  \bibfield  {author} {\bibinfo {author} {\bibfnamefont {M.}~\bibnamefont {Bender}}, \bibinfo {author} {\bibfnamefont {P.-H.}\ \bibnamefont {Heenen}},\ and\ \bibinfo {author} {\bibfnamefont {P.-G.}\ \bibnamefont {Reinhard}},\ }\bibfield  {title} {\bibinfo {title} {{Self-consistent mean-field models for nuclear structure}},\ }\href {https://doi.org/10.1103/RevModPhys.75.121} {\bibfield  {journal} {\bibinfo  {journal} {Rev. Mod. Phys.}\ }\textbf {\bibinfo {volume} {75}},\ \bibinfo {pages} {121} (\bibinfo {year} {2003})}\BibitemShut {NoStop}%
\bibitem [{\citenamefont {Col\`{o}}(2020)}]{Colo2020Adv.Phys.X5_1740061}%
  \BibitemOpen
  \bibfield  {author} {\bibinfo {author} {\bibfnamefont {G.}~\bibnamefont {Col\`{o}}},\ }\bibfield  {title} {\bibinfo {title} {{Nuclear density functional theory}},\ }\href {https://doi.org/10.1080/23746149.2020.1740061} {\bibfield  {journal} {\bibinfo  {journal} {Adv. Phys.:X}\ }\textbf {\bibinfo {volume} {5}},\ \bibinfo {pages} {1740061} (\bibinfo {year} {2020})}\BibitemShut {NoStop}%
\bibitem [{\citenamefont {Naito}\ \emph {et~al.}(2019)\citenamefont {Naito}, \citenamefont {Ohashi},\ and\ \citenamefont {Liang}}]{Naito2019J.Phys.B52_245003}%
  \BibitemOpen
  \bibfield  {author} {\bibinfo {author} {\bibfnamefont {T.}~\bibnamefont {Naito}}, \bibinfo {author} {\bibfnamefont {D.}~\bibnamefont {Ohashi}},\ and\ \bibinfo {author} {\bibfnamefont {H.}~\bibnamefont {Liang}},\ }\bibfield  {title} {\bibinfo {title} {{Improvement of functionals in density functional theory by the inverse Kohn--Sham method and density functional perturbation theory}},\ }\href {https://doi.org/10.1088/1361-6455/ab4eef} {\bibfield  {journal} {\bibinfo  {journal} {J. Phys. B}\ }\textbf {\bibinfo {volume} {52}},\ \bibinfo {pages} {245003} (\bibinfo {year} {2019})}\BibitemShut {NoStop}%
\bibitem [{\citenamefont {Accorto}\ \emph {et~al.}(2020)\citenamefont {Accorto}, \citenamefont {Brandolini}, \citenamefont {Marino}, \citenamefont {Porro}, \citenamefont {Scalesi}, \citenamefont {Col\`o}, \citenamefont {Roca-Maza},\ and\ \citenamefont {Vigezzi}}]{Accorto2020Phys.Rev.C101_024315}%
  \BibitemOpen
  \bibfield  {author} {\bibinfo {author} {\bibfnamefont {G.}~\bibnamefont {Accorto}}, \bibinfo {author} {\bibfnamefont {P.}~\bibnamefont {Brandolini}}, \bibinfo {author} {\bibfnamefont {F.}~\bibnamefont {Marino}}, \bibinfo {author} {\bibfnamefont {A.}~\bibnamefont {Porro}}, \bibinfo {author} {\bibfnamefont {A.}~\bibnamefont {Scalesi}}, \bibinfo {author} {\bibfnamefont {G.}~\bibnamefont {Col\`o}}, \bibinfo {author} {\bibfnamefont {X.}~\bibnamefont {Roca-Maza}},\ and\ \bibinfo {author} {\bibfnamefont {E.}~\bibnamefont {Vigezzi}},\ }\bibfield  {title} {\bibinfo {title} {{First step in the nuclear inverse Kohn-Sham problem: From densities to potentials}},\ }\href {https://doi.org/10.1103/PhysRevC.101.024315} {\bibfield  {journal} {\bibinfo  {journal} {Phys. Rev. C}\ }\textbf {\bibinfo {volume} {101}},\ \bibinfo {pages} {024315} (\bibinfo {year} {2020})}\BibitemShut {NoStop}%
\bibitem [{\citenamefont {Marino}\ \emph {et~al.}(2021)\citenamefont {Marino}, \citenamefont {Barbieri}, \citenamefont {Carbone}, \citenamefont {Col\`o}, \citenamefont {Lovato}, \citenamefont {Pederiva}, \citenamefont {Roca-Maza},\ and\ \citenamefont {Vigezzi}}]{Marino2021Phys.Rev.C104_024315}%
  \BibitemOpen
  \bibfield  {author} {\bibinfo {author} {\bibfnamefont {F.}~\bibnamefont {Marino}}, \bibinfo {author} {\bibfnamefont {C.}~\bibnamefont {Barbieri}}, \bibinfo {author} {\bibfnamefont {A.}~\bibnamefont {Carbone}}, \bibinfo {author} {\bibfnamefont {G.}~\bibnamefont {Col\`o}}, \bibinfo {author} {\bibfnamefont {A.}~\bibnamefont {Lovato}}, \bibinfo {author} {\bibfnamefont {F.}~\bibnamefont {Pederiva}}, \bibinfo {author} {\bibfnamefont {X.}~\bibnamefont {Roca-Maza}},\ and\ \bibinfo {author} {\bibfnamefont {E.}~\bibnamefont {Vigezzi}},\ }\bibfield  {title} {\bibinfo {title} {{Nuclear energy density functionals grounded in \textit{ab initio} calculations}},\ }\href {https://doi.org/10.1103/PhysRevC.104.024315} {\bibfield  {journal} {\bibinfo  {journal} {Phys. Rev. C}\ }\textbf {\bibinfo {volume} {104}},\ \bibinfo {pages} {024315} (\bibinfo {year} {2021})}\BibitemShut {NoStop}%
\bibitem [{\citenamefont {Skyrme}(1958)}]{Skyrme1958Nucl.Phys.9_615}%
  \BibitemOpen
  \bibfield  {author} {\bibinfo {author} {\bibfnamefont {T.~H.~R.}\ \bibnamefont {Skyrme}},\ }\bibfield  {title} {\bibinfo {title} {{The effective nuclear potential}},\ }\href {https://doi.org/10.1016/0029-5582(58)90345-6} {\bibfield  {journal} {\bibinfo  {journal} {Nucl. Phys.}\ }\textbf {\bibinfo {volume} {9}},\ \bibinfo {pages} {615} (\bibinfo {year} {1958})}\BibitemShut {NoStop}%
\bibitem [{\citenamefont {Meng}\ \emph {et~al.}(2006)\citenamefont {Meng}, \citenamefont {Toki}, \citenamefont {Zhou}, \citenamefont {Zhang}, \citenamefont {Long},\ and\ \citenamefont {Geng}}]{Meng2006Prog.Part.Nucl.Phys.57_470}%
  \BibitemOpen
  \bibfield  {author} {\bibinfo {author} {\bibfnamefont {J.}~\bibnamefont {Meng}}, \bibinfo {author} {\bibfnamefont {H.}~\bibnamefont {Toki}}, \bibinfo {author} {\bibfnamefont {S.~G.}\ \bibnamefont {Zhou}}, \bibinfo {author} {\bibfnamefont {S.~Q.}\ \bibnamefont {Zhang}}, \bibinfo {author} {\bibfnamefont {W.~H.}\ \bibnamefont {Long}},\ and\ \bibinfo {author} {\bibfnamefont {L.~S.}\ \bibnamefont {Geng}},\ }\bibfield  {title} {\bibinfo {title} {{Relativistic continuum Hartree Bogoliubov theory for ground-state properties of exotic nuclei}},\ }\href {https://doi.org/10.1016/j.ppnp.2005.06.001} {\bibfield  {journal} {\bibinfo  {journal} {Prog. Part. Nucl. Phys.}\ }\textbf {\bibinfo {volume} {57}},\ \bibinfo {pages} {470} (\bibinfo {year} {2006})}\BibitemShut {NoStop}%
\bibitem [{\citenamefont {Nik\v{s}i\'{c}}\ \emph {et~al.}(2011)\citenamefont {Nik\v{s}i\'{c}}, \citenamefont {Vretenar},\ and\ \citenamefont {Ring}}]{Niksic2011Prog.Part.Nucl.Phys.66_519}%
  \BibitemOpen
  \bibfield  {author} {\bibinfo {author} {\bibfnamefont {T.}~\bibnamefont {Nik\v{s}i\'{c}}}, \bibinfo {author} {\bibfnamefont {D.}~\bibnamefont {Vretenar}},\ and\ \bibinfo {author} {\bibfnamefont {P.}~\bibnamefont {Ring}},\ }\bibfield  {title} {\bibinfo {title} {{Relativistic nuclear energy density functionals: Mean-field and beyond}},\ }\href {https://doi.org/10.1016/j.ppnp.2011.01.055} {\bibfield  {journal} {\bibinfo  {journal} {Prog. Part. Nucl. Phys.}\ }\textbf {\bibinfo {volume} {66}},\ \bibinfo {pages} {519} (\bibinfo {year} {2011})}\BibitemShut {NoStop}%
\bibitem [{\citenamefont {Wiringa}\ \emph {et~al.}(1995)\citenamefont {Wiringa}, \citenamefont {Stoks},\ and\ \citenamefont {Schiavilla}}]{AV18}%
  \BibitemOpen
  \bibfield  {author} {\bibinfo {author} {\bibfnamefont {R.~B.}\ \bibnamefont {Wiringa}}, \bibinfo {author} {\bibfnamefont {V.~G.~J.}\ \bibnamefont {Stoks}},\ and\ \bibinfo {author} {\bibfnamefont {R.}~\bibnamefont {Schiavilla}},\ }\bibfield  {title} {\bibinfo {title} {{Accurate nucleon-nucleon potential with charge-independence breaking}},\ }\href {https://doi.org/10.1103/PhysRevC.51.38} {\bibfield  {journal} {\bibinfo  {journal} {Phys. Rev. C}\ }\textbf {\bibinfo {volume} {51}},\ \bibinfo {pages} {38} (\bibinfo {year} {1995})}\BibitemShut {NoStop}%
\bibitem [{\citenamefont {Lalazissis}\ \emph {et~al.}(2005)\citenamefont {Lalazissis}, \citenamefont {Nik\v{s}i\'{c}}, \citenamefont {Vretenar},\ and\ \citenamefont {Ring}}]{ddme2}%
  \BibitemOpen
  \bibfield  {author} {\bibinfo {author} {\bibfnamefont {G.~A.}\ \bibnamefont {Lalazissis}}, \bibinfo {author} {\bibfnamefont {T.}~\bibnamefont {Nik\v{s}i\'{c}}}, \bibinfo {author} {\bibfnamefont {D.}~\bibnamefont {Vretenar}},\ and\ \bibinfo {author} {\bibfnamefont {P.}~\bibnamefont {Ring}},\ }\bibfield  {title} {\bibinfo {title} {{New relativistic mean-field interaction with density-dependent meson-nucleon couplings}},\ }\href {https://doi.org/10.1103/PhysRevC.71.024312} {\bibfield  {journal} {\bibinfo  {journal} {Phys. Rev. C}\ }\textbf {\bibinfo {volume} {71}},\ \bibinfo {pages} {024312} (\bibinfo {year} {2005})}\BibitemShut {NoStop}%
\bibitem [{\citenamefont {Piekarewicz}\ and\ \citenamefont {Williams}(1993)}]{PhysRevC.47.R2462}%
  \BibitemOpen
  \bibfield  {author} {\bibinfo {author} {\bibfnamefont {J.}~\bibnamefont {Piekarewicz}}\ and\ \bibinfo {author} {\bibfnamefont {A.~G.}\ \bibnamefont {Williams}},\ }\bibfield  {title} {\bibinfo {title} {{Momentum dependence of the $ \rho $-$ \omega $ mixing amplitude in a hadronic model}},\ }\href {https://doi.org/10.1103/PhysRevC.47.R2462} {\bibfield  {journal} {\bibinfo  {journal} {Phys. Rev. C}\ }\textbf {\bibinfo {volume} {47}},\ \bibinfo {pages} {R2462} (\bibinfo {year} {1993})}\BibitemShut {NoStop}%
\bibitem [{\citenamefont {Berger}\ \emph {et~al.}(1991)\citenamefont {Berger}, \citenamefont {Girod},\ and\ \citenamefont {Gogny}}]{BGG91}%
  \BibitemOpen
  \bibfield  {author} {\bibinfo {author} {\bibfnamefont {J.~F.}\ \bibnamefont {Berger}}, \bibinfo {author} {\bibfnamefont {M.}~\bibnamefont {Girod}},\ and\ \bibinfo {author} {\bibfnamefont {D.}~\bibnamefont {Gogny}},\ }\bibfield  {title} {\bibinfo {title} {{Time-dependent quantum collective dynamics applied to nuclear fission}},\ }\href {https://doi.org/10.1016/0010-4655(91)90263-K} {\bibfield  {journal} {\bibinfo  {journal} {Comput. Phys. Commun.}\ }\textbf {\bibinfo {volume} {63}},\ \bibinfo {pages} {365} (\bibinfo {year} {1991})}\BibitemShut {NoStop}%
\bibitem [{\citenamefont {Younes}\ \emph {et~al.}(2019)\citenamefont {Younes}, \citenamefont {Gogny},\ and\ \citenamefont {Berger}}]{YGB19}%
  \BibitemOpen
  \bibfield  {author} {\bibinfo {author} {\bibfnamefont {W.}~\bibnamefont {Younes}}, \bibinfo {author} {\bibfnamefont {D.~M.}\ \bibnamefont {Gogny}},\ and\ \bibinfo {author} {\bibfnamefont {J.-F.}\ \bibnamefont {Berger}},\ }\href {https://doi.org/10.1007/978-3-030-04424-4} {\emph {\bibinfo {title} {{A Microscopic Theory of Fission Dynamics Based on the Generator Coordinate Method}}}},\ Lecture Notes in Physics\ (\bibinfo  {publisher} {Springer Nature Switzerland AG},\ \bibinfo {address} {Cham, Switzerland},\ \bibinfo {year} {2019})\BibitemShut {NoStop}%
\bibitem [{\citenamefont {Nik\v{s}i\'{c}}\ \emph {et~al.}(2002)\citenamefont {Nik\v{s}i\'{c}}, \citenamefont {Vretenar}, \citenamefont {Finelli},\ and\ \citenamefont {Ring}}]{ddme1}%
  \BibitemOpen
  \bibfield  {author} {\bibinfo {author} {\bibfnamefont {T.}~\bibnamefont {Nik\v{s}i\'{c}}}, \bibinfo {author} {\bibfnamefont {D.}~\bibnamefont {Vretenar}}, \bibinfo {author} {\bibfnamefont {P.}~\bibnamefont {Finelli}},\ and\ \bibinfo {author} {\bibfnamefont {P.}~\bibnamefont {Ring}},\ }\bibfield  {title} {\bibinfo {title} {{Relativistic Hartree-Bogoliubov model with density-dependent meson-nucleon couplings}},\ }\href {https://doi.org/10.1103/PhysRevC.66.024306} {\bibfield  {journal} {\bibinfo  {journal} {Phys. Rev. C}\ }\textbf {\bibinfo {volume} {66}},\ \bibinfo {pages} {024306} (\bibinfo {year} {2002})}\BibitemShut {NoStop}%
\bibitem [{\citenamefont {Nik\v{s}i\'{c}}\ \emph {et~al.}(2014)\citenamefont {Nik\v{s}i\'{c}}, \citenamefont {Paar}, \citenamefont {Vretenar},\ and\ \citenamefont {Ring}}]{dirhb}%
  \BibitemOpen
  \bibfield  {author} {\bibinfo {author} {\bibfnamefont {T.}~\bibnamefont {Nik\v{s}i\'{c}}}, \bibinfo {author} {\bibfnamefont {N.}~\bibnamefont {Paar}}, \bibinfo {author} {\bibfnamefont {D.}~\bibnamefont {Vretenar}},\ and\ \bibinfo {author} {\bibfnamefont {P.}~\bibnamefont {Ring}},\ }\bibfield  {title} {\bibinfo {title} {{DIRHB---A relativistic self-consistent mean-field framework for atomic nuclei}},\ }\href {https://doi.org/10.1016/j.cpc.2014.02.027} {\bibfield  {journal} {\bibinfo  {journal} {Comput. Phys. Commun.}\ }\textbf {\bibinfo {volume} {185}},\ \bibinfo {pages} {1808} (\bibinfo {year} {2014})}\BibitemShut {NoStop}%
\bibitem [{\citenamefont {Lalazissis}\ \emph {et~al.}(1998)\citenamefont {Lalazissis}, \citenamefont {Vretenar}, \citenamefont {P\"{o}schl},\ and\ \citenamefont {Ring}}]{LVPR98}%
  \BibitemOpen
  \bibfield  {author} {\bibinfo {author} {\bibfnamefont {G.~A.}\ \bibnamefont {Lalazissis}}, \bibinfo {author} {\bibfnamefont {D.}~\bibnamefont {Vretenar}}, \bibinfo {author} {\bibfnamefont {W.}~\bibnamefont {P\"{o}schl}},\ and\ \bibinfo {author} {\bibfnamefont {P.}~\bibnamefont {Ring}},\ }\bibfield  {title} {\bibinfo {title} {{Relativistic Hartree-Bogoliubov description of the neutron drip-line in light nuclei}},\ }\href {https://doi.org/10.1016/S0375-9474(98)00009-8} {\bibfield  {journal} {\bibinfo  {journal} {Nucl. Phys. A}\ }\textbf {\bibinfo {volume} {632}},\ \bibinfo {pages} {363} (\bibinfo {year} {1998})}\BibitemShut {NoStop}%
\bibitem [{\citenamefont {Serra}\ and\ \citenamefont {Ring}(2002)}]{SeRi02}%
  \BibitemOpen
  \bibfield  {author} {\bibinfo {author} {\bibfnamefont {M.}~\bibnamefont {Serra}}\ and\ \bibinfo {author} {\bibfnamefont {P.}~\bibnamefont {Ring}},\ }\bibfield  {title} {\bibinfo {title} {{Relativistic Hartree-Bogoliubov theory for finite nuclei}},\ }\href {https://doi.org/10.1103/PhysRevC.65.064324} {\bibfield  {journal} {\bibinfo  {journal} {Phys. Rev. C}\ }\textbf {\bibinfo {volume} {65}},\ \bibinfo {pages} {064324} (\bibinfo {year} {2002})}\BibitemShut {NoStop}%
\bibitem [{\citenamefont {Kucharek}\ and\ \citenamefont {Ring}(1991)}]{KuRi91}%
  \BibitemOpen
  \bibfield  {author} {\bibinfo {author} {\bibfnamefont {H.}~\bibnamefont {Kucharek}}\ and\ \bibinfo {author} {\bibfnamefont {P.}~\bibnamefont {Ring}},\ }\bibfield  {title} {\bibinfo {title} {{Relativistic field theory of superfluidity in nuclei}},\ }\href {https://doi.org/10.1007/BF01282930} {\bibfield  {journal} {\bibinfo  {journal} {Z. Phys. A}\ }\textbf {\bibinfo {volume} {339}},\ \bibinfo {pages} {23} (\bibinfo {year} {1991})}\BibitemShut {NoStop}%
\bibitem [{\citenamefont {Kurasawa}\ and\ \citenamefont {Suzuki}(2000)}]{KuSu00}%
  \BibitemOpen
  \bibfield  {author} {\bibinfo {author} {\bibfnamefont {H.}~\bibnamefont {Kurasawa}}\ and\ \bibinfo {author} {\bibfnamefont {T.}~\bibnamefont {Suzuki}},\ }\bibfield  {title} {\bibinfo {title} {{Effects of the neutron spin-orbit density on the nuclear charge density in relativistic models}},\ }\href {https://doi.org/10.1103/PhysRevC.62.054303} {\bibfield  {journal} {\bibinfo  {journal} {Phys. Rev. C}\ }\textbf {\bibinfo {volume} {62}},\ \bibinfo {pages} {054303} (\bibinfo {year} {2000})}\BibitemShut {NoStop}%
\bibitem [{\citenamefont {Perdrisat}\ \emph {et~al.}(2007)\citenamefont {Perdrisat}, \citenamefont {Punjabi},\ and\ \citenamefont {Vanderhaeghen}}]{PPV07}%
  \BibitemOpen
  \bibfield  {author} {\bibinfo {author} {\bibfnamefont {C.~F.}\ \bibnamefont {Perdrisat}}, \bibinfo {author} {\bibfnamefont {V.}~\bibnamefont {Punjabi}},\ and\ \bibinfo {author} {\bibfnamefont {M.}~\bibnamefont {Vanderhaeghen}},\ }\bibfield  {title} {\bibinfo {title} {{Nucleon electromagnetic form factors}},\ }\href {https://doi.org/10.1016/j.ppnp.2007.05.001} {\bibfield  {journal} {\bibinfo  {journal} {Prog. Part. Nucl. Phys.}\ }\textbf {\bibinfo {volume} {59}},\ \bibinfo {pages} {694} (\bibinfo {year} {2007})}\BibitemShut {NoStop}%
\bibitem [{\citenamefont {Horowitz}\ and\ \citenamefont {Piekarewicz}(2012)}]{Horowitz2012Phys.Rev.C86_045503}%
  \BibitemOpen
  \bibfield  {author} {\bibinfo {author} {\bibfnamefont {C.~J.}\ \bibnamefont {Horowitz}}\ and\ \bibinfo {author} {\bibfnamefont {J.}~\bibnamefont {Piekarewicz}},\ }\bibfield  {title} {\bibinfo {title} {{Impact of spin-orbit currents on the electroweak skin of neutron-rich nuclei}},\ }\href {https://doi.org/10.1103/PhysRevC.86.045503} {\bibfield  {journal} {\bibinfo  {journal} {Phys. Rev. C}\ }\textbf {\bibinfo {volume} {86}},\ \bibinfo {pages} {045503} (\bibinfo {year} {2012})}\BibitemShut {NoStop}%
\bibitem [{\citenamefont {Kurasawa}\ and\ \citenamefont {Suzuki}(2019)}]{KuSu19}%
  \BibitemOpen
  \bibfield  {author} {\bibinfo {author} {\bibfnamefont {H.}~\bibnamefont {Kurasawa}}\ and\ \bibinfo {author} {\bibfnamefont {T.}~\bibnamefont {Suzuki}},\ }\bibfield  {title} {\bibinfo {title} {{The $ n $th-order moment of the nuclear charge density and contribution from the neutrons}},\ }\href {https://doi.org/10.1093/ptep/ptz121} {\bibfield  {journal} {\bibinfo  {journal} {Prog. Theor. Exp. Phys.}\ }\textbf {\bibinfo {volume} {2019}},\ \bibinfo {pages} {113D01} (\bibinfo {year} {2019})}\BibitemShut {NoStop}%
\bibitem [{\citenamefont {Reinhard}\ and\ \citenamefont {Nazarewicz}(2021)}]{ReNa21}%
  \BibitemOpen
  \bibfield  {author} {\bibinfo {author} {\bibfnamefont {P.-G.}\ \bibnamefont {Reinhard}}\ and\ \bibinfo {author} {\bibfnamefont {W.}~\bibnamefont {Nazarewicz}},\ }\bibfield  {title} {\bibinfo {title} {{Nuclear charge densities in spherical and deformed nuclei: Toward precise calculations of charge radii}},\ }\href {https://doi.org/10.1103/PhysRevC.103.054310} {\bibfield  {journal} {\bibinfo  {journal} {Phys. Rev. C}\ }\textbf {\bibinfo {volume} {103}},\ \bibinfo {pages} {054310} (\bibinfo {year} {2021})}\BibitemShut {NoStop}%
\bibitem [{\citenamefont {Naito}\ \emph {et~al.}(2021)\citenamefont {Naito}, \citenamefont {Col\`o}, \citenamefont {Liang},\ and\ \citenamefont {Roca-Maza}}]{Naito2021Phys.Rev.C104_024316}%
  \BibitemOpen
  \bibfield  {author} {\bibinfo {author} {\bibfnamefont {T.}~\bibnamefont {Naito}}, \bibinfo {author} {\bibfnamefont {G.}~\bibnamefont {Col\`o}}, \bibinfo {author} {\bibfnamefont {H.}~\bibnamefont {Liang}},\ and\ \bibinfo {author} {\bibfnamefont {X.}~\bibnamefont {Roca-Maza}},\ }\bibfield  {title} {\bibinfo {title} {{Second and fourth moments of the charge density and neutron-skin thickness of atomic nuclei}},\ }\href {https://doi.org/10.1103/PhysRevC.104.024316} {\bibfield  {journal} {\bibinfo  {journal} {Phys. Rev. C}\ }\textbf {\bibinfo {volume} {104}},\ \bibinfo {pages} {024316} (\bibinfo {year} {2021})}\BibitemShut {NoStop}%
\bibitem [{\citenamefont {Gentile}\ and\ \citenamefont {Crawford}(2011)}]{Ge11}%
  \BibitemOpen
  \bibfield  {author} {\bibinfo {author} {\bibfnamefont {T.~R.}\ \bibnamefont {Gentile}}\ and\ \bibinfo {author} {\bibfnamefont {C.~B.}\ \bibnamefont {Crawford}},\ }\bibfield  {title} {\bibinfo {title} {{Neutron charge radius and the neutron electric form factor}},\ }\href {https://doi.org/10.1103/PhysRevC.83.055203} {\bibfield  {journal} {\bibinfo  {journal} {Phys. Rev. C}\ }\textbf {\bibinfo {volume} {83}},\ \bibinfo {pages} {055203} (\bibinfo {year} {2011})}\BibitemShut {NoStop}%
\bibitem [{\citenamefont {Greiner}\ and\ \citenamefont {Reinhardt}(2009)}]{GreinerQED}%
  \BibitemOpen
  \bibfield  {author} {\bibinfo {author} {\bibfnamefont {W.}~\bibnamefont {Greiner}}\ and\ \bibinfo {author} {\bibfnamefont {J.}~\bibnamefont {Reinhardt}},\ }\href {https://doi.org/10.1007/978-3-540-87561-1} {\emph {\bibinfo {title} {{Quantum Electrodynamics}}}}\ (\bibinfo  {publisher} {Springer-Verlag},\ \bibinfo {address} {Berlin Heidelberg},\ \bibinfo {year} {2009})\BibitemShut {NoStop}%
\bibitem [{\citenamefont {Huang}\ \emph {et~al.}(2021)\citenamefont {Huang}, \citenamefont {Wang}, \citenamefont {Kondev}, \citenamefont {Audi},\ and\ \citenamefont {Naimi}}]{AME2020-1}%
  \BibitemOpen
  \bibfield  {author} {\bibinfo {author} {\bibfnamefont {W.~J.}\ \bibnamefont {Huang}}, \bibinfo {author} {\bibfnamefont {M.}~\bibnamefont {Wang}}, \bibinfo {author} {\bibfnamefont {F.~G.}\ \bibnamefont {Kondev}}, \bibinfo {author} {\bibfnamefont {G.}~\bibnamefont {Audi}},\ and\ \bibinfo {author} {\bibfnamefont {S.}~\bibnamefont {Naimi}},\ }\bibfield  {title} {\bibinfo {title} {{The AME 2020 atomic mass evaluation (I). Evaluation of input data, and adjustment procedures}},\ }\href {https://doi.org/10.1088/1674-1137/abddb0} {\bibfield  {journal} {\bibinfo  {journal} {Chin. Phys. C}\ }\textbf {\bibinfo {volume} {45}},\ \bibinfo {pages} {030002} (\bibinfo {year} {2021})}\BibitemShut {NoStop}%
\bibitem [{\citenamefont {Wang}\ \emph {et~al.}(2021)\citenamefont {Wang}, \citenamefont {Huang}, \citenamefont {Kondev}, \citenamefont {Audi},\ and\ \citenamefont {Naimi}}]{AME2020-2}%
  \BibitemOpen
  \bibfield  {author} {\bibinfo {author} {\bibfnamefont {M.}~\bibnamefont {Wang}}, \bibinfo {author} {\bibfnamefont {W.~J.}\ \bibnamefont {Huang}}, \bibinfo {author} {\bibfnamefont {F.~G.}\ \bibnamefont {Kondev}}, \bibinfo {author} {\bibfnamefont {G.}~\bibnamefont {Audi}},\ and\ \bibinfo {author} {\bibfnamefont {S.}~\bibnamefont {Naimi}},\ }\bibfield  {title} {\bibinfo {title} {{The AME 2020 atomic mass evaluation (II). Tables, graphs and references}},\ }\href {https://doi.org/10.1088/1674-1137/abddaf} {\bibfield  {journal} {\bibinfo  {journal} {Chin. Phys. C}\ }\textbf {\bibinfo {volume} {45}},\ \bibinfo {pages} {030003} (\bibinfo {year} {2021})}\BibitemShut {NoStop}%
\bibitem [{\citenamefont {Angeli}\ and\ \citenamefont {Marinova}(2013)}]{Ang13}%
  \BibitemOpen
  \bibfield  {author} {\bibinfo {author} {\bibfnamefont {I.}~\bibnamefont {Angeli}}\ and\ \bibinfo {author} {\bibfnamefont {K.~P.}\ \bibnamefont {Marinova}},\ }\bibfield  {title} {\bibinfo {title} {{Table of experimental nuclear ground state charge radii: An update}},\ }\href {https://doi.org/10.1016/j.adt.2011.12.006} {\bibfield  {journal} {\bibinfo  {journal} {At. Data Nucl. Data Tables}\ }\textbf {\bibinfo {volume} {99}},\ \bibinfo {pages} {69} (\bibinfo {year} {2013})}\BibitemShut {NoStop}%
\bibitem [{\citenamefont {Li}\ \emph {et~al.}(2021)\citenamefont {Li}, \citenamefont {Luo},\ and\ \citenamefont {Wang}}]{Li21}%
  \BibitemOpen
  \bibfield  {author} {\bibinfo {author} {\bibfnamefont {T.}~\bibnamefont {Li}}, \bibinfo {author} {\bibfnamefont {Y.}~\bibnamefont {Luo}},\ and\ \bibinfo {author} {\bibfnamefont {N.}~\bibnamefont {Wang}},\ }\bibfield  {title} {\bibinfo {title} {{Compilation of recent nuclear ground state charge radius measurements and tests for models}},\ }\href {https://doi.org/10.1016/j.adt.2021.101440} {\bibfield  {journal} {\bibinfo  {journal} {At. Data Nucl. Data Tables}\ }\textbf {\bibinfo {volume} {140}},\ \bibinfo {pages} {101440} (\bibinfo {year} {2021})}\BibitemShut {NoStop}%
\bibitem [{\citenamefont {Sommer}\ \emph {et~al.}(2022)\citenamefont {Sommer}, \citenamefont {K\"onig}, \citenamefont {Rossi}, \citenamefont {Everett}, \citenamefont {Garand}, \citenamefont {de~Groote}, \citenamefont {Holt}, \citenamefont {Imgram}, \citenamefont {Incorvati}, \citenamefont {Kalman}, \citenamefont {Klose}, \citenamefont {Lantis}, \citenamefont {Liu}, \citenamefont {Miller}, \citenamefont {Minamisono}, \citenamefont {Miyagi}, \citenamefont {Nazarewicz}, \citenamefont {N\"ortersh\"auser}, \citenamefont {Pineda}, \citenamefont {Powel}, \citenamefont {Reinhard}, \citenamefont {Renth}, \citenamefont {Romero-Romero}, \citenamefont {Roth}, \citenamefont {Schwenk}, \citenamefont {Sumithrarachchi},\ and\ \citenamefont {Teigelh\"ofer}}]{Sommer22}%
  \BibitemOpen
  \bibfield  {author} {\bibinfo {author} {\bibfnamefont {F.}~\bibnamefont {Sommer}}, \bibinfo {author} {\bibfnamefont {K.}~\bibnamefont {K\"onig}}, \bibinfo {author} {\bibfnamefont {D.~M.}\ \bibnamefont {Rossi}}, \bibinfo {author} {\bibfnamefont {N.}~\bibnamefont {Everett}}, \bibinfo {author} {\bibfnamefont {D.}~\bibnamefont {Garand}}, \bibinfo {author} {\bibfnamefont {R.~P.}\ \bibnamefont {de~Groote}}, \bibinfo {author} {\bibfnamefont {J.~D.}\ \bibnamefont {Holt}}, \bibinfo {author} {\bibfnamefont {P.}~\bibnamefont {Imgram}}, \bibinfo {author} {\bibfnamefont {A.}~\bibnamefont {Incorvati}}, \bibinfo {author} {\bibfnamefont {C.}~\bibnamefont {Kalman}}, \bibinfo {author} {\bibfnamefont {A.}~\bibnamefont {Klose}}, \bibinfo {author} {\bibfnamefont {J.}~\bibnamefont {Lantis}}, \bibinfo {author} {\bibfnamefont {Y.}~\bibnamefont {Liu}}, \bibinfo {author} {\bibfnamefont {A.~J.}\ \bibnamefont {Miller}}, \bibinfo {author} {\bibfnamefont {K.}~\bibnamefont {Minamisono}}, \bibinfo {author} {\bibfnamefont {T.}~\bibnamefont {Miyagi}}, \bibinfo {author} {\bibfnamefont {W.}~\bibnamefont {Nazarewicz}}, \bibinfo {author} {\bibfnamefont {W.}~\bibnamefont {N\"ortersh\"auser}}, \bibinfo {author} {\bibfnamefont {S.~V.}\ \bibnamefont {Pineda}}, \bibinfo {author} {\bibfnamefont {R.}~\bibnamefont {Powel}}, \bibinfo {author} {\bibfnamefont {P.-G.}\ \bibnamefont {Reinhard}}, \bibinfo {author} {\bibfnamefont {L.}~\bibnamefont {Renth}}, \bibinfo {author} {\bibfnamefont {E.}~\bibnamefont {Romero-Romero}}, \bibinfo {author} {\bibfnamefont {R.}~\bibnamefont {Roth}}, \bibinfo {author} {\bibfnamefont {A.}~\bibnamefont {Schwenk}}, \bibinfo {author} {\bibfnamefont {C.}~\bibnamefont {Sumithrarachchi}},\ and\ \bibinfo {author} {\bibfnamefont {A.}~\bibnamefont {Teigelh\"ofer}},\ }\bibfield  {title} {\bibinfo {title} {{Charge Radii of $^{55,56}\mathrm{Ni}$ Reveal a Surprisingly Similar Behavior at $N=28$ in Ca and Ni Isotopes}},\ }\href {https://doi.org/10.1103/PhysRevLett.129.132501} {\bibfield  {journal} {\bibinfo  {journal} {Phys. Rev. Lett.}\ }\textbf {\bibinfo {volume} {129}},\ \bibinfo {pages} {132501} (\bibinfo {year} {2022})}\BibitemShut {NoStop}%
\bibitem [{\citenamefont {Tanimura}\ and\ \citenamefont {Cheoun}(2024)}]{TaCh24}%
  \BibitemOpen
  \bibfield  {author} {\bibinfo {author} {\bibfnamefont {Y.}~\bibnamefont {Tanimura}}\ and\ \bibinfo {author} {\bibfnamefont {M.-K.}\ \bibnamefont {Cheoun}},\ }\bibfield  {title} {\bibinfo {title} {{Effects of center-of-mass correction and nucleon anomalous magnetic moments on nuclear charge radii}},\ }\href {https://doi.org/10.1103/PhysRevC.109.054323} {\bibfield  {journal} {\bibinfo  {journal} {Phys. Rev. C}\ }\textbf {\bibinfo {volume} {109}},\ \bibinfo {pages} {054323} (\bibinfo {year} {2024})}\BibitemShut {NoStop}%
\bibitem [{\citenamefont {Lalazissis}\ \emph {et~al.}(1997)\citenamefont {Lalazissis}, \citenamefont {K\"onig},\ and\ \citenamefont {Ring}}]{NL3}%
  \BibitemOpen
  \bibfield  {author} {\bibinfo {author} {\bibfnamefont {G.~A.}\ \bibnamefont {Lalazissis}}, \bibinfo {author} {\bibfnamefont {J.}~\bibnamefont {K\"onig}},\ and\ \bibinfo {author} {\bibfnamefont {P.}~\bibnamefont {Ring}},\ }\bibfield  {title} {\bibinfo {title} {{New parametrization for the Lagrangian density of relativistic mean field theory}},\ }\href {https://doi.org/10.1103/PhysRevC.55.540} {\bibfield  {journal} {\bibinfo  {journal} {Phys. Rev. C}\ }\textbf {\bibinfo {volume} {55}},\ \bibinfo {pages} {540} (\bibinfo {year} {1997})}\BibitemShut {NoStop}%
\bibitem [{\citenamefont {Chen}\ and\ \citenamefont {Piekarewicz}(2014)}]{fsugold2}%
  \BibitemOpen
  \bibfield  {author} {\bibinfo {author} {\bibfnamefont {W.-C.}\ \bibnamefont {Chen}}\ and\ \bibinfo {author} {\bibfnamefont {J.}~\bibnamefont {Piekarewicz}},\ }\bibfield  {title} {\bibinfo {title} {{Building relativistic mean field models for finite nuclei and neutron stars}},\ }\href {https://doi.org/10.1103/PhysRevC.90.044305} {\bibfield  {journal} {\bibinfo  {journal} {Phys. Rev. C}\ }\textbf {\bibinfo {volume} {90}},\ \bibinfo {pages} {044305} (\bibinfo {year} {2014})}\BibitemShut {NoStop}%
\bibitem [{\citenamefont {M\"{o}ller}\ \emph {et~al.}(2016)\citenamefont {M\"{o}ller}, \citenamefont {Sierk}, \citenamefont {Ichikawa},\ and\ \citenamefont {Sagawa}}]{frdm12}%
  \BibitemOpen
  \bibfield  {author} {\bibinfo {author} {\bibfnamefont {P.}~\bibnamefont {M\"{o}ller}}, \bibinfo {author} {\bibfnamefont {A.~J.}\ \bibnamefont {Sierk}}, \bibinfo {author} {\bibfnamefont {T.}~\bibnamefont {Ichikawa}},\ and\ \bibinfo {author} {\bibfnamefont {H.}~\bibnamefont {Sagawa}},\ }\bibfield  {title} {\bibinfo {title} {{Nuclear ground-state masses and deformations: FRDM (2012)}},\ }\href {https://doi.org/10.1016/j.adt.2015.10.002} {\bibfield  {journal} {\bibinfo  {journal} {At. Data Nucl. Data Tables}\ }\textbf {\bibinfo {volume} {109}},\ \bibinfo {pages} {1} (\bibinfo {year} {2016})}\BibitemShut {NoStop}%
\bibitem [{\citenamefont {Naito}\ \emph {et~al.}(2023{\natexlab{b}})\citenamefont {Naito}, \citenamefont {Oishi}, \citenamefont {Sagawa},\ and\ \citenamefont {Wang}}]{Naito2023Phys.Rev.C107_054307}%
  \BibitemOpen
  \bibfield  {author} {\bibinfo {author} {\bibfnamefont {T.}~\bibnamefont {Naito}}, \bibinfo {author} {\bibfnamefont {T.}~\bibnamefont {Oishi}}, \bibinfo {author} {\bibfnamefont {H.}~\bibnamefont {Sagawa}},\ and\ \bibinfo {author} {\bibfnamefont {Z.}~\bibnamefont {Wang}},\ }\bibfield  {title} {\bibinfo {title} {{Comparative study on charge radii and their kinks at magic numbers}},\ }\href {https://doi.org/10.1103/PhysRevC.107.054307} {\bibfield  {journal} {\bibinfo  {journal} {Phys. Rev. C}\ }\textbf {\bibinfo {volume} {107}},\ \bibinfo {pages} {054307} (\bibinfo {year} {2023}{\natexlab{b}})}\BibitemShut {NoStop}%
\bibitem [{\citenamefont {Tanimura}\ and\ \citenamefont {Hagino}(2012)}]{TaHa12}%
  \BibitemOpen
  \bibfield  {author} {\bibinfo {author} {\bibfnamefont {Y.}~\bibnamefont {Tanimura}}\ and\ \bibinfo {author} {\bibfnamefont {K.}~\bibnamefont {Hagino}},\ }\bibfield  {title} {\bibinfo {title} {{Description of single-$\ensuremath{\Lambda}$ hypernuclei with a relativistic point-coupling model}},\ }\href {https://doi.org/10.1103/PhysRevC.85.014306} {\bibfield  {journal} {\bibinfo  {journal} {Phys. Rev. C}\ }\textbf {\bibinfo {volume} {85}},\ \bibinfo {pages} {014306} (\bibinfo {year} {2012})}\BibitemShut {NoStop}%
\bibitem [{\citenamefont {B\"urvenich}\ \emph {et~al.}(2002)\citenamefont {B\"urvenich}, \citenamefont {Madland}, \citenamefont {Maruhn},\ and\ \citenamefont {Reinhard}}]{BMMR02}%
  \BibitemOpen
  \bibfield  {author} {\bibinfo {author} {\bibfnamefont {T.}~\bibnamefont {B\"urvenich}}, \bibinfo {author} {\bibfnamefont {D.~G.}\ \bibnamefont {Madland}}, \bibinfo {author} {\bibfnamefont {J.~A.}\ \bibnamefont {Maruhn}},\ and\ \bibinfo {author} {\bibfnamefont {P.-G.}\ \bibnamefont {Reinhard}},\ }\bibfield  {title} {\bibinfo {title} {{Nuclear ground state observables and QCD scaling in a refined relativistic point coupling model}},\ }\href {https://doi.org/10.1103/PhysRevC.65.044308} {\bibfield  {journal} {\bibinfo  {journal} {Phys. Rev. C}\ }\textbf {\bibinfo {volume} {65}},\ \bibinfo {pages} {044308} (\bibinfo {year} {2002})}\BibitemShut {NoStop}%
\bibitem [{\citenamefont {Naito}\ \emph {et~al.}(2024)\citenamefont {Naito}, \citenamefont {Col\`{o}}, \citenamefont {Hatsuda}, \citenamefont {Liang}, \citenamefont {Roca-Maza},\ and\ \citenamefont {Sagawa}}]{NaGi24}%
  \BibitemOpen
  \bibfield  {author} {\bibinfo {author} {\bibfnamefont {T.}~\bibnamefont {Naito}}, \bibinfo {author} {\bibfnamefont {G.}~\bibnamefont {Col\`{o}}}, \bibinfo {author} {\bibfnamefont {T.}~\bibnamefont {Hatsuda}}, \bibinfo {author} {\bibfnamefont {H.}~\bibnamefont {Liang}}, \bibinfo {author} {\bibfnamefont {X.}~\bibnamefont {Roca-Maza}},\ and\ \bibinfo {author} {\bibfnamefont {H.}~\bibnamefont {Sagawa}},\ }\bibfield  {title} {\bibinfo {title} {{Possible inconsistency between phenomenological and theoretical determinations of charge symmetry breaking in nuclear energy density functionals}},\ }\href {https://doi.org/10.1393/ncc/i2024-24052-9} {\bibfield  {journal} {\bibinfo  {journal} {Nuovo Cim. C}\ }\textbf {\bibinfo {volume} {47}},\ \bibinfo {pages} {52} (\bibinfo {year} {2024})}\BibitemShut {NoStop}%
\bibitem [{\citenamefont {Botta}\ \emph {et~al.}(2017)\citenamefont {Botta}, \citenamefont {Bressani},\ and\ \citenamefont {Feliciello}}]{Botta17}%
  \BibitemOpen
  \bibfield  {author} {\bibinfo {author} {\bibfnamefont {E.}~\bibnamefont {Botta}}, \bibinfo {author} {\bibfnamefont {T.}~\bibnamefont {Bressani}},\ and\ \bibinfo {author} {\bibfnamefont {A.}~\bibnamefont {Feliciello}},\ }\bibfield  {title} {\bibinfo {title} {{On the binding energy and the charge symmetry breaking in $A\leq 16$ $\Lambda$-hypernuclei}},\ }\href {https://doi.org/10.1016/j.nuclphysa.2017.02.005} {\bibfield  {journal} {\bibinfo  {journal} {Nucl. Phys. A}\ }\textbf {\bibinfo {volume} {960}},\ \bibinfo {pages} {165} (\bibinfo {year} {2017})}\BibitemShut {NoStop}%
\bibitem [{\citenamefont {Sun}\ \emph {et~al.}(2025)\citenamefont {Sun}, \citenamefont {Tanimura}, \citenamefont {Sagawa},\ and\ \citenamefont {Hiyama}}]{STSH25}%
  \BibitemOpen
  \bibfield  {author} {\bibinfo {author} {\bibfnamefont {T.-T.}\ \bibnamefont {Sun}}, \bibinfo {author} {\bibfnamefont {Y.}~\bibnamefont {Tanimura}}, \bibinfo {author} {\bibfnamefont {H.}~\bibnamefont {Sagawa}},\ and\ \bibinfo {author} {\bibfnamefont {E.}~\bibnamefont {Hiyama}},\ }\bibfield  {title} {\bibinfo {title} {{Charge symmetry breaking in hypernuclei within RMF model}},\ }\href {https://doi.org/10.1016/j.physletb.2025.139460} {\bibfield  {journal} {\bibinfo  {journal} {Phys. Lett. B}\ }\textbf {\bibinfo {volume} {865}},\ \bibinfo {pages} {139460} (\bibinfo {year} {2025})}\BibitemShut {NoStop}%
\bibitem [{\citenamefont {Dalitz}\ and\ \citenamefont {Von~Hippel}(1964)}]{Dalitz64}%
  \BibitemOpen
  \bibfield  {author} {\bibinfo {author} {\bibfnamefont {R.~H.}\ \bibnamefont {Dalitz}}\ and\ \bibinfo {author} {\bibfnamefont {F.}~\bibnamefont {Von~Hippel}},\ }\bibfield  {title} {\bibinfo {title} {{Electromagnetic $\Lambda$-$\Sigma^0$ mixing and charge symmetry for the $\Lambda$-Hyperon}},\ }\href {https://doi.org/10.1016/0031-9163(64)90617-1} {\bibfield  {journal} {\bibinfo  {journal} {Phys. Lett.}\ }\textbf {\bibinfo {volume} {10}},\ \bibinfo {pages} {153} (\bibinfo {year} {1964})}\BibitemShut {NoStop}%
\bibitem [{\citenamefont {Adhikari}\ \emph {et~al.}(2022)\citenamefont {Adhikari}, \citenamefont {Albataineh}, \citenamefont {Androic}, \citenamefont {Aniol}, \citenamefont {Armstrong}, \citenamefont {Averett}, \citenamefont {Ayerbe~Gayoso}, \citenamefont {Barcus}, \citenamefont {Bellini}, \citenamefont {Beminiwattha}, \citenamefont {Benesch}, \citenamefont {Bhatt}, \citenamefont {Bhatta~Pathak}, \citenamefont {Bhetuwal}, \citenamefont {Blaikie}, \citenamefont {Boyd}, \citenamefont {Campagna}, \citenamefont {Camsonne}, \citenamefont {Cates}, \citenamefont {Chen}, \citenamefont {Clarke}, \citenamefont {Cornejo}, \citenamefont {Covrig~Dusa}, \citenamefont {Dalton}, \citenamefont {Datta}, \citenamefont {Deshpande}, \citenamefont {Dutta}, \citenamefont {Feldman}, \citenamefont {Fuchey}, \citenamefont {Gal}, \citenamefont {Gaskell}, \citenamefont {Gautam}, \citenamefont {Gericke}, \citenamefont {Ghosh}, \citenamefont {Halilovic}, \citenamefont {Hansen}, \citenamefont {Hauenstein}, \citenamefont {Henry}, \citenamefont {Horowitz}, \citenamefont {Jantzi}, \citenamefont {Jian}, \citenamefont {Johnston}, \citenamefont {Jones}, \citenamefont {Karki}, \citenamefont {Kakkar}, \citenamefont {Katugampola}, \citenamefont {Keppel}, \citenamefont {King}, \citenamefont {King}, \citenamefont {Knauss}, \citenamefont {Kumar}, \citenamefont {Kutz}, \citenamefont {Lashley-Colthirst}, \citenamefont {Leverick}, \citenamefont {Liu}, \citenamefont {Liyange}, \citenamefont {Malace}, \citenamefont {Mammei}, \citenamefont {Mammei}, \citenamefont {McCaughan}, \citenamefont {McNulty}, \citenamefont {Meekins}, \citenamefont {Metts}, \citenamefont {Michaels}, \citenamefont {Mihovilovic}, \citenamefont {Mondal}, \citenamefont {Napolitano}, \citenamefont {Nikolaev}, \citenamefont {Rashad}, \citenamefont {Owen}, \citenamefont {Palatchi}, \citenamefont {Pan}, \citenamefont {Pandey}, \citenamefont {Park}, \citenamefont {Paschke}, \citenamefont {Petrusky}, \citenamefont {Pitt}, \citenamefont {Premathilake}, \citenamefont {Puckett}, \citenamefont {Quinn}, \citenamefont {Radloff}, \citenamefont {Rahman}, \citenamefont {Rathnayake}, \citenamefont {Reed}, \citenamefont {Reimer}, \citenamefont {Richards}, \citenamefont {Riordan}, \citenamefont {Roblin}, \citenamefont {Seeds}, \citenamefont {Shahinyan}, \citenamefont {Souder}, \citenamefont {Tang}, \citenamefont {Thiel}, \citenamefont {Tian}, \citenamefont {Urciuoli}, \citenamefont {Wertz}, \citenamefont {Wojtsekhowski}, \citenamefont {Xiong}, \citenamefont {Yale}, \citenamefont {Ye}, \citenamefont {Zec}, \citenamefont {Zhang}, \citenamefont {Zhang},\ and\ \citenamefont {Zheng}}]{crex}%
  \BibitemOpen
  \bibfield  {author} {\bibinfo {author} {\bibfnamefont {D.}~\bibnamefont {Adhikari}}, \bibinfo {author} {\bibfnamefont {H.}~\bibnamefont {Albataineh}}, \bibinfo {author} {\bibfnamefont {D.}~\bibnamefont {Androic}}, \bibinfo {author} {\bibfnamefont {K.}~\bibnamefont {Aniol}}, \bibinfo {author} {\bibfnamefont {D.~S.}\ \bibnamefont {Armstrong}}, \bibinfo {author} {\bibfnamefont {T.}~\bibnamefont {Averett}}, \bibinfo {author} {\bibfnamefont {C.}~\bibnamefont {Ayerbe~Gayoso}}, \bibinfo {author} {\bibfnamefont {S.}~\bibnamefont {Barcus}}, \bibinfo {author} {\bibfnamefont {V.}~\bibnamefont {Bellini}}, \bibinfo {author} {\bibfnamefont {R.~S.}\ \bibnamefont {Beminiwattha}}, \bibinfo {author} {\bibfnamefont {J.~F.}\ \bibnamefont {Benesch}}, \bibinfo {author} {\bibfnamefont {H.}~\bibnamefont {Bhatt}}, \bibinfo {author} {\bibfnamefont {D.}~\bibnamefont {Bhatta~Pathak}}, \bibinfo {author} {\bibfnamefont {D.}~\bibnamefont {Bhetuwal}}, \bibinfo {author} {\bibfnamefont {B.}~\bibnamefont {Blaikie}}, \bibinfo {author} {\bibfnamefont {J.}~\bibnamefont {Boyd}}, \bibinfo {author} {\bibfnamefont {Q.}~\bibnamefont {Campagna}}, \bibinfo {author} {\bibfnamefont {A.}~\bibnamefont {Camsonne}}, \bibinfo {author} {\bibfnamefont {G.~D.}\ \bibnamefont {Cates}}, \bibinfo {author} {\bibfnamefont {Y.}~\bibnamefont {Chen}}, \bibinfo {author} {\bibfnamefont {C.}~\bibnamefont {Clarke}}, \bibinfo {author} {\bibfnamefont {J.~C.}\ \bibnamefont {Cornejo}}, \bibinfo {author} {\bibfnamefont {S.}~\bibnamefont {Covrig~Dusa}}, \bibinfo {author} {\bibfnamefont {M.~M.}\ \bibnamefont {Dalton}}, \bibinfo {author} {\bibfnamefont {P.}~\bibnamefont {Datta}}, \bibinfo {author} {\bibfnamefont {A.}~\bibnamefont {Deshpande}}, \bibinfo {author} {\bibfnamefont {D.}~\bibnamefont {Dutta}}, \bibinfo {author} {\bibfnamefont {C.}~\bibnamefont {Feldman}}, \bibinfo {author} {\bibfnamefont {E.}~\bibnamefont {Fuchey}}, \bibinfo {author} {\bibfnamefont {C.}~\bibnamefont {Gal}}, \bibinfo {author} {\bibfnamefont {D.}~\bibnamefont {Gaskell}}, \bibinfo {author} {\bibfnamefont {T.}~\bibnamefont {Gautam}}, \bibinfo {author} {\bibfnamefont {M.}~\bibnamefont {Gericke}}, \bibinfo {author} {\bibfnamefont {C.}~\bibnamefont {Ghosh}}, \bibinfo {author} {\bibfnamefont {I.}~\bibnamefont {Halilovic}}, \bibinfo {author} {\bibfnamefont {J.-O.}\ \bibnamefont {Hansen}}, \bibinfo {author} {\bibfnamefont {F.}~\bibnamefont {Hauenstein}}, \bibinfo {author} {\bibfnamefont {W.}~\bibnamefont {Henry}}, \bibinfo {author} {\bibfnamefont {C.~J.}\ \bibnamefont {Horowitz}}, \bibinfo {author} {\bibfnamefont {C.}~\bibnamefont {Jantzi}}, \bibinfo {author} {\bibfnamefont {S.}~\bibnamefont {Jian}}, \bibinfo {author} {\bibfnamefont {S.}~\bibnamefont {Johnston}}, \bibinfo {author} {\bibfnamefont {D.~C.}\ \bibnamefont {Jones}}, \bibinfo {author} {\bibfnamefont {B.}~\bibnamefont {Karki}}, \bibinfo {author} {\bibfnamefont {S.}~\bibnamefont {Kakkar}}, \bibinfo {author} {\bibfnamefont {S.}~\bibnamefont {Katugampola}}, \bibinfo {author} {\bibfnamefont {C.~E.}\ \bibnamefont {Keppel}}, \bibinfo {author} {\bibfnamefont {P.~M.}\ \bibnamefont {King}}, \bibinfo {author} {\bibfnamefont {D.~E.}\ \bibnamefont {King}}, \bibinfo {author} {\bibfnamefont {M.}~\bibnamefont {Knauss}}, \bibinfo {author} {\bibfnamefont {K.~S.}\ \bibnamefont {Kumar}}, \bibinfo {author} {\bibfnamefont {T.}~\bibnamefont {Kutz}}, \bibinfo {author} {\bibfnamefont {N.}~\bibnamefont {Lashley-Colthirst}}, \bibinfo {author} {\bibfnamefont {G.}~\bibnamefont {Leverick}}, \bibinfo {author} {\bibfnamefont {H.}~\bibnamefont {Liu}}, \bibinfo {author} {\bibfnamefont {N.}~\bibnamefont {Liyange}}, \bibinfo {author} {\bibfnamefont {S.}~\bibnamefont {Malace}}, \bibinfo {author} {\bibfnamefont {J.}~\bibnamefont {Mammei}}, \bibinfo {author} {\bibfnamefont {R.}~\bibnamefont {Mammei}}, \bibinfo {author} {\bibfnamefont {M.}~\bibnamefont {McCaughan}}, \bibinfo {author} {\bibfnamefont {D.}~\bibnamefont {McNulty}}, \bibinfo {author} {\bibfnamefont {D.}~\bibnamefont {Meekins}}, \bibinfo {author} {\bibfnamefont {C.}~\bibnamefont {Metts}}, \bibinfo {author} {\bibfnamefont {R.}~\bibnamefont {Michaels}}, \bibinfo {author} {\bibfnamefont {M.}~\bibnamefont {Mihovilovic}}, \bibinfo {author} {\bibfnamefont {M.~M.}\ \bibnamefont {Mondal}}, \bibinfo {author} {\bibfnamefont {J.}~\bibnamefont {Napolitano}}, \bibinfo {author} {\bibfnamefont {D.}~\bibnamefont {Nikolaev}}, \bibinfo {author} {\bibfnamefont {M.~N.~H.}\ \bibnamefont {Rashad}}, \bibinfo {author} {\bibfnamefont {V.}~\bibnamefont {Owen}}, \bibinfo {author} {\bibfnamefont {C.}~\bibnamefont {Palatchi}}, \bibinfo {author} {\bibfnamefont {J.}~\bibnamefont {Pan}}, \bibinfo {author} {\bibfnamefont {B.}~\bibnamefont {Pandey}}, \bibinfo {author} {\bibfnamefont {S.}~\bibnamefont {Park}}, \bibinfo {author} {\bibfnamefont {K.~D.}\ \bibnamefont {Paschke}}, \bibinfo {author} {\bibfnamefont {M.}~\bibnamefont {Petrusky}}, \bibinfo {author} {\bibfnamefont {M.~L.}\ \bibnamefont {Pitt}}, \bibinfo {author} {\bibfnamefont {S.}~\bibnamefont {Premathilake}}, \bibinfo {author} {\bibfnamefont {A.~J.~R.}\ \bibnamefont {Puckett}}, \bibinfo {author} {\bibfnamefont {B.}~\bibnamefont {Quinn}}, \bibinfo {author} {\bibfnamefont {R.}~\bibnamefont {Radloff}}, \bibinfo {author} {\bibfnamefont {S.}~\bibnamefont {Rahman}}, \bibinfo {author} {\bibfnamefont {A.}~\bibnamefont {Rathnayake}}, \bibinfo {author} {\bibfnamefont {B.~T.}\ \bibnamefont {Reed}}, \bibinfo {author} {\bibfnamefont {P.~E.}\ \bibnamefont {Reimer}}, \bibinfo {author} {\bibfnamefont {R.}~\bibnamefont {Richards}}, \bibinfo {author} {\bibfnamefont {S.}~\bibnamefont {Riordan}}, \bibinfo {author} {\bibfnamefont {Y.}~\bibnamefont {Roblin}}, \bibinfo {author} {\bibfnamefont {S.}~\bibnamefont {Seeds}}, \bibinfo {author} {\bibfnamefont {A.}~\bibnamefont {Shahinyan}}, \bibinfo {author} {\bibfnamefont {P.~A.}\ \bibnamefont {Souder}}, \bibinfo {author} {\bibfnamefont {L.}~\bibnamefont {Tang}}, \bibinfo {author} {\bibfnamefont {M.}~\bibnamefont {Thiel}}, \bibinfo {author} {\bibfnamefont {Y.}~\bibnamefont {Tian}}, \bibinfo {author} {\bibfnamefont {G.~M.}\ \bibnamefont {Urciuoli}}, \bibinfo {author} {\bibfnamefont {E.~W.}\ \bibnamefont {Wertz}}, \bibinfo {author} {\bibfnamefont {B.}~\bibnamefont {Wojtsekhowski}}, \bibinfo {author} {\bibfnamefont {W.}~\bibnamefont {Xiong}}, \bibinfo {author} {\bibfnamefont {B.}~\bibnamefont {Yale}}, \bibinfo {author} {\bibfnamefont {T.}~\bibnamefont {Ye}}, \bibinfo {author} {\bibfnamefont {A.}~\bibnamefont {Zec}}, \bibinfo {author} {\bibfnamefont {W.}~\bibnamefont {Zhang}}, \bibinfo {author} {\bibfnamefont {J.}~\bibnamefont {Zhang}},\ and\ \bibinfo {author} {\bibfnamefont {X.}~\bibnamefont {Zheng}} (\bibinfo {collaboration} {PREX and CREX Collaborations}),\ }\bibfield  {title} {\bibinfo {title} {{New Measurements of the Beam-Normal Single Spin Asymmetry in Elastic Electron Scattering over a Range of Spin-0 Nuclei}},\ }\href {https://doi.org/10.1103/PhysRevLett.128.142501} {\bibfield  {journal} {\bibinfo  {journal} {Phys. Rev. Lett.}\ }\textbf {\bibinfo {volume} {128}},\ \bibinfo {pages} {142501} (\bibinfo {year} {2022})}\BibitemShut {NoStop}%
\bibitem [{\citenamefont {Roca-Maza}\ \emph {et~al.}(2011)\citenamefont {Roca-Maza}, \citenamefont {Centelles}, \citenamefont {Vi\~nas},\ and\ \citenamefont {Warda}}]{RCVW11}%
  \BibitemOpen
  \bibfield  {author} {\bibinfo {author} {\bibfnamefont {X.}~\bibnamefont {Roca-Maza}}, \bibinfo {author} {\bibfnamefont {M.}~\bibnamefont {Centelles}}, \bibinfo {author} {\bibfnamefont {X.}~\bibnamefont {Vi\~nas}},\ and\ \bibinfo {author} {\bibfnamefont {M.}~\bibnamefont {Warda}},\ }\bibfield  {title} {\bibinfo {title} {{Neutron Skin of $^{208}\mathrm{Pb}$, Nuclear Symmetry Energy, and the Parity Radius Experiment}},\ }\href {https://doi.org/10.1103/PhysRevLett.106.252501} {\bibfield  {journal} {\bibinfo  {journal} {Phys. Rev. Lett.}\ }\textbf {\bibinfo {volume} {106}},\ \bibinfo {pages} {252501} (\bibinfo {year} {2011})}\BibitemShut {NoStop}%
\end{thebibliography}%
\end{document}